\documentclass{fac}

\usepackage{psfrag}
\usepackage{epsfig}
\usepackage{graphicx}
\usepackage{amsfonts}
\usepackage{comment}
\usepackage{amssymb}
\usepackage{amsmath}
\usepackage{subfigure}
\usepackage{paralist}
\usepackage{mdwlist}
\usepackage{rotating}
\usepackage{algorithm}
\usepackage{algorithmic}
\usepackage{rotating}
\usepackage{color}
\usepackage{tcolorbox}
\usepackage{multicol}

\ifprodtf \else \usepackage{latexsym}\fi

\newcommand\black{\ensuremath{\blacktriangleright}}
\newcommand\white{\ensuremath{\vartriangleright}}

\newif\ifamsfontsloaded
\ifprodtf
  \newcommand\whbl{\white\kern-.1em--\kern-.1em\black}
  \newcommand\blwh{\black\kern-.1em--\kern-.1em\white}
  \newcommand\blbl{\black\kern-.1em--\kern-.1em\black}
  \newcommand\whwh{\white\kern-.1em--\kern-.1em\white}
  \amsfontsloadedtrue
\else
  \checkfont{msam10}
  \iffontfound
    \IfFileExists{amssymb.sty}
      {\usepackage{amssymb}\amsfontsloadedtrue
       \newcommand\whbl{\white\kern-.125em--\kern-.125em\black}%
       \newcommand\blwh{\black\kern-.125em--\kern-.125em\white}%
       \newcommand\blbl{\black\kern-.125em--\kern-.125em\black}%
       \newcommand\whwh{\white\kern-.125em--\kern-.125em\white}}
      {}
  \fi
\fi

\newcommand{\qed}{\hspace*{\fill}$\Box$}

\newtheorem{example}{Example}

\title[Formal Methods for Characterization and Analysis of Quality
Specifications in Component-based Systems]
      {Formal Methods for Characterization and Analysis of Quality
Specifications in Component-based Systems}

\author[Aritra Hazra]
    {Aritra Hazra\\
     Department of Computer Science and Engineering, Indian Institute of Technology Kharagpur, India.}

\correspond{Aritra Hazra, Department of Computer Science and Engineering, Indian Institute of Technology Kharagpur, Paschim Medinipur, West Bengal, India, PIN -- 721302.
            E-mail: aritrah@cse.iitkgp.ac.in}

\pubyear{2000}
\pagerange{\pageref{firstpage}--\pageref{lastpage}}

\begin{document}
\label{firstpage}

\makecorrespond

\maketitle

\begin{abstract}
Component-based design paradigm is of paramount importance due to prolific 
growth in the complexity of modern-day systems. Since the components are 
developed primarily by multi-party vendors and often assembled to realize the 
overall system, it is an onus of the designer to certify both the functional 
and non-functional requirements of such systems. Several of the earlier works 
concentrated on formally analyzing the behavioral correctness, safety, 
security, reliability and robustness of such compositional systems. However, 
the assurance for quality measures of such systems is also considered as an 
important parameter for their acceptance. Formalization of quality measures is 
still at an immature state and often dictated by the user satisfaction. This 
paper presents a novel compositional framework for reliable quality analysis of 
component-based systems from the formal quality specifications of its 
constituent components. The proposed framework enables elegant and generic 
computation methods for quality attributes of various component-based system 
structures. In addition to this, we provide a formal query-driven quality 
assessment and design exploration framework which enables the designer to 
explore various component structures and operating setups and finally converge 
into better acceptable systems. A detailed case-study is presented over a 
component-based system structure to show the efficacy and practicality of our 
proposed framework.
\end{abstract}


\section{Introduction} \label{sec:introduction}
Component-based design engineering is of paramount importance across all 
engineering disciplines where complex systems can be obtained by assembling 
components as basic building blocks~\cite{FV1999ICCD, KSLB2003IEEE, LS2011DATE, 
R2009BOOK}. Such paradigms are becoming prevalent due to increasing complexity 
of hardware and software systems and components requiring procurement from 
multi-party vendors for product completion. Components are design abstractions 
that ignore implementation details and are conceptualized formally using 
behavioral, interaction and execution models. These models are, then, 
hierarchically composed to build the overall system architecture where the 
formalism for the composite model is attributed from the notion of transfer 
functions, interfaces and contracts~\cite{BBBBS2011MEMICS, BBBCJNS2011IEEESW, 
BBBBSH2011MEMOCODE, S2012FTEDA}. Component-based engineering broadly involves 
model-based development of systems~\cite{S2003IEEESW, SK1997IEEECOMP}, 
platform-based design methodologies~\cite{SP2022BOOK} and developing software 
modules, supported by a large number of existing tools and standards, 
involving object-oriented languages (such as C++, System-C) and modeling 
languages (such as Stateflow/Statechart, UML, SysML)~\cite{B2006SE, 
GLMS2002BOOK, H1987SCP}.

The computer scientists and researchers have responded to the challenge of 
designing complex systems adopting various formal methods and compositional 
algebra based frameworks~\cite{S2012FTEDA, SBBB2004BOOKCHAP}. Typically, the 
component interactions and concurrency have been modeled formally using 
automated compositional reasoning and connector algebra~\cite{BS2008CONCUR, 
BS2008FMSD, SC2013IEEECOMP}. With the wider adoption of such structured 
frameworks for system modeling, it also becomes a primary concern for the 
designers to ensure the correct operability of composite systems. Over the past 
few decades, model-based testing~\cite{UL2007BOOK, ZSM2011BOOK} and model 
checking paradigms are explored to ensure the functional correctness, timing and 
safety behaviors~\cite{CGKPV2018BOOK, CHVB2018BOOK}. Additionally, several 
formal methods are proposed to enforce correct-by-construction approaches while 
designing system-level composite architectures~\cite{GS2002FMCO, PBBJS2015FAC}. 
In recent times, the certification regime has also been extended towards formal 
assurance of the non-functional requirements as well, such as 
power~\cite{HGDP2013TVLSI, HMDPHBM2013TCAD}, reliability~\cite{GHD2015VLSID, 
HDC2016JAL, HGVCD2013ESL} and security~\cite{CS2010JCS, HHS2008ICARS, 
KRRH2019BOOKCHAPTER}, for component-based systems. Though the composition 
modeling is primarily formalized by the help of behavioral component models, 
formal component interactions and architectural properties, but there is a gap 
in modeling and assessment of system-level quality from the quality standards of 
its constituent components. 

Quality measures of a system (often considered as a non-functional attribute) 
primarily reflects how perfectly an operation can be performed by the system and 
it is often attributed from the composed quality goals attained by its 
constituent components. It leads to the satisfaction of the users who are being 
serviced by the output of that system. A real-life example can be witnessed in 
case of video rendering (one can consider YouTube as an example) during the 
transmission/streaming where multiple levels of video quality can be produced. 
Typically, a high quality video input is being processed in multiple operating 
modes and the output video is produced in various quality levels depending on 
the bandwidth of the transmission and receiver channel. The two interesting 
non-functional behaviors in such a procedure is the reliability (or the 
availability) of the output video and also its quality parameters. The notion of 
reliability and availability in system design has been extensively 
studied~\cite{KK2007BOOK}. However, at recent times, there is a growing need to 
incorporate the quality measures also into the main-stream formal system design 
process. The existing formalism for component-based rigorous system 
designs~\cite{S2012FTEDA} do not consider into account quality attributes and 
associated specifications into their present modeling setup, hence the notion 
and formal treatment of quality measures in component-based systems are still 
remaining at a premature stage.


As per the best of our knowledge, there are very few works in formal system 
quality analysis. Some of the early works have subjectively quantified the 
notion of quality of systems and informally established the same by interaction 
of components and by following strict designing principles~\cite{CLWK2000APSEC}. 
In embedded system design parlance, architectural quality assurance frameworks, 
leveraging na\"{i}ve formal verification and model-based techniques, are 
formalized in~\cite{J2018THESIS, LLVC2018IINTEC}. The notion of symbolic quality 
and its control was introduced in~\cite{CFLS2005ICES}, where the authors propose 
an optimal quality schedule to ensure Quality-of-Service (QoS). On the same 
line, the work presented in~\cite{CFSS2008RTS} proposes a fine grain symbolic 
quality control method for multimedia applications using speed diagrams. The 
proposed methods takes as input an application software composed of actions, 
whose execution times are unknown increasing functions considering quality level 
parameters (considered as integers). The quality controller is able to compute 
adequate action schedules and choose pre-defined quality levels, in order to 
meet QoS requirements for a given platform. However, all these existing works do 
not provide the formal basis for designing and reasoning system involving 
component-level quality compositions; rather these methods invokes a quality 
manager to change action quality levels based on the knowledge of control 
constraints and finally computing a set of optimal schedules improving overall 
system performance.

Formal characterization of system quality has several deeper underpinnings in 
relation with the underlying system behavior. The choice of quality may have 
certain dependence with the system reliability, since some good quality 
component failure may degrade the overall quality of the system as well. So, 
formalizing the notions of quality should also incorporate the component-based 
system reliability formalism into its characterization. Existing literature 
provides several stand-alone techniques for calculating the reliability of 
systems based on symbolic/algebraic compositions of component-level reliability 
expressions from its given component structures~\cite{KK2007BOOK}. However, 
there is a lack of effort in formally capturing quality measures for a design in 
similar terms, and thereby building a suitable framework to assess the 
compositional quality of component-based systems. This article is an enabler in 
this direction.

\begin{figure}
\centering
\includegraphics[width=\textwidth]{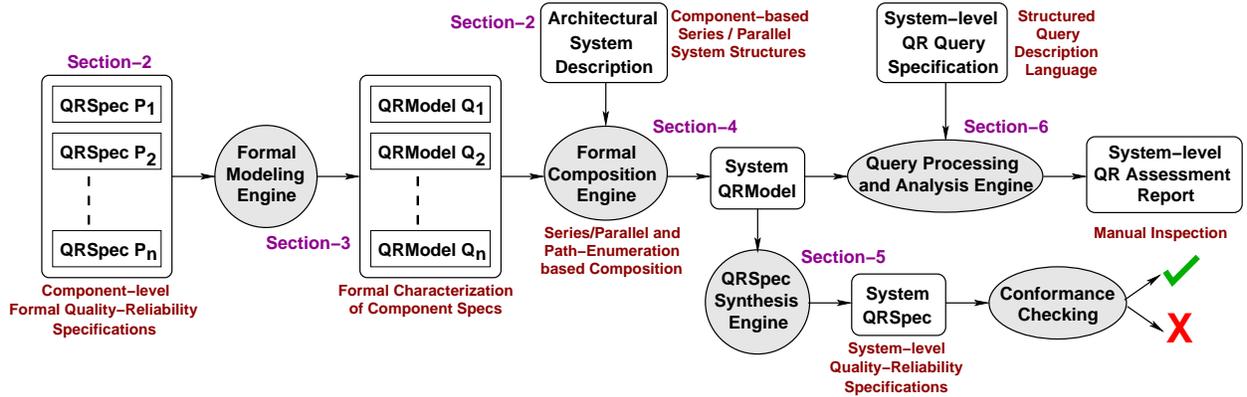}
\caption{The Overall Framework of Characterization and Assessment of Quality 
Specifications}
\label{fig:framework}
\end{figure}

Formally, we define the quality of a component as a set of random variables each
associated with a probability value. We assume that for a given minimum input
quality value/level, the corresponding minimum output quality value/level for
any component is associated with a reliability value, which dictates its
probability of successful execution without compromising the mentioned quality.
The set of quality levels with different reliability values for a component
arises due to a bound on limited execution time and cost of the component
comprising the system~\cite{CFSS2008RTS}. Apart from the failure that can occur
in a component structure, its operation can also be manually suspended by 
controlling it (for example, switching it off or executing it in a low 
functional/power modes) systematically. Such control of component suspension is 
very important at present context, considering the criticality of power, 
security and other performance criteria under restricted design setup. Clearly, 
a component has the output quality zero when its reliability is also zero 
(irrespective of input quality), meaning that failure is imminent, or its 
operation has been suspended (controlled). So, it is imperative to formulate 
algebra for component quality and reliability composition in order to reason 
about the quality of a system constructed hierarchically from set of component 
structures.

Given a set of formal quality and reliability specifications for a set of 
components, it is non-trivial to define the quality and reliability measures of 
the composite system structure. The composition is not similar to reliability 
analysis, since there is a choice here to maximize the overall quality during 
the compositional exploration considering the functional reliability of its 
constituent components (and their failures). This choice is complicated because 
there can be multiple ways during system execution to produce the range of 
quality expressions. For example, when several components, each having a set of 
quality specifications, are all connected in parallel, the overall quality 
measure of such a setup may be generated by -- (i) computing the highest 
quality value/level among the operational components every time, (ii) assuming 
a pre-selected order among the operability of the constituent components, or 
(iii) reporting the quality from a fixed component (for simplicity) and the 
quality becomes zero when that component fails -- thereby keeping the 
possibility for varied range of behavioral options. On the other hand, for 
series composition, the quality value is dictated by the series of components 
and becomes zero if any one of the components fails.  Apart from such choices, 
the option to suspend/control the component execution in any operating mode 
brings inherent challenges in assessment of overall system quality attributes. 

This work proposes novel methods for formal characterization and assessment of 
quality specifications for component-based systems that are built involving 
of series and parallel compositions of component structures at an architectural 
level. Figure~\ref{fig:framework} presents the overall flow (steps) of the work 
presented in this article.
To start with, the component-level quality and reliability specifications 
({\tt QRSpec}) are drafted, which are then formally characterized into a set of 
component-level {\tt QRModel}. Next, depending on the system structure, a 
generic series and parallel composition algebra is followed to formally compose 
these component models into a composite system-level {\tt QRModel}. Now, the 
system level (architectural) {\tt QRSpec} can also be reverse-synthesized, 
which may help in -- (i) understanding the non-functional performance boundaries 
(and extremities) of the system under various component-level operating 
conditions, and (ii) performing a conformance check with respect to the given 
system configuration, in case the system specifications are known a priori, to 
compare and know the functional coverage of the system. However, {\tt QRSpec} 
only defines an abstraction in terms of specifying the underlying detailed {\tt 
QRModel} of the system. Many other interesting observations can be gleaned from 
{\tt QRModel} that are not reflected directly through {\tt QRSpec}, and these 
will be useful for the designers, who may want to explore various structures and 
combinations of available component to converge into the best possible system 
architecture. For that, the designer are given an option to formulate their 
queries over the system attributes systematically using a proposed structured 
query description language ({\tt SQDL}), using which the information gets 
automatically extracted from the {\tt QRModel} so that system-level quality and 
reliability (non-functional) assessments can be performed.
In particular, the key contributions out of this work are:
\begin{itemize}
 \item We define the specification formalism for component-level and 
system-level quality attributes.

 \item We introduce a uniform formal characterization of quality configurations 
for components and subsystems.

 \item We propose a novel series/parallel quality composition mechanism to 
hierarchically derive quality configurations for any given system architecture 
from its component-level quality models.

 \item We derive the (abstracted) system quality specification from the formal 
representation of system quality measures through reverse-synthesis to check 
for conformance and estimate coverage.

 \item We design a query-driven framework to aid designers for better system 
exploration and quality assessment.

 \item We provide a detailed case-study to show the efficacy and practicality 
of our proposed formal framework.
\end{itemize}
The rest of the paper is organized as follows\footnote{Sections are also 
highlighted in Figure~\ref{fig:framework} to provide a meaningful organization 
to this article.}.
Section~\ref{sec:formal_model} describes the formal modeling of component-based 
systems and the notion of quality specifications for the components and 
subsystems. Section~\ref{sec:qual_config} presents the formal characterization 
of quality measures from quality specifications. Section~\ref{sec:qual_compose} 
introduces the primitive composition techniques for series and parallel system 
structures and formulate generic composition principles for systems. 
Section~\ref{sec:qual_conformance} elaborates on the synthesis of quality 
specifications from formal models and conformance checking part. 
Section~\ref{sec:qual_query_analysis} proposes the query processing 
and analysis platform for system-level assessment and design exploration. 
Section~\ref{sec:case_study} illustrates our proposed framework empirically 
over few case-studies of composite structures. Finally, 
Section~\ref{sec:conclusion} concludes the work presented in this article.

\section{Formal Model and Specification} \label{sec:formal_model}
Component-based design approaches follow the development of an overall system 
starting from unit-level component structures and laying their specifications. 
In this section, we first formalize the description of such component-based 
systems and the notion of formal quality specifications.

\subsection{Component-based Systems} \label{subsec:system}
Several components may be connected with each other in series and parallel 
structures to form a system. Formally, such a component-based system, 
$\Upsilon$, can be expressed as,
\[ \Upsilon = \langle \mathcal{I}, \mathcal{O}, \mathcal{C}, \mathcal{V},
\mathcal{E}, \mathcal{L} \rangle,\ \ \mbox{where:} \]
\begin{itemize}
 \item $\mathcal{I} = \{ I_1, I_2, \ldots, I_u \}$ denotes the set of $u$ input
nodes of the system.

 \item $\mathcal{O} = \{ O_1, O_2, \ldots, O_v \}$ denotes the set of $v$ output
nodes of the system.

 \item $\mathcal{C} = \{ C_1, C_2, \ldots, C_n \}$ denotes the set of $n$
components of the system, where each component $C_i$ (with a single input and 
single output) has their own non-functional specification, $\mathcal{P}_i$ 
(defined in next subsection).

 \item $\mathcal{V} = \{ V_1, V_2, \ldots, V_z \}$ denotes the set of 
$z$ vertices ($z \geq n$).

 \item $\mathcal{E} \subseteq (\mathcal{I} \cup \mathcal{V}) \times (\mathcal{V}
\cup \mathcal{O})$ denotes the set of edges representing the connectors among
the vertices including the input/output nodes of the system.

 \item $\mathcal{L}: \mathcal{V} \rightarrow \mathcal{C}$ is a function that
labels each vertices with a component. There may be multiple vertices labeled
with a same component indicating that the component is used in multiple places 
in the system.
\end{itemize}
The following example of a component-based system illustrates the above 
mentioned formalism in details.
\begin{example} \label{ex:comp-based_system}
Consider the example of a series/parallel component-based system, 
$\Upsilon_{sp}$, given in Figure~\ref{fig:system_example}. Here, 
$\mathcal{C} = \{C_1, C_2, C_3\}$, $\mathcal{I} = \{I_1\}$ and $\mathcal{O} = 
\{O_1\}$. The overall system architecture can be represented using a directed 
acyclic graph with the set of vertices, $\mathcal{V} = \{V_1, V_2, V_3, V_4\}$, 
and the set of edges, $\mathcal{E} = \{(I_1, V_1); (I_1, V_2); (I_1, V_3); (V_3, 
V_4);$ $(V_1, O_1); (V_2, O_1); (V_4, O_1)\}$. Moreover, the vertices are 
labeled using three components forming the system architecture as, 
$\mathcal{L}(V_1) = C_1$, $\mathcal{L}(V_2) = C_2$, $\mathcal{L}(V_3) = C_3$, 
and $\mathcal{L}(V_4) = C_2$.
\begin{figure}
    \centering
    \includegraphics[scale=0.5]{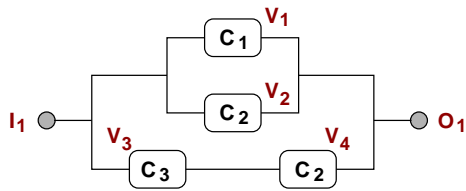}
    \caption{Component-based System Structure, $\Upsilon_{sp}$, having Series 
and Parallel Component Organization}
    \label{fig:system_example}
\end{figure}
\qed
\end{example}
A few pertinent points to note here are as follows.
\begin{itemize}
 \item The series and parallel system structure is useful in designing many 
safety-critical systems where the same input-output design functionality can 
get realized via redundant parallel paths in order to tolerate some 
intermediate component failures. Typically, the functionalities from a set of 
parallel executions/paths are converged using a voting mechanism 
(choosing consistent outcomes from $m$-out-of-$n$ paths). For example, in 
Figure~\ref{fig:system_example}, three parallel paths, i.e. via $C_1$, via 
$C_2$, or via $C_3-C_2$, may yield the same input-output functionality, though 
their way of implementation may vary (due to procurement of each component from 
different sources). From a functional perspective, we may vote and produce 
output by taking consistent outcomes from $2$-out-of-$3$ paths\footnote{To 
reduce clutter and without loss of generality, we have not explicitly shown the 
voting component in the system structure.}. However, whenever the parallel 
paths converge, the overall output quality may come from either choosing the 
maximum quality across all paths, or choosing quality values from some 
pre-defined ordering of paths (we discuss this in details in the following 
sections). Further, we assume that the voting component is fully reliable and 
do not degrade the quality parameters while bypassing through it. In practical 
cases, we can take quality and reliability specifications also for the voter 
and treat it like another component being placed in series whenever a set of 
parallel paths converge.

 \item When the same component is placed in multiple positions of the overall 
system, it indicates different instantiation (in case of hardware systems) or 
different invocations (in case of software systems) of the same object (defined 
once). So, going by this analogy, in Figure~\ref{fig:system_example}, where 
there is two installation of same component $C_2$, if $C_2$ fails, then the
operating probability of $C_2$ becomes $0$ and hence two paths ($I_1 
\rightarrow V_3 \rightarrow V_4 \rightarrow O_1$ and $I_1 \rightarrow V_2 
\rightarrow O_1$) fail together (out of three possible) from input to output.

 \item On the other hand, whenever there is a need to place a duplicate but 
identical component in the other places, a different component name (with same 
specifications though) needs to used in every other constellations. For 
example, 
it may be the case in Figure~\ref{fig:system_example} that $C_1$ and $C_3$ are 
identical components in terms of their functionality, but separately defined.

 \item In practical conditions, sometimes to save power and other design 
overhead, some redundant components may be {\em suspended or switched off} 
(usually controlled by the designer). Component failure and suspension are 
different, because in first case, if the component has an operating reliability 
of $r$, then it has its failure probability of $(1-r)$; whereas the latter 
makes the component to bypass assuming a reliability of $1$ for the suspended 
component. For example, in Figure~\ref{fig:system_example}, if we suspend 
$C_3$, then the path via $I_1 \rightarrow V_3 \rightarrow V_4 \rightarrow O_1$ 
is blocked due to suspension of $C_3$, but $C_3$ will have reliability as $1$.
\end{itemize}

\subsection{Quality and Reliability Specifications} \label{subsec:qual_spec}
Typically, in component-based system design, the quality and reliability 
attributes, often known as {\em non-functional} requirements, are laid from an 
architectural level of the system and are analyzed over the system structures. 
We term such non-functional quality and reliability specifications as {\tt 
QRSpec}. First, let us formally define such a high-level {\tt QRSpec} of a 
component as well as an overall system.

\subsubsection{Component-level Specifications} \label{subsubsec:comp_qual_spec}
Every unit-level component within a system operates in multiple modes (each 
mode having a given operational reliability value between $[0,1]$) provided it 
does not undergo any failure\footnote{We treat component failure also as an 
operational mode with reliability being $0$.}. Now, corresponding to every 
quality level of the input for a component, it produces the output in some 
defined quality levels depending on its mode of operation.
Formally, we represent the {\tt QRSpec} for a unit-level component, $C_i$, as:
\[ \mathcal{P}_i = \langle \{C_i\}, M_i, Z_i, Q_i^I, Q_i^O \rangle, \quad 
\text{where:} \]
\begin{itemize}
 \item $\{C_i\}$ is the participating component.

 \item $M_i = \{m_i^0, m_i^1, m_i^2, \ldots, m_i^{d_i}\}$, denotes the 
set of $(d_i+1)$ operational modes of $C_i$ including $m_i^0$ as the failure 
mode ($d_i \in \mathbb{N}$).

 \item $Z_i: M_i \rightarrow \mathbb{R}^{[0,1]}$, is a function that associates 
a reliability value (within $[0,1]$) to each operational mode of the component.
Here, $Z_i(m_i^0) = 0$ and $\forall k\ (1 \le k \le d_i),\ Z_i(m_i^k) > 0$.

 \item $Q_i^I = \{ q_i^1, q_i^2, \ldots, q_i^{l_i} \}$, denotes the
set of $l_i$ non-negative input quality values (or levels), where $\forall 
j\ (1 \le j < l_i),\ q_i^j \in \mathbb{R}^{+}$ and $q_i^j > q_i^{j+1}$.

 \item $Q_i^O: M_i \times Q_i^I \rightarrow \mathbb{R}^{+}$, is a function that 
maps each input quality level to an output quality (a non-negative real number)
corresponding to every operational modes.\\
It may be noted that, $\forall k\ (1 \le k \le d_i),\ \forall j\ (1 \le j \le 
l_i),\ Q_i^O(m_i^k,q_i^j) \le q_i^j$ and $Q_i^O(m_i^0,q_i^j) = 0$, as $m_i^0$ 
indicates the mode where $C_i$ failed completely. Moreover, if the input quality 
($q$) falls below $q_i^{l_i}$, then the output quality becomes zero, i.e. 
$Q_i^O(m_i^k,q) = 0$, when $q < q_i^{l_i}$.
\end{itemize}
Here, every operating mode for a component has a reliability and can produces 
different quality output values based on given input quality ranges. The 
following example illustrates the component {\tt QRSpec} formalism.
\begin{example} \label{ex:comp_qual_spec}
Let the {\tt QRSpec} for a component, $C_1$, is given as, 
$\mathcal{P}_1 = \langle M_1, Z_1, Q_1^I, Q_1^O \rangle$, where:
\begin{itemize}
 \item $C_1$ operates in two operational modes along with a 
permanent failure mode. Therefore, $M_1 = \{ m_1^0, m_1^1, m_1^2\}$.

 \item The operational reliability values are, $Z_1(m_1^1) = 0.8$, $Z_1(m_1^2) 
= 0.7$, and $Z_1(m_1^0) = 0$.

 \item The input quality levels are specified as, $Q_1^I = \{50, 30, 20\}$.
 
 \item The corresponding output quality values are,
 $Q_1^O(m_1^1,50) = 40$, $Q_1^O(m_1^1,30) = 25$, $Q_1^O(m_1^1,20) = 10$ and
 $Q_1^O(m_1^2,50) = 35$, $Q_1^O(m_1^2,30) = 25$, $Q_1^O(m_1^2,20) = 10$.\\
It intuitively means that, the output quality levels maintained by $C_1$ at 
mode $m_1^1$ are at least $40$, $25$ and $10$ when the input quality value is 
at least $50$, within $[30,50)$ and within $[10,30)$, respectively.\\ 
Implicitly, for any $q \in \mathbb{R}^+$, $Q_1^O(m_1^0,q) = 0$ and when $q < 
20$, $Q_1^O(m_1^1,q) = 0$ and $Q_1^O(m_1^2,q) = 0$.
\end{itemize}
Similarly, let the quality specification for $C_2$ is given as, $\mathcal{P}_2 
= \langle M_2, Z_2, Q_2^I, Q_2^O \rangle$, where:
$M_2 = \{ m_2^0, m_2^1 \}$;
$Z_2(m_2^1) = 0.95$ and $Z_2(m_2^0) = 0$;
$Q_2^I = \{ 40, 10\}$;
$Q_1^O(m_2^1,40) = 30$ and $Q_2^O(m_2^1,10) = 10$.

Also, let the quality specification for $C_3$, is given as, $\mathcal{P}_3 = 
\langle M_3, Z_3, Q_3^I,  Q_3^O \rangle$, where:
$M_3 = \{ m_3^0, m_3^1,  m_3^2\}$;
$Z_3(m_3^1) = 0.9$, $Z_3(m_3^2) = 0.8$ and $Z_3(m_3^0) = 0$;
$Q_3^I = \{50, 20, 10\}$;
$Q_3^O(m_3^1,50) = 45$, $Q_3^O(m_3^1,20) = 20$, $Q_3^O(m_3^1,10) = 5$ and 
$Q_3^O(m_3^2,50) = 40$, $Q_3^O(m_3^2,20) = 15$, $Q_3^O(m_3^2,10) = 5$.
\qed
\end{example}

\subsubsection{System-level Specifications} \label{subsubsec:sys_qual_spec}
A system architecture is built using a set of inter-connected components which 
are glued with each other in a series/parallel manner. Formally, we define the 
{\tt QRSpec} for an overall system, $\Upsilon$, as\footnote{It may be 
noted that, component-level {\tt QRSpec} can also be viewed as a system-level 
{\tt QRSpec} where $C_{\Upsilon} = \{ C_i \}$, i.e., the system, $\Upsilon$,
comprises of a single component, $C_i$. {\em Hence, from the next section 
onwards, we do not differentiate between component and system as such, since 
their formal modeling and treatment remains uniform.}}:
\[ \mathcal{P}_{\Upsilon} = \langle C_{\Upsilon}, M_{\Upsilon}, Z_{\Upsilon}, 
Q_{\Upsilon}^I, Q_{\Upsilon}^O \rangle, \quad \text{where:} \]
\begin{itemize}
 \item $C_{\Upsilon} = \{ C_1, C_2, \ldots, C_n\}$, is the set of $n$ components 
used to form the system, $\Upsilon$.\\
The quality specification for each component, $C_i$ ($1 \leq i \leq n$), is denoted 
by, $\mathcal{P}_i = \langle M_i, Z_i, Q_i^I, Q_i^O \rangle$.

 \item $M_{\Upsilon} = (M_1 \times M_2 \times \cdots \times M_n)$,
denotes the set of $d_{\Upsilon}$ operational modes of $\Upsilon$, where each
mode is an $n$-tuple and hence $d_{\Upsilon} = \prod_{i=1}^{n} (d_i+1)$.

 \item $Z_{\Upsilon}: M_{\Upsilon} \rightarrow \mathbb{R}^{[0,1]}$, is a 
function that associates a reliability value (within $[0,1]$) to each 
operational mode of the system.

 \item $Q_{\Upsilon}^I \subseteq \bigcup\limits_{i=1}^{n} Q_i^I$, denotes the 
set of non-negative input quality values (levels);

 \item $Q_{\Upsilon}^O: M_{\Upsilon} \times Q_{\Upsilon}^I \rightarrow
\mathbb{R}^{+}$, is a function that maps each input quality level to an output
quality (a non-negative real number) for every operational modes.
\end{itemize}
The following example demonstrates the {\tt QRSpec} of two elementary systems, 
$\Upsilon_P$ and $\Upsilon_S$, each having two components connected simply in 
parallel and series, respectively.

\begin{example} \label{ex:system_qual_spec}
First, let us consider another elementary system, $\Upsilon_S$, comprised of 
two 
components, $C_2$ and $C_3$ (as introduced in Example~\ref{ex:comp_qual_spec}), 
where $C_2$ is connected with $C_3$ in series (refer to 
Figure~\ref{fig:series_parallel}(a)). Now, the {\tt QRSpec} for $\Upsilon_S$ is 
given as, $\mathcal{P}_{\Upsilon_S} = \langle C_{\Upsilon_S}, M_{\Upsilon_S}, 
Z_{\Upsilon_S}, Q_{\Upsilon_S}^I, Q_{\Upsilon_S}^O \rangle$, where:
\begin{itemize}
 \item The participating components are, $C_{\Upsilon_S} = \{ C_2, C_3 \}$ 
with {\tt QRSpec}, $\mathcal{P}_2$ and $\mathcal{P}_3$, respectively.

 \item The {\em three} operational modes of $\Upsilon_S$ are, $M_{\Upsilon_S} = 
\{ m_{\Upsilon_S}^0, m_{\Upsilon_S}^1, m_{\Upsilon_S}^2 \}$, where:
\[ m_{\Upsilon_S}^0 \equiv (m_3^0,m_2^0) \equiv (m_3^0,m_2^1) \equiv 
(m_3^1,m_2^0) \equiv (m_3^2,m_2^0), \quad m_{\Upsilon_S}^1 \equiv 
(m_3^1,m_2^1), \quad m_{\Upsilon_S}^2 \equiv (m_3^2,m_2^1). \]
Effectively, only two modes, $(m_3^1,m_2^1)$ and $(m_3^2,m_2^1)$, are 
non-failure (operating) modes and the rest are all failure modes.

 \item Since $C_2$ and $C_3$ are in series, so the reliability corresponding 
to every mode of $M_{\Upsilon_S}$ is be computed as, $Z_{\Upsilon_S}\big{(} 
(m_3^j,m_2^k) \big{)} = Z_1(m_3^j) . Z_2(m_2^k)\ (0 \leq j \leq 2 \text{ and 
} 0 \leq k \leq 1)$ and shown in Table~\ref{tab:system_series_qual_rel}.

 \item The input quality levels are, $Q_{\Upsilon_S}^I = \{ 50, 20 \}$ (due to 
series connected components).

 \item The output quality value function ($Q_{\Upsilon_S}^O$), with respect to 
the input quality values/levels for every operational mode of $\Upsilon_S$ 
is also given in Table~\ref{tab:system_series_qual_rel}\footnote{To 
intuitively explain (detailed methodology is presented later) the last row of 
Table~\ref{tab:system_series_qual_rel}, note that the system mode, 
$m_{\Upsilon_S}^2 \equiv (m_3^2,m_2^1)$, is dictated by the operating modes of 
$C_3$ and $C_2$. The output quality values corresponding to input quality level 
{\em at least $50$} and {\em within $[20,50)$} can be computed as, 
$Q_{\Upsilon_S}^O\big{(}(m_3^2,m_2^1),50\big{)} = 
Q_2^O\big{(}m_2^1,Q_3^O(m_3^2,50)\big{)} = Q_2^O\big{(}m_2^1, 40\big{)} = 
30$, and similarly, $Q_{\Upsilon_S}^O\big{(}(m_3^2,m_2^1),20\big{)} = 
Q_2^O\big{(}m_2^1,Q_3^O(m_3^2,20)\big{)} = Q_2^O\big{(}m_2^1, 15\big{)} = 
Q_2^O\big{(}m_2^1, 10\big{)} = 10$ (since $C_2$ is placed after $C_3$ in 
series).}.
\end{itemize}
\begin{figure}
\centering
\includegraphics[scale=0.5]{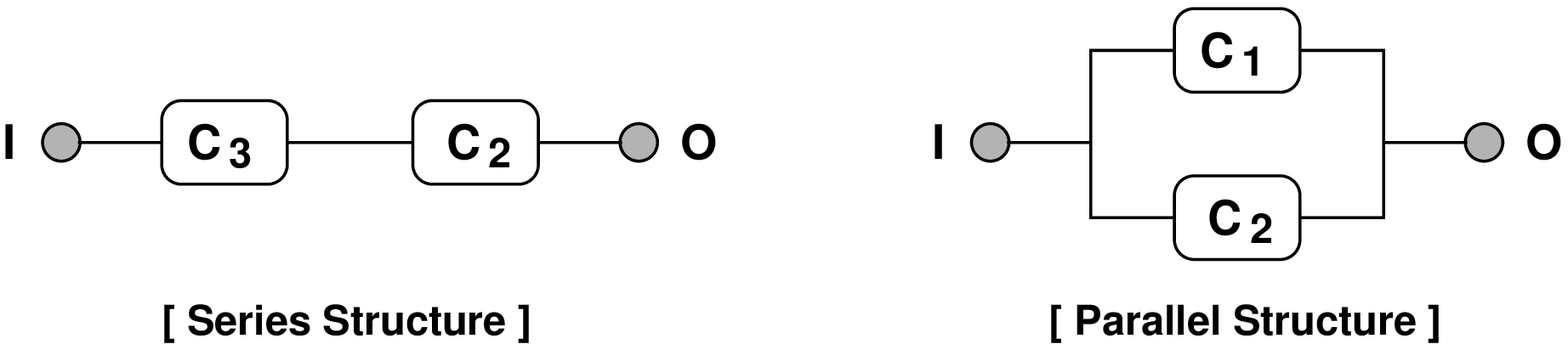}
\caption{{\bf (a)} 2-Component Series System, $\Upsilon_S$ \qquad {\bf (b)} 
2-Component Parallel System, $\Upsilon_P$}
\label{fig:series_parallel}
\end{figure}
\begin{table}
{\scriptsize \sf
    \begin{minipage}[b]{0.47\hsize}\centering
        \begin{tabular}{cccc}
            \cline{1-4}
            System Mode & Reliability & Input Quality & Output Quality\\
            \cline{1-4}
            $m_{\Upsilon_S}^0\equiv(m_3^0, m_2^0)$ & $0.000$ & $\langle 0 
\rangle$ & $\langle 0 \rangle$\\
            $m_{\Upsilon_S}^0\equiv(m_3^0, m_2^1)$ & $0.000$  & $\langle 0 
\rangle$ & $\langle 0 \rangle$\\
            $m_{\Upsilon_S}^0\equiv(m_3^1, m_2^0)$ & $0.000$  & $\langle 0 
\rangle$ & $\langle 0 \rangle$\\
            $m_{\Upsilon_S}^1\equiv(m_3^1, m_2^1)$ & $0.855$  & $\langle 50, 20 
\rangle$ & $\langle 30, 10 \rangle$\\
            $m_{\Upsilon_S}^0\equiv(m_3^2, m_2^0)$ & $0.000$  & $\langle 0 
\rangle$ & $\langle 0 \rangle$\\
            $m_{\Upsilon_S}^2\equiv(m_3^2, m_2^1)$ & $0.760$  & $\langle 50, 20 
\rangle$ & $\langle 30, 10 \rangle$\\
            \cline{1-4}
            \end{tabular}
        \caption{\small Mode-wise Reliability and Quality Values for $\Upsilon_S$}
        \label{tab:system_series_qual_rel}
    \end{minipage}
    \hfill
    \begin{minipage}[b]{0.47\hsize}\centering
        \begin{tabular}{cccc}
            \cline{1-4}
            System Mode & Reliability & Input Quality & Output Quality\\
            \cline{1-4}
            $m_{\Upsilon_P}^0\equiv(m_1^0, m_2^0)$ & $0.000$ & $\langle 0 
\rangle$ & $\langle 0 \rangle$\\
            $m_{\Upsilon_P}^1\equiv(m_1^0, m_2^1)$ & $0.950$  & $\langle 40, 10 
\rangle$ & $\langle 30, 10 \rangle$\\
            $m_{\Upsilon_P}^2\equiv(m_1^1, m_2^0)$ & $0.800$  & $\langle 50, 
30, 
20 \rangle$ & $\langle 40, 25, 10 \rangle$\\
            $m_{\Upsilon_P}^3\equiv(m_1^1, m_2^1)$ & $0.990$  & $\langle 50, 
40, 
30, 10 \rangle$ & $\langle 40, 30, 25, 10 \rangle$\\
            $m_{\Upsilon_P}^4\equiv(m_1^2, m_2^0)$ & $0.700$  & $\langle 50, 
30, 
20 \rangle$ & $\langle 35, 25, 10 \rangle$\\
            $m_{\Upsilon_P}^5\equiv(m_1^2, m_2^1)$ & $0.985$  & $\langle 50, 
40, 
30, 10 \rangle$ & $\langle 35, 30, 25, 10 \rangle$\\
            \cline{1-4}
            \end{tabular}
        \caption{\small Mode-wise Reliability and Quality Values for $\Upsilon_P$}
        \label{tab:system_parallel_qual_rel}
    \end{minipage}
}
\end{table}
Now, let us consider an elementary system, $\Upsilon_P$, comprised of two 
components, $C_1$ and $C_2$ (as introduced in Example~\ref{ex:comp_qual_spec}), 
where $C_1$ is connected with $C_2$ in parallel (refer to 
Figure~\ref{fig:series_parallel}(b)). Now, the {\tt QRSpec} for $\Upsilon_P$ is 
given as, $\mathcal{P}_{\Upsilon_P} = \langle C_{\Upsilon_P}, M_{\Upsilon_P}, 
Z_{\Upsilon_P}, Q_{\Upsilon_P}^I, Q_{\Upsilon_P}^O \rangle$, where:
\begin{itemize}
 \item The participating components are, $C_{\Upsilon_P} = \{ C_1, C_2 \}$ 
with {\tt QRSpec}, $\mathcal{P}_1$ and $\mathcal{P}_2$, respectively.

 \item The {\em six} operational modes of $\Upsilon_P$ are,
$M_{\Upsilon_P} = \{ m_{\Upsilon_P}^0, m_{\Upsilon_P}^1, m_{\Upsilon_P}^2, 
m_{\Upsilon_P}^3, m_{\Upsilon_P}^4, m_{\Upsilon_P}^5 \}$, where: 
\[ m_{\Upsilon_P}^0 \equiv (m_1^0,m_2^0),\ m_{\Upsilon_P}^1 \equiv 
(m_1^0,m_2^1),\ m_{\Upsilon_P}^2 \equiv (m_1^1,m_2^0),\ m_{\Upsilon_P}^3 \equiv 
(m_1^1,m_2^1),\ m_{\Upsilon_P}^4 \equiv (m_1^2,m_2^0),\ m_{\Upsilon_P}^5 \equiv 
(m_1^2,m_2^1). \]

 \item Since $C_1$ and $C_2$ are in parallel, so the reliability corresponding 
to every mode of $M_{\Upsilon_P}$ is be computed as, $Z_{\Upsilon_P}\big{(} 
(m_1^j,m_2^k) \big{)} = \Big{[} 1 - \big{(}1 -  Z_1(m_1^j)\big{)} . \big{(}1 - 
Z_2(m_2^k)\big{)} \Big{]}\ (0 \leq j \leq 2 \text{ and } 0 \leq k \leq 1)$ and shown in
Table~\ref{tab:system_parallel_qual_rel}.

 \item The input quality levels are, $Q_{\Upsilon_P}^I = \{50, 40, 30, 20, 
10\}$ (due to parallel connected components).

 \item The output quality value function ($Q_{\Upsilon_P}^O$), with respect to 
the input quality values/levels for every operational mode of $\Upsilon_P$ 
is also given in Table~\ref{tab:system_parallel_qual_rel}\footnote{To 
intuitively explain (detailed methodology is presented later) the last row of 
Table~\ref{tab:system_parallel_qual_rel}, note that the system mode, 
$m_{\Upsilon_P}^5 \equiv (m_1^2,m_2^1)$, is dictated by the operating modes of 
$C_1$ and $C_2$. The (best possible) output quality value with respect to input 
quality level {\em at least $40$} can be computed as, 
$Q_{\Upsilon_P}^O\big{(}(m_1^2,m_2^1),40\big{)} = {\tt MAX}[Q_1^O(m_1^2,40), 
Q_2^O(m_2^1,40)] = {\tt MAX}[Q_1^O(m_1^2,30), Q_2^O(m_2^1,40)] = {\tt 
MAX}[25,30] = 30$. Similarly, we can also find all the (best possible) output 
quality values, $35$, $30$, $25$ and $10$ for the corresponding input quality 
values, which are {\em at least $50$}, {\em within $[40,50)$}, {\em within
$[30,40)$} and {\em within $[10,30)$}, respectively.}.
\qed
\end{itemize}
\end{example}
It is worthy to note here is that, the architectural {\tt QRSpec} of a system 
derives its operating modes and quality-reliability attributes with respect to 
component failures as well as component suspensions. Though the above example 
only deals with component failures (not provisionize component suspensions), 
however as pointed at the end of Section~\ref{subsec:system} (and also in 
Section~\ref{sec:introduction}), the designer may explore various quality 
attributes of a system also involving suspended components. Exploration of such 
orchestrations in component behaviors needs a generic compositional framework 
through which one can automatically derive out the system-level {\tt QRSpec} 
from the {\tt QRSpec} of its component modules involving both failure and 
suspended components. In subsequent sections, we present the generic procedures 
to formally derive system-level {\tt QRSpec} from component {\tt QRSpec}.

\section{Formal Characterization of Quality Configurations} 
\label{sec:qual_config}
In this section, we present a formal characterization of the system-level 
{\em quality measures} under various operating configurations of its 
constituent components as state transition models, also termed as {\tt 
QRModel}, which enables further compositional interpretation of the {\tt 
QRSpec}.
Formally, we characterize the quality measure under different operating 
configurations of a subsystem, $\Upsilon$ (having one or more components), 
with the {\tt QRSpec}, $\mathcal{P}_{\Upsilon} = 
\langle C_{\Upsilon}, M_{\Upsilon}, Z_{\Upsilon}, Q_{\Upsilon}^I,  
Q_{\Upsilon}^O \rangle$, in terms of the following {\tt QRModel} as,
\[ \mathcal{Q}_{\Upsilon} = \langle C_{\Upsilon}, Q_{\Upsilon}^I, R_{\Upsilon}, 
S_{\Upsilon}, s_{\Upsilon}^1, \Gamma_{\Upsilon}, Q_{\Upsilon}^O, T_{\Upsilon},
{\tt Expr}_{\Upsilon} \rangle, \quad \text{where:} \]
\begin{itemize}
 \item $C_{\Upsilon} = \{ C_1, C_2, \ldots, C_n \}$ is the ordered set of $n$ 
participating components in the system, $\Upsilon$.
 
 \item $Q_{\Upsilon}^I = \{ q_{\Upsilon}^1, q_{\Upsilon}^2, \ldots, 
q_{\Upsilon}^{l_{\Upsilon}} \}$, denotes the
set of $l_{\Upsilon}$ non-negative input quality values (or levels), where 
$\forall j\ (1 \le j < l_{\Upsilon}),\ q_{\Upsilon}^j \in \mathbb{R}^{+}$ and 
$q_{\Upsilon}^j > q_{\Upsilon}^{j+1}$.

 \item $R_{\Upsilon} = \{ r_{\Upsilon,1}, r_{\Upsilon,2}, \ldots, 
r_{\Upsilon,e_{\Upsilon}} \}$ is the set of $e_{\Upsilon} (= \sum_{i=1}^{n}d_i)$
symbolic reliability variables, such that each $r_{\Upsilon,k}$, corresponding
to the operating mode $m_y^t$ of $C_y$, has the reliability value of
$Z_y(m_y^t)$, where $1 \leq \sum_{i=1}^{y-1}d_{i} < k \leq \sum_{i=1}^{y}d_{i}
\leq e_{\Upsilon}$ and $t = k - \sum_{i=1}^{y-1}d_{i}$.

 \item $S_{\Upsilon} = \{ s_{\Upsilon}^1, s_{\Upsilon}^2, \ldots, 
s_{\Upsilon}^{n_{\Upsilon}} \}$ is the set of $n_{\Upsilon}$ operational states
characterizing the quality measures. Here, $n_{\Upsilon} =
\prod_{i=1}^n (2^{d_i+1}-1)$, where each $C_i$ can operate in $d_i$ number of
modes ($1 \leq i \leq n$). 

 \item $s_{\Upsilon}^1 \in S_{\Upsilon}$ is the start/initial operating state.

 \item $\Gamma_{\Upsilon}: S_{\Upsilon} \rightarrow \{ {\tt 1, 0, X, Y}
\}^{e_{\Upsilon}}$ denotes the $e_{\Upsilon}$-length configuration function of 
a state corresponding to $d_{\Upsilon}$ operational modes in $M_{\Upsilon}$ 
(excluding failure mode, $m_{\Upsilon}^0 \in M_{\Upsilon}$, which can be 
expressed as the state having all $0$'s in the $e_{\Upsilon}$-length
configuration).  Here, $e_{\Upsilon} = \sum_{i=1}^{n}d_i$, where $d_i$ is the 
number of operating modes (excluding the failure mode) of $C_i$ ($1 \leq i 
\leq n$). Typically, $C_i$ shall execute in the operating mode, $m_i^j$, when 
all modes $m_i^{k}\ (1 \leq k < j \leq d_i)$ have failed.

For a state, $s_{\Upsilon}^j \in S_{\Upsilon}$, 
$\Gamma_{\Upsilon}(s_{\Upsilon}^j)[k] \in \{ {\tt 1, 0, X, Y} \}$ and
$\Gamma_{\Upsilon}(s_{\Upsilon}^j)[k..k'] \in \{ {\tt 1, 0, X, Y} 
\}^{(k'-k+1)}$ ($1 \leq k < k' \leq e_{\Upsilon}$) represent 
the $k^{th}$ and $k$-to-$k'$ configuration location(s) of $s_{\Upsilon}^j$, 
respectively. Formally,
\[
 \Gamma_{\Upsilon}(s_{\Upsilon}^j)[k] =
\left\{ \begin{array}{rl}
{\tt 1}, & \mbox{indicating $C_y$ is operating in mode $m_y^t$ }\\
{\tt 0}, & \mbox{indicating $C_y$ is has failed in mode $m_y^t$}\\
{\tt X}, & \mbox{indicating $C_y$ is not availing mode $m_y^t$}\\
{\tt Y}, & \mbox{indicating $C_y$ has suspended mode $m_y^t$}
\end{array}\right.
\mbox{where } 1 \leq \sum_{i=1}^{y-1}d_{i} < k \leq \sum_{i=1}^{y}d_{i} \leq
e_{\Upsilon} \mbox{ and } t = k - \sum_{i=1}^{y-1}d_{i}
\]
Additionally, when $\Gamma_{\Upsilon}(s_{\Upsilon}^j)[k] = {\tt 1}$, we
have, $\Gamma_{\Upsilon}(s_{\Upsilon}^j)[k'] = {\tt 0} / {\tt Y}$ ($\forall k'$
such that $(\sum_{i=1}^{y-1}d_{i}) < k' < k$), and
$\Gamma_{\Upsilon}(s_{\Upsilon}^j)[k''] = {\tt X}$ ($\forall k''$ such that
$k < k'' \leq (\sum_{i=1}^{y}d_{i})$).
Moreover, $\Gamma_{\Upsilon}(s_{\Upsilon}^j) \equiv 
\Gamma_{\Upsilon}(s_{\Upsilon}^j)[1..e_{\Upsilon}]$.

 \item $Q_{\Upsilon}^O: S_{\Upsilon} \times Q_{\Upsilon}^I \rightarrow 
\mathbb{R}^{+}$, is a function that 
maps each input quality level to an output quality (a non-negative real number)
corresponding to every operational modes.\\
It may be noted that, $\forall k\ (1 \le k \le n_{\Upsilon}),\ \forall j\ (1 \le 
j \le l_{\Upsilon}),\ Q_{\Upsilon}^O(s_{\Upsilon}^k,q_{\Upsilon}^j) \le 
q_{\Upsilon}^j$ and $Q_{\Upsilon}^O(s,q_{\Upsilon}^j) = 
Q_{\Upsilon}^O(s',q_{\Upsilon}^j) = 
0$, whenever $\Gamma_{\Upsilon}(s) = {\tt 0}^{e_{\Upsilon^{}}}$ and 
$\Gamma_{\Upsilon}(s') = {\tt Y}^{e_{\Upsilon}^{}}$ indicating the operating
modes where $C_{\Upsilon}$ completely failed or fully suspended, respectively. 
Moreover, if the input quality ($q$) falls below $q_{\Upsilon}^{l_{\Upsilon}}$, 
then the output quality becomes zero, i.e. $Q_{\Upsilon}^O(s_{\Upsilon}^k,q) = 
0$, whenever $q < q_{\Upsilon}^{l_{\Upsilon}}$.

 \item $T_{\Upsilon} = T_{\Upsilon}^F \cup T_{\Upsilon}^C$ denotes the set of 
transitions from one state to another and $T_{\Upsilon} \subseteq S_{\Upsilon} 
\times S_{\Upsilon}$. $T_{\Upsilon}^F$ and $T_{\Upsilon}^C$ are the
transitions indicating {\em failure} and {\em suspension} of an operating mode, 
respectively.
\begin{itemize}
\item The transition, $T_{\Upsilon}^F(s_{\Upsilon}^j, s_{\Upsilon}^k)$, from 
the state $s_{\Upsilon}^j \in S_{\Upsilon}$ to the state $s_{\Upsilon}^k \in 
S_{\Upsilon}$ ($1 \leq j,k \leq n_{\Upsilon}$) is {\em permissible} only 
when either of the following two conditions happen:
\begin{enumerate}[(a)]
 \item $\exists w'\ (1 \leq w' \leq e_{\Upsilon})$, such that 
$\Gamma_{\Upsilon}(s_{\Upsilon}^j)[w'] = {\tt 1}$ and 
$\Gamma_{\Upsilon}(s_{\Upsilon}^k)[w'] = {\tt 0}$, whereas $\forall w\ (1 \leq 
w \leq e_{\Upsilon} \text{ and } w \neq w')$, 
$\Gamma_{\Upsilon}(s_{\Upsilon}^j)[w] = \Gamma_{\Upsilon}(s_{\Upsilon}^k)[w]$.

 \item $\exists w'\ (2 \leq w' \leq e_{\Upsilon})$, such that 
$\Gamma_{\Upsilon}(s_{\Upsilon}^j)[w'-1] = {\tt 1},\ 
\Gamma_{\Upsilon}(s_{\Upsilon}^j)[w'] = {\tt X}$ and 
$\Gamma_{\Upsilon}(s_{\Upsilon}^k)[w'-1] = {\tt 0},\
\Gamma_{\Upsilon}(s_{\Upsilon}^k)[w'] = {\tt 1}$, whereas $\forall w\ (1 \leq w 
\leq e_{\Upsilon} \text{ and } w \neq w' \text{ or } w \neq w'-1)$, 
$\Gamma_{\Upsilon}(s_{\Upsilon}^j)[w] = \Gamma_{\Upsilon}(s_{\Upsilon}^k)[w]$.
\end{enumerate}

\item The transition, $T_{\Upsilon}^C(s_{\Upsilon}^j, s_{\Upsilon}^k)$, from 
the state $s_{\Upsilon}^j \in S_{\Upsilon}$ to the state $s_{\Upsilon}^k \in 
S_{\Upsilon}$ ($1 \leq j,k \leq n_{\Upsilon}$) is {\em permissible} only 
when either of the following two conditions happen:
\begin{enumerate}[(a)]
 \item $\exists w'\ (1 \leq w' \leq e_{\Upsilon})$, such that 
$\Gamma_{\Upsilon}(s_{\Upsilon}^j)[w'] = {\tt 1}$ 
and $\Gamma_{\Upsilon}(s_{\Upsilon}^k)[w'] = {\tt Y}$, whereas $\forall w\ (1 
\leq w \leq e_{\Upsilon} \text{ and } w \neq w')$, 
$\Gamma_{\Upsilon}(s_{\Upsilon}^j)[w] = 
\Gamma_{\Upsilon}(s_{\Upsilon}^k)[w]$.

 \item $\exists w'\ (2 \leq w' \leq e_{\Upsilon})$, such that 
$\Gamma_{\Upsilon}(s_{\Upsilon}^j)[w'-1] = {\tt 1},\ 
\Gamma_{\Upsilon}(s_{\Upsilon}^j)[w'] = {\tt X}$ and 
$\Gamma_{\Upsilon}(s_{\Upsilon}^k)[w'-1] = {\tt Y},\ 
\Gamma_{\Upsilon}(s_{\Upsilon}^k)[w'] = {\tt 1}$, whereas $\forall w\ (1 \leq w 
\leq e_{\Upsilon} \text{ and } w \neq w')$, 
$\Gamma_{\Upsilon}(s_{\Upsilon}^j)[w] = \Gamma_{\Upsilon}(s_{\Upsilon}^k)[w]$.
\end{enumerate}
\end{itemize}

 \item ${\tt Expr}_{\Upsilon}$ is a function that produces an algebraic 
operational probability expression\footnote{The need to upkeep such algebraic 
expressions, in its complete form without calculating values instantly, is to 
enable seamless reliability composition (which will be explained in later 
sections) where same component participates (i.e. instantiated or invoked) 
multiple times in a component-based system with series/parallel system setup.} 
containing the symbolic reliability variables from $R_{\Upsilon}$ based on the 
configuration of an input state, $s_{\Upsilon}^j \in S_{\Upsilon}$ ($1 \leq j 
\leq n_{\Upsilon}$). Formally,
\[
\forall j\ (1 \leq j \leq n_{\Upsilon}),\ {\tt Expr}_{\Upsilon} (s_{\Upsilon}^j) 
= \prod_{k=1}^{e_{\Upsilon}} {\tt rel}_{\Upsilon}^j[k], \quad \mbox{ where, } 
{\tt rel}_{\Upsilon}^j[k] =
\left\{ \begin{array}{rl}
  (r_{\Upsilon,k}^{}), & \mbox{when } \Gamma_{\Upsilon}(s_{\Upsilon}^j)[k] = 
{\tt 1}\\
  (1 - r_{\Upsilon,k}^{}), & \mbox{when } \Gamma_{\Upsilon}(s_{\Upsilon}^j)[k] 
= {\tt 0}\\
  1, & \mbox{when } \Gamma_{\Upsilon}(s_{\Upsilon}^j)[k] = {\tt X} \mbox{ or } 
{\tt Y}
\end{array}\right.
\]
Substituting the reliability values for each $r_{\Upsilon,j}^{} \in R_{\Upsilon}$
($1 \leq j \leq e_{\Upsilon}$) in the expression ${\tt Expr}_{\Upsilon}$, we
can get the successful operating probability value for the mode of operation
in state $s_{\Upsilon}^j$.
\end{itemize}
The following example illustrates such formal {\tt QRModel} characterization in 
details.
\begin{example} \label{ex:comp_qual_measure}
Consider the three components, $C_1$, $C_2$ and $C_3$, having the {\tt QRSpec}, 
$\mathcal{P}_1$, $\mathcal{P}_2$ and $\mathcal{P}_3$, as presented in 
Example~\ref{ex:comp_qual_spec}. The {\tt QRModel} for these {\em three} 
components, which are represented by the state transition diagram and the 
configuration details corresponding to every state, are illustrated in 
Figure~\ref{fig:comp_qual} with Table~\ref{tab:comp_qual}.
For $C_1$, we have the {\tt QRModel},
$\mathcal{Q}_1 = \langle \{C_1\}, Q_1^I, R_1, S_1, s_1^1, \Gamma_1, Q_1^O, 
T_1^F \cup T_1^C, {\tt Expr}_1 \rangle$, where:
\begin{itemize}
 \item The input quality levels are specified as, $Q_1^I = \{50,30,20\}$.
 
 \item The symbolic reliability variables are, $R_1=\{r_{1,1},r_{1,2}\}$.
 
 \item The $7$ operational states are, $S_1 = \{s_1^1, s_1^2, \ldots, s_1^7 \}$.
 
 \item The $2$-length operational state configurations are, $\Gamma_1(s_1^1) = 
{\tt 1X},\ \ldots,\ \Gamma_1(s_1^4) = {\tt 00},\ \ldots,\ \Gamma_1(s_1^7) = 
{\tt YY}$.

 \item The output quality are defined as, $Q_1^O(s_1^1,50) = 40,\ 
Q_1^O(s_1^1,30) = 25,\ Q_1^O(s_1^1,20) = 10$.\\
All output quality values at all other operating states are defined similarly.

 \item The failure state transitions are, $T_1^F = \{ (s_1^1, s_1^2); (s_1^2, 
s_1^4); (s_1^3, s_1^6) \}$.

 \item The suspend state transitions are, $T_1^C = \{ (s_1^1, s_1^3); (s_1^2, 
s_1^5); (s_1^3, s_1^7) \}$.

 \item The operational probability expressions are, ${\tt Expr}_1(s_1^1) = 
(r_{1,1}),\ {\tt Expr}_1(s_1^2) = (1 - r_{1,1})(r_{1,2}),\ \ldots$ so on.
\end{itemize}
\begin{figure}
\centering
\includegraphics[width=0.85\textwidth]{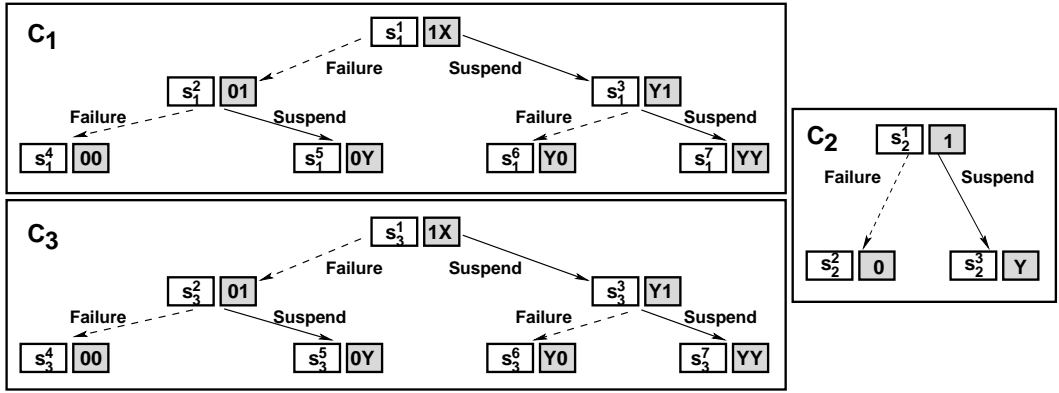}
\caption{State Transition Diagram corresponding to {\tt QRModel}, 
$\mathcal{Q}_i$, for three Components, $C_i$ ($1 \leq i \leq 3$)}
\label{fig:comp_qual}
\end{figure}
\begin{table}
{\sf
\centering
 \begin{tabular}{ccccccc}
    \cline{1-7}
    Component & State & Operating Mode & Input Quality & Output Quality & 
\multicolumn{2}{c}{Operating Mode Probability}\\
    \cline{6-7}
    Name & ID & Configuration & Values (Levels) & Values (Levels) & Algebraic 
Expression & Value\\
    \cline{1-7}
          & $s_1^1$ & {\tt 1X} & $\langle 50,30,20 \rangle$ & $\langle 40,25,10 
\rangle$ & $r_{1,1}$ & $0.80$\\
          & $s_1^2$ & {\tt 01} & $\langle 50,30,20 \rangle$ & $\langle 35,25,10 
\rangle$ & $(1-r_{1,1}).r_{1,2}$ & $0.14$\\
          & $s_1^3$ & {\tt Y1} & $\langle 50,30,20 \rangle$ & $\langle 35,25,10 
\rangle$& $r_{1,2}$ & $0.70$\\
    $C_1$ & $s_1^4$ & {\tt 00} & $\langle 0 \rangle$ & $\langle 0 \rangle$ & 
$(1-r_{1,1}).(1-r_{1,2})$ & $0.06$\\
          & $s_1^5$ & {\tt 0Y} & $\langle 0 \rangle$ & $\langle 0 \rangle$ & 
$(1-r_{1,1})$ & $0.20$\\
          & $s_1^6$ & {\tt Y0} & $\langle 0 \rangle$ & $\langle 0 \rangle$ & 
$(1-r_{1,2})$ & $0.30$\\
          & $s_1^7$ & {\tt YY} & $\langle 0 \rangle$ & $\langle 0 \rangle$ & 
$1$ & $1.00$\\
    \cline{1-7}
          & $s_2^1$ & {\tt 1} & $\langle 40,10 \rangle$ & $\langle 30,10 
\rangle$ & $r_{2,1}$ & $0.95$\\
    $C_2$ & $s_2^2$ & {\tt 0} & $\langle 0 \rangle$ & $\langle 0 \rangle$ & 
$(1-r_{2,1})$ & $0.05$\\
          & $s_2^3$ & {\tt Y} & $\langle 0 \rangle$ & $\langle 0 \rangle$ & 
$1$ & $1.00$\\
    \cline{1-7}
          & $s_3^1$ & {\tt 1X} & $\langle 50,20,10 \rangle$ & $\langle 45,20,5 
\rangle$ & $r_{3,1}$ & $0.90$\\
          & $s_3^2$ & {\tt 01} & $\langle 50,20,10 \rangle$ & $\langle 40,15,5 
\rangle$ & $(1-r_{3,1}).r_{3,2}$ & $0.08$\\
          & $s_3^3$ & {\tt Y1} & $\langle 50,20,10 \rangle$ & $\langle 40,15,5 
\rangle$& $r_{3,2}$ & $0.80$\\
    $C_3$ & $s_3^4$ & {\tt 00} & $\langle 0 \rangle$ & $\langle 0 \rangle$ & 
$(1-r_{3,1}).(1-r_{3,2})$ & $0.02$\\
          & $s_3^5$ & {\tt 0Y} & $\langle 0 \rangle$ & $\langle 0 \rangle$ & 
$(1-r_{3,1})$ & $0.10$\\
          & $s_3^6$ & {\tt Y0} & $\langle 0 \rangle$ & $\langle 0 \rangle$ & 
$(1-r_{3,2})$ & $0.20$\\
          & $s_3^7$ & {\tt YY} & $\langle 0 \rangle$ & $\langle 0 \rangle$ & 
$1$ & $1.00$\\
    \cline{1-7}
    \end{tabular}
 \caption{{\tt QRModel} Configuration Details, $\mathcal{Q}_i$, for Three 
Components, $C_i$ ($1 \leq i \leq 3$)}
 \label{tab:comp_qual}
}
\end{table}
The intuitive explanation of such a {\tt QRModel} for $C_1$ is as follows. 
$C_1$ can operate in two modes with reliability, $r_{1,1} = Z_1(m_1^1) = 0.8$ 
and $r_{1,2} = Z_1(m_1^2) = 0.7$. Whenever $C_1$ is operating in $m_1^1$ with 
the reliability $0.8$ (as well as the operational probability = $0.8$), the 
output quality values are $40$, $25$ and $10$ units when the input quality 
levels are at least $50$, within $[30,50)$ and within $[20,30)$ units, 
respectively. Now, when $C_1$ fails in the first mode of operation and moves to 
operate in mode $m_1^2$, then the reliability becomes $0.7$ (however, the 
operational probability becomes = $(1-0.8) \times 0.7 = 0.14$ as derived from 
${\tt Expr}_1(s_1^2)$) in the next/second operational mode, the output quality 
values are $35$, $25$ and $10$ units when the input quality levels are at least 
$50$, within $[30,50)$ and within $[20,30)$ units, respectively. Finally, when 
$C_1$ completely fails to operate in any of these two modes and enters $m_1^0$ 
(having the operating probability = $(1-0.8) \times (1-0.7) = 0.06$ derived from 
${\tt Expr}_1(s_1^4)$), the output quality is $0$ (zero) irrespective of its 
input quality levels. In addition to this, when $C_1$ is suspended (controlled) 
at mode $m_1^1$, then it automatically enters/operates at $m_1^2$ with the 
reliability $0.7$ (i.e., having the operational probability = $1.0 \times 0.7 = 
0.7$), and the output quality values are $35$, $25$ and $10$ units when the 
input quality levels are at least $50$, within $[30,50)$ and within $[20,30)$ 
units, respectively. When $C_1$ is suspended in both its operating mode, the 
output quality is $0$ (zero) irrespective of the input quality levels, since it 
is not operating at all.
\qed
\end{example}
It may be noted that the formal representation of the quality measures in 
the component-level and the system-level are generic and representation-wise 
similar. Since the underlying representation of the component {\tt QRModel} is 
a generic state-transition system, so our proposed compositional framework will 
formally derive composite state-transition model to represent system {\tt 
QRModel}, as described in the next section in details.

\section{A Generic Quality Composition Framework} \label{sec:qual_compose}
This section presents the formal approaches to (a) hierarchically compose 
(series/parallel) formal quality measures ({\tt QRModel}) of components and 
sub-systems, and (b) then obtain system {\tt QRSpec} directly from such 
formally represented composed quality measures. The first part will be enabled 
by the proposed algebra for quality composition of series and parallel 
component structures and then extend our approach to show its applicability 
over generic component structures as well.

\subsection{Problem Statement} \label{subsec:problem}
The formal problem statement for quality assessment of a generic 
component-based system is described as:
\begin{description}
 \item[{\em Given} --] A component-based system, $\Upsilon = \langle 
\mathcal{I}, \mathcal{O}, \mathcal{C}, \mathcal{V}, \mathcal{E}, \mathcal{L}
\rangle$ with $n$ connected subsystem structures, where each $C_i \in
\mathcal{C}$ has the {\tt QRModel}, $\mathcal{Q}_i = \langle \{C_i\}, Q_i^I, 
R_i, S_i,
s_i^1, \Gamma_i, Q_i^O, T_i, {\tt Expr}_i \rangle$.

 \item[{\em Objective} --] Derive the composed {\tt QRModel}, 
$\mathcal{Q}_\Upsilon = \langle C_{\Upsilon}, Q_\Upsilon^I, R_\Upsilon, 
S_\Upsilon, s_{\Upsilon}^1, \Gamma_\Upsilon, Q_\Upsilon^O, T_\Upsilon, {\tt 
Expr}_\Upsilon \rangle$, for the overall system structure, $\Upsilon$.
\end{description}
The generic framework to the above problem can be obtained by hierarchical
compositions of elementary series and parallel component structures in
component-based systems. Therefore, given the two subsystem structures, 
$\Upsilon_i$  and $\Upsilon_j$, having their {\tt QRModel} as, 
$\mathcal{Q}_{\Upsilon_i} = \langle C_{\Upsilon_i}, Q_{\Upsilon_i}^I, 
R_{\Upsilon_i}, S_{\Upsilon_i}, s_{\Upsilon_i}^1, \Gamma_{\Upsilon_i}, 
Q_{\Upsilon_i}^O, T_{\Upsilon_i}, {\tt Expr}_{\Upsilon_i} \rangle$ and 
$\mathcal{Q}_{\Upsilon_j} = \langle C_{\Upsilon_j}, Q_{\Upsilon_j}^I, 
R_{\Upsilon_j}, S_{\Upsilon_j}, s_{\Upsilon_j}^1, \Gamma_{\Upsilon_j}, 
Q_{\Upsilon_j}^O, T_{\Upsilon_j}, {\tt Expr}_{\Upsilon_j} \rangle$
respectively, we formulate the following two primitive quality
composition sub-problems to establish the generic compositional framework.
\begin{description}
 \item[Series Composition Problem:] Derive the compositional {\tt QRModel}, 
$\mathcal{Q}_{\Upsilon_S} = \mathcal{Q}_{\Upsilon_i}\ \circ\ 
\mathcal{Q}_{\Upsilon_j}$, when $\Upsilon_i$  and $\Upsilon_j$ are connected
in series ($\Upsilon_i$  followed by $\Upsilon_j$) forming the system,
$\Upsilon_S \equiv {\Upsilon_i}\ \circ\ {\Upsilon_j}$.

 \item[Parallel Composition Problem:] Derive the compositional {\tt QRModel}, 
$\mathcal{Q}_{\Upsilon_P} = \mathcal{Q}_{\Upsilon_i}\ ||\ 
\mathcal{Q}_{\Upsilon_j}$, when $\Upsilon_i$  and $\Upsilon_j$ are connected
in parallel forming the system,
$\Upsilon_P \equiv {\Upsilon_i}\ ||\ {\Upsilon_j}$.
\end{description}

\subsection{Approaches for Quality Composition} \label{subsec:appr_qual_comp}
The composition of quality measures for the series and parallel subsystems can
be carried out using various principles. Here, we introduce the compositional 
algebra for the primitive (series and parallel) operations.

\subsubsection{Series Composition}
Let the two subsystem structures, $\Upsilon_i$ and $\Upsilon_j$ have the 
respective {\tt QRModel} as,
\[ \mathcal{Q}_{\Upsilon_i} = \langle C_{\Upsilon_i}, Q_{\Upsilon_i}^I, 
R_{\Upsilon_i}, S_{\Upsilon_i}, s_{\Upsilon_i}^1, \Gamma_{\Upsilon_i}, 
Q_{\Upsilon_i}^O, T_{\Upsilon_i}, {\tt Expr}_{\Upsilon_i} \rangle \text{ and } 
\mathcal{Q}_{\Upsilon_j} = \langle C_{\Upsilon_j}, Q_{\Upsilon_j}^I, 
R_{\Upsilon_j}, S_{\Upsilon_j}, s_{\Upsilon_j}^1, \Gamma_{\Upsilon_j}, 
Q_{\Upsilon_j}^O, T_{\Upsilon_j}, {\tt Expr}_{\Upsilon_j} \rangle. \]
These are appended in series forming the composite system, $\Upsilon_S \equiv 
\Upsilon_i\ \circ\ \Upsilon_j$, whose {\tt QRModel}, can be derived as follows: 
\quad
$\mathcal{Q}_{\Upsilon_i}\ \circ\ \mathcal{Q}_{\Upsilon_j} \equiv
\mathcal{Q}_{\Upsilon_S} = \langle C_{\Upsilon_S}, Q_{\Upsilon_S}^I, 
R_{\Upsilon_S}, S_{\Upsilon_S}, s_{\Upsilon_S}^1, \Gamma_{\Upsilon_S}, 
Q_{\Upsilon_S}^O, T_{\Upsilon_S}, {\tt Expr}_{\Upsilon_S} \rangle$, where:
\begin{itemize}
 \item $C_{\Upsilon_S} = C_{\Upsilon_i} \cup C_{\Upsilon_j}$, denotes the set 
of participating components in $\Upsilon_S$.
 
 \item $Q_{\Upsilon_S}^I = Q_{\Upsilon_i}^I = \{ q_{\Upsilon_i}^1, 
q_{\Upsilon_i}^2, \ldots, q_{\Upsilon_i}^{l_{\Upsilon_i}} \}$ 
denotes the set of $l_{\Upsilon_S} = l_{\Upsilon_i}$ input quality values 
(levels) used by 
$C_{\Upsilon_i}$.

 \item $R_{\Upsilon_S} = R_{\Upsilon_i} \cup R_{\Upsilon_j}$,
denotes the combined set of all symbolic reliability variables with 
$|R_{\Upsilon_S}| = e_{\Upsilon_S}$.
%

 \item $S_{\Upsilon_S}$ is the set of states (with $s_{\Upsilon_S}^1 \in 
S_{\Upsilon_S}$ being the start state) representing state configurations of 
$\mathcal{Q}_{\Upsilon_S}$.

 \item $\Gamma_{\Upsilon_S}: S_{\Upsilon_S} \rightarrow \{ {\tt 1, 0, X, Y}
\}^{e_{\Upsilon_S}^{}}$ is the $e_{\Upsilon_S}$-length state configuration, 
such that for a state, $s_{\Upsilon_S}^k \in S_{\Upsilon_S}$, the composed 
state configuration, $s_{\Upsilon_S}^k$ is obtained from states 
$s_{\Upsilon_i}^a \in S_{\Upsilon_i}\ (1 \leq a \leq n_{\Upsilon_i})$ 
and $s_{\Upsilon_j}^b \in S_{\Upsilon_j}\ (1 \leq b \leq 
n_{\Upsilon_j})$, denoted as 
$\Gamma_{\Upsilon_S}(s_{\Upsilon_S}^k) = 
\Gamma_{\Upsilon_i}(s_{\Upsilon_i}^a) \odot 
\Gamma_{\Upsilon_j}(s_{\Upsilon_j}^b)$, 
as follows: \quad ($e_{\Upsilon_i} \leq e_{\Upsilon_S} \leq e_{\Upsilon_i} + 
e_{\Upsilon_j}$)
\begin{itemize}
 \item $\Gamma_{\Upsilon_S}(s_{\Upsilon_S}^k)[1..e_{\Upsilon_i}] = 
\Gamma_{\Upsilon_i}(s_{\Upsilon_i}^a)[1..e_{\Upsilon_i}]$, and

 \item $\forall v\ (1 \leq v \leq  e_{\Upsilon_j}),\ \exists u\ (u \leq v)$, 
$\Gamma_{\Upsilon_S}(s_{\Upsilon_S}^k)[e_{\Upsilon_i}+u] = 
\Gamma_{\Upsilon_j}(s_{\Upsilon_j}^b)[v]\ (1 \leq u 
\leq e_{\Upsilon_S} - e_{\Upsilon_i})$ when $r_{\Upsilon_j,v}^{} \in 
R_{\Upsilon_j}$ but $r_{\Upsilon_j,v}^{} \notin R_{\Upsilon_i}$.\\
Intuitively, the last rule prevents the composed configuration to create 
duplicate configuration entries in case of multiple occurrence of the same 
component in both $C_{\Upsilon_i}$ and $C_{\Upsilon_j}$ subsystems.
\end{itemize}

 \item $Q_{\Upsilon_S}^O: S_{\Upsilon_S} \times Q_{\Upsilon_S}^I \rightarrow
\mathbb{R}^{+}$ indicates the function to compute the output quality value.\\
For a state, $s_{\Upsilon_S}^k \in S_{\Upsilon_S}$, where
$\Gamma_{\Upsilon_S}(s_{\Upsilon_S}^k) = 
\Gamma_{\Upsilon_i}(s_{\Upsilon_i}^a) \odot
\Gamma_{\Upsilon_j}(s_{\Upsilon_j}^b)\ (s_{\Upsilon_i}^a \in 
S_{\Upsilon_i},\ s_{\Upsilon_j}^b \in S_{\Upsilon_j})$, we have:\\
$\forall k_i\ (1 \leq k_i \leq l_{\Upsilon_i}),\ \exists k_j\ (1 \leq k_j \leq 
l_{\Upsilon_j})$ such that 
$Q_{\Upsilon_S}^O(s_{\Upsilon_S}^k, q_{\Upsilon_i}^{k_i}) = Q_{\Upsilon_j}^O 
(s_{\Upsilon_j}^b, q_{\Upsilon_j}^{k_j})$ and 
$Q_{\Upsilon_i}^O (s_{\Upsilon_i}^a, q_{\Upsilon_i}^{k_i}) \geq 
q_{\Upsilon_j}^{k_j}$, as well as $\nexists k'\ (1 \leq k'
\leq l_{\Upsilon_j})$ so that $q_{\Upsilon_j}^{k_j} < q_{\Upsilon_j}^{k'} 
\leq Q_{\Upsilon_i}^O(s_{\Upsilon_i}^a, q_{\Upsilon_i}^{k_i})$.

 \item $T_{\Upsilon_S} \subseteq S_{\Upsilon_S} \times S_{\Upsilon_S}$ denotes 
the transition relation and let $T_{\Upsilon_S} = T_{\Upsilon_S}^F \cup
T_{\Upsilon_S}^C$, where $T_{\Upsilon_S}^F$ and $T_{\Upsilon_S}^C$ are the set 
of failure and suspend transitions, respectively. Suppose, 
$T_{\Upsilon_i}(s_{\Upsilon_i}^a,s_{\Upsilon_i}^{a'})$ 
and $T_{\Upsilon_j}(s_{\Upsilon_j}^b,s_{\Upsilon_j}^{b'})$ are transition 
in $\mathcal{Q}_{\Upsilon_i}$ and  
$\mathcal{Q}_{\Upsilon_j}$, respectively where $s_{\Upsilon_i}^a, 
s_{\Upsilon_i}^{a'} \in S_{\Upsilon_i}\ (1 \leq a,a' \leq 
n_{\Upsilon_i})$ and $s_{\Upsilon_j}^b, s_{\Upsilon_j}^{b'} \in 
S_{\Upsilon_j}\ (1 \leq b,b' \leq n_{\Upsilon_j})$. For 
$s_{\Upsilon_S}^k \in S_{\Upsilon_S}$, if
$\Gamma_{\Upsilon_S}(s_{\Upsilon_S}^k) = 
\Gamma_{\Upsilon_i}(s_{\Upsilon_i}^a) \odot 
\Gamma_{\Upsilon_j}(s_{\Upsilon_j}^b)$, 
then $T_{\Upsilon_S}(s_{\Upsilon_S}^k,s_{\Upsilon_S}^{k'})$ is a permissible 
transition ($\exists s_{\Upsilon_S}^{k'} \in S_{\Upsilon_S}$) such that one of 
the following three conditions hold:
\begin{enumerate}[(i)]
\item $\Gamma_{\Upsilon_S}(s_{\Upsilon_S}^{k'}) = 
\Gamma_{\Upsilon_i}(s_{\Upsilon_i}^a) \odot
\Gamma_{\Upsilon_j}(s_{\Upsilon_j}^{b'})$, or

\item $\Gamma_{\Upsilon_S}(s_{\Upsilon_S}^{k'}) = 
\Gamma_{\Upsilon_i}(s_{\Upsilon_i}^{a'})
\odot \Gamma_{\Upsilon_j}(s_{\Upsilon_j}^b)$, or

\item $\Gamma_{\Upsilon_S}(s_{\Upsilon_S}^{k'}) = 
\Gamma_{\Upsilon_i}(s_{\Upsilon_i}^{a'}) \odot
\Gamma_{\Upsilon_j}(s_{\Upsilon_j}^{b'})$ along with either of the 
following 
two constraints:\\
(a) $T_{\Upsilon_i}(s_{\Upsilon_i}^a, s_{\Upsilon_i}^{a'}) \in 
T_{\Upsilon_i}^F$; $T_{\Upsilon_j}(s_{\Upsilon_j}^b, 
s_{\Upsilon_j}^{b'}) \in T_{\Upsilon_j}^F$, {\bf ~~or~~}
(b) $T_{\Upsilon_i}(s_{\Upsilon_i}^a, s_{\Upsilon_i}^{a'}) \in 
T_{\Upsilon_i}^C$; $T_{\Upsilon_j}(s_{\Upsilon_j}^b, 
s_{\Upsilon_j}^{b'}) \in T_{\Upsilon_j}^C$.
\end{enumerate}
Further, it is important to note the categorization of transitions (failure or 
suspended) here as follows:
\begin{itemize}
 \item $T_{\Upsilon_S}(s_{\Upsilon_S}^k,s_{\Upsilon_S}^{k'}) \in 
T_{\Upsilon_S}^F$ (is a failure transition), whenever either of the following 
happens:
 \begin{compactenum}[(a)]
  \item $T_{\Upsilon_i}(s_{\Upsilon_j}^b, s_{\Upsilon_j}^{b'}) \in 
T_{\Upsilon_j}^F$ and Condition-(i) / Condition-(iii-a) 
is satisfied from above, or

  \item $T_{\Upsilon_i}(s_{\Upsilon_i}^a, s_{\Upsilon_i}^{a'}) \in 
T_{\Upsilon_i}^F$ and Condition-(ii) / Condition-(iii-a) 
is satisfied from above.
 \end{compactenum}

 \item $T_{\Upsilon_S}(s_{\Upsilon_S}^k,s_{\Upsilon_S}^{k'}) \in 
T_{\Upsilon_S}^C$ (is a suspend transition), whenever either of the following 
happens:
 \begin{compactenum}[(a)]
  \item $T_{\Upsilon_i}(s_{\Upsilon_j}^b, s_{\Upsilon_j}^{b'}) \in 
T_{\Upsilon_j}^C$ and Condition-(i) / Condition-(iii-b) 
is satisfied from above, or

  \item $T_{\Upsilon_i}(s_{\Upsilon_i}^a, s_{\Upsilon_i}^{a'}) \in 
T_{\Upsilon_i}^C$ and Condition-(ii) / Condition-(iii-b) 
is satisfied from above.
 \end{compactenum}
\end{itemize}

 \item ${\tt Expr}_{\Upsilon_S}$ indicates the function for deriving algebraic
reliability expression. For a state, $s_{\Upsilon_S}^k \in S_{\Upsilon_S}$, 
where $\Gamma_{\Upsilon_S}(s_{\Upsilon_S}^k) = 
\Gamma_{\Upsilon_i}(s_{\Upsilon_i}^a) \odot
\Gamma_{\Upsilon_j}(s_{\Upsilon_j}^b)$ with $s_{\Upsilon_i}^a \in 
S_{\Upsilon_i},\ s_{\Upsilon_j}^b \in S_{\Upsilon_j}$, we have:
\[
{\tt Expr}_{\Upsilon_S}(s_{\Upsilon_S}^k) \leftarrow \big{[} {\tt 
Expr}_{\Upsilon_i}(s_{\Upsilon_i}^a) \big{]} . \big{[} {\tt 
Expr}_{\Upsilon_j}(s_{\Upsilon_j}^b) \big{]}\ 
\Big{|}_{r_{\Upsilon_i,k_i}^p = r_{\Upsilon_i,k_i}^{} \mbox{ and } 
r_{\Upsilon_j,k_j}^p = r_{\Upsilon_j,k_j}^{}} (1 \leq k_i \leq 
e_{\Upsilon_i},\ 1 \leq k_j \leq e_{\Upsilon_j}, \text{ and } p \in 
\mathbb{N})
\]
which means that, after making the simplified expression of ${\tt 
Expr}_{\Upsilon_S}(s_{\Upsilon_S}^k)$, all the common symbolic reliability 
variables, leading to the raise in their power of terms, are {\em normalized}. 
Intuitively, this step takes care of the effect of having multiple 
instantiation/invocation of the same component inside the system structure 
while calculating operating probability from the generated 
expressions~\cite{KK2007BOOK}.
\end{itemize}
The following example demonstrates the {\tt QRModel} derivation for a series 
system from the given {\tt QRModel} for each of its constituent 
components/subsystems.
\begin{example} \label{ex:series_comp}
Let us revisit the 2-component series system ($\Upsilon_S$) shown in 
Figure~\ref{fig:series_parallel}(a) and determine the {\tt QRModel} for this, 
say $\mathcal{Q}_{\Upsilon_S}$, comprising of the component $C_3$ followed by 
the component $C_2$. Formally, $\mathcal{Q}_{\Upsilon_S} = \mathcal{Q}_3\ 
\circ\ \mathcal{Q}_2$, where $\mathcal{Q}_3$ and $\mathcal{Q}_2$ are already 
defined in Example~\ref{ex:comp_qual_measure} (with Figure~\ref{fig:comp_qual} 
and Table~\ref{tab:comp_qual}).
\begin{figure}
\centering
\includegraphics[width=0.9\textwidth]{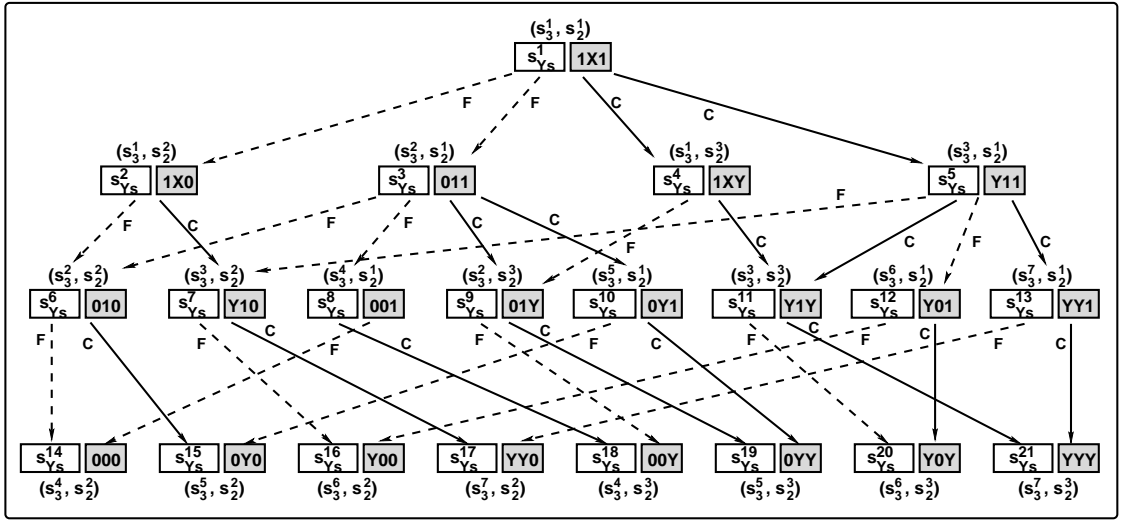}
\caption{State Transition Diagram for the Example {\tt QRModel} after Series 
Composition: $\mathcal{Q}_{\Upsilon_S} = \mathcal{Q}_3\ \circ\ \mathcal {Q}_2$}
\label{fig:compose_qual_ser}
\end{figure}
\begin{table}
{\sf
\centering
 \begin{tabular}{ccccccc}
    \cline{1-7}
    State & Composite & Operating Mode & Input Quality & Output Quality & 
\multicolumn{2}{c}{Operating Mode Probability}\\
    \cline{6-7}
    ID & State Pair & Configuration & Values (Levels) & Values (Levels) & 
Algebraic Expression & Value\\
    \cline{1-7}
    $s_{\Upsilon_S}^1$ & $(s_3^1,s_2^1)$ & {\tt 1X1} & $\langle 50,20 \rangle$ & 
$\langle 30,10 \rangle$ & $r_{3,1}.r_{2,1}$ & $0.855$\\
    $s_{\Upsilon_S}^2$ & $(s_3^1,s_2^2)$ & {\tt 1X0} & $\langle 0 \rangle$ & 
$\langle 0 \rangle$ & $r_{3,1}.(1-r_{2,1})$ & $0.045$\\
    $s_{\Upsilon_S}^3$ & $(s_3^2,s_2^1)$ & {\tt 011} & $\langle 50,20 \rangle$ & 
$\langle 30,10 \rangle$& $(1-r_{3,1}).2_{3,2}.r_{2,1}$ & $0.076$\\
    $s_{\Upsilon_S}^4$ & $(s_3^1,s_2^3)$ & {\tt 1XY} & $\langle 0 \rangle$ & 
$\langle 0 \rangle$ & $r_{3,1}$ & $0.900$\\
    $s_{\Upsilon_S}^5$ & $(s_3^3,s_2^1)$ & {\tt Y11} & $\langle 50,20 \rangle$ 
& $\langle 30,10 \rangle$ & $r_{3,2}.r_{2,1}$ & $0.760$\\
    $s_{\Upsilon_S}^6$ & $(s_3^2,s_2^2)$ & {\tt 010} & $\langle 0 \rangle$ & 
$\langle 0 \rangle$ & $(1-r_{3,1}).r_{3,2}.(1-r_{2,1})$ & $0.004$\\
    $s_{\Upsilon_S}^7$ & $(s_3^3,s_2^2)$ & {\tt Y10} & $\langle 0 \rangle$ & 
$\langle 0 \rangle$ & $r_{3,2}.(1-r_{2,1})$ & $0.040$\\
    $s_{\Upsilon_S}^8$ & $(s_3^4,s_2^1)$ & {\tt 001} & $\langle 0 \rangle$ & 
$\langle 0 \rangle$ & $(1-r_{3,1}).(1-r_{3,2}).r_{2,1}$ & $0.019$\\
    $s_{\Upsilon_S}^9$ & $(s_3^2,s_2^3)$ & {\tt 01Y} & $\langle 0 \rangle$ & 
$\langle 0 \rangle$ & $(1-r_{3,1}).r_{3,2}$ & $0.080$\\
    $s_{\Upsilon_S}^{10}$ & $(s_3^5,s_2^1)$ & {\tt 0Y1} & $\langle 0 \rangle$ & 
$\langle 0 \rangle$ & $(1-r_{3,1}).r_{2,1}$ & $0.095$\\
    $s_{\Upsilon_S}^{11}$ & $(s_3^3,s_2^3)$ & {\tt Y1Y} & $\langle 0 \rangle$ & 
$\langle 0 \rangle$ & $r_{3,2}$ & $0.800$\\
    $s_{\Upsilon_S}^{12}$ & $(s_3^6,s_2^1)$ & {\tt Y01} & $\langle 0 \rangle$ & 
$\langle 0 \rangle$ & $(1-r_{3,2}).r_{2,1}$ & $0.190$\\
    $s_{\Upsilon_S}^{13}$ & $(s_3^7,s_2^1)$ & {\tt YY1} & $\langle 0 \rangle$ & 
$\langle 0 \rangle$ & $r_{2,1}$ & $0.950$\\
    $s_{\Upsilon_S}^{14}$ & $(s_3^4,s_2^2)$ & {\tt 000} & $\langle 0 \rangle$ & 
$\langle 0 \rangle$ & $(1-r_{3,1}),(1-r_{3,2}).(1-r_{2,1})$ & $0.001$\\
    $s_{\Upsilon_S}^{15}$ & $(s_3^5,s_2^2)$ & {\tt 0Y0} & $\langle 0 \rangle$ & 
$\langle 0 \rangle$ & $(1-r_{3,1}).(1-r_{2,1})$ & $0.005$\\
    $s_{\Upsilon_S}^{16}$ & $(s_3^6,s_2^2)$ & {\tt Y00} & $\langle 0 \rangle$ & 
$\langle 0 \rangle$ & $(1-r_{3,2}).(1-r_{2,1})$ & $0.010$\\
    $s_{\Upsilon_S}^{17}$ & $(s_3^7,s_2^2)$ & {\tt YY0} & $\langle 0 \rangle$ & 
$\langle 0 \rangle$ & $(1-r_2{,1})$ & $0.050$\\
    $s_{\Upsilon_S}^{18}$ & $(s_3^4,s_2^3)$ & {\tt 00Y} & $\langle 0 \rangle$ & 
$\langle 0 \rangle$ & $(1-r_{3,1}).(1-r_{3,2})$ & $0.020$\\
    $s_{\Upsilon_S}^{19}$ & $(s_3^5,s_2^3)$ & {\tt 0YY} & $\langle 0 \rangle$ & 
$\langle 0 \rangle$ & $(1-r_{3,1})$ & $0.100$\\
    $s_{\Upsilon_S}^{20}$ & $(s_3^6,s_2^3)$ & {\tt Y0Y} & $\langle 0 \rangle$ & 
$\langle 0 \rangle$ & $(1-r_{3,2})$ & $0.200$\\
    $s_{\Upsilon_S}^{21}$ & $(s_3^7,s_2^3)$ & {\tt YYY} & $\langle 0 \rangle$ & 
$\langle 0 \rangle$ & $1$ & $1.000$\\
    \cline{1-7}
    \end{tabular}
 \caption{{\tt QRModel} Configuration Details after the Example Series 
Composition: $\mathcal{Q}_{\Upsilon_S} = \mathcal{Q}_3\ \circ\ \mathcal 
{Q}_2$}
 \label{tab:compose_qual_ser}
}
\end{table}
Here for $\Upsilon_S$, the derived {\tt QRModel}, 
$\mathcal{Q}_{\Upsilon_S}$, which is represented by the state transition diagram 
and the configuration details corresponding to every state, is expressed in 
Figure~\ref{fig:compose_qual_ser} with Table~\ref{tab:compose_qual_ser}.
For $\Upsilon_S$, we have:
\begin{itemize}
 \item The set of $2$ participating components as, $C_{\Upsilon_S} = \{ C_3, 
C_2\}$
 
 \item The set of $3$ input quality levels as, $Q_{\Upsilon_S}^I = 
\{50,20,10\}$.
 
 \item The set of $3$ symbolic reliability variables as, $R_{\Upsilon_S} = \{ 
r_{3,1}, r_{3,2}, r_{2,1} \}$ with values, $0.9, 0.8, 0.95$, respectively.
 
 \item The set of $21$ states, $S_{\Upsilon_S} = \{ s_{\Upsilon_S}^1, 
s_{\Upsilon_S}^2, \ldots, s_{\Upsilon_S}^{21} \}$.

 \item The state configurations are given as,\\ 
$\Gamma_{\Upsilon_S}(s_{\Upsilon_S}^1) = \Gamma_{\Upsilon_3}(s_3^1) \odot 
\Gamma_{\Upsilon_2}(s_2^1) = {\tt 1X1}, \quad 
\Gamma_{\Upsilon_S}(s_{\Upsilon_S}^2) = \Gamma_{\Upsilon_3}(s_3^1) \odot 
\Gamma_{\Upsilon_2}(s_2^2) = {\tt 1X0},\ \ldots$ so on.

 \item All derived output quality values are given in 
Figure~\ref{fig:compose_qual_ser}. To exemplify the derivation,\\
$Q_{\Upsilon_S}^O(s_{\Upsilon_S}^1,50) = Q_2^O(s_2^1,Q_3^O(s_3^1,50)) = 
Q_2^O(s_2^1,45) = Q_2^O(s_2^1,40) = 30$ \quad (since, $Q_3^O(s_3^1,50) = 45 > 
40$).

 \item All state transitions (both failure and suspension) are shown in 
Figure~\ref{fig:compose_qual_ser}. To exemplify the derivation,
\begin{itemize}
 \item The failure transitions are, $T_{\Upsilon_S}^F = \big{\{} 
(s_{\Upsilon_S}^1, s_{\Upsilon_S}^2), (s_{\Upsilon_S}^1, s_{\Upsilon_S}^3), 
\ldots \big{\}}$.\\
(marked with dotted edges and labeled using {\sf F})

 \item The suspend transitions are, $T_{\Upsilon_S}^C = \big{\{} 
(s_{\Upsilon_S}^1, 
s_{\Upsilon_S}^4), (s_{\Upsilon_S}^1, s_{\Upsilon_S}^5), \ldots \big{\}}$.\\
(marked with solid edges and labeled using {\sf C})
\end{itemize}

 \item All algebraic operational probability expressions are shown in 
Figure~\ref{fig:compose_qual_ser}. To exemplify the derivation,\\
${\tt Expr}_{\Upsilon_S}(s_{\Upsilon_S}^1) = r_{3,1}.r_{2,1};\ {\tt 
Expr}_{\Upsilon_S}(s_{\Upsilon_S}^2) = r_{3,1}.(1-r_{2,1});\ {\tt 
Expr}_{\Upsilon_S}(s_{\Upsilon_S}^3) = (1-r_{3,1}).r_{3,2}.r_{2,1};\ \ldots$ so 
on.
\end{itemize}
There may be multiple choices from an operational state upon its failure or 
suspend and such a choice can lead to any one among these options {\em 
non-deterministically}.
\qed
\end{example}

\subsubsection{Parallel Composition}
Let the two subsystem structures, $\Upsilon_i$ and $\Upsilon_j$ have the 
respective {\tt QRModel} as,
\[ \mathcal{Q}_{\Upsilon_i} = \langle C_{\Upsilon_i}, Q_{\Upsilon_i}^I, 
R_{\Upsilon_i}, S_{\Upsilon_i}, s_{\Upsilon_i}^1, \Gamma_{\Upsilon_i}, 
Q_{\Upsilon_i}^O, T_{\Upsilon_i}, {\tt Expr}_{\Upsilon_i} \rangle \text{ and } 
\mathcal{Q}_{\Upsilon_j} = \langle C_{\Upsilon_j}, Q_{\Upsilon_j}^I, 
R_{\Upsilon_j}, S_{\Upsilon_j}, s_{\Upsilon_j}^1, \Gamma_{\Upsilon_j}, 
Q_{\Upsilon_j}^O, T_{\Upsilon_j}, {\tt Expr}_{\Upsilon_j} \rangle. \]
These are connected in parallel forming the composite system, $\Upsilon_P \equiv 
\Upsilon_i\ ||\ \Upsilon_j$, whose {\tt QRModel}, can be derived as follows: 
\quad
$\mathcal{Q}_{\Upsilon_i}\ ||\ \mathcal{Q}_{\Upsilon_j} \equiv
\mathcal{Q}_{\Upsilon_P} = \langle C_{\Upsilon_P}, Q_{\Upsilon_P}^I, 
R_{\Upsilon_P}, S_{\Upsilon_P}, s_{\Upsilon_P}^1, \Gamma_{\Upsilon_P}, 
Q_{\Upsilon_P}^O, T_{\Upsilon_P}, {\tt Expr}_{\Upsilon_P} \rangle$, where:
\begin{itemize}
 \item $C_{\Upsilon_P} = C_{\Upsilon_i} \cup C_{\Upsilon_j}$, denotes the set 
of participating components in $\Upsilon_P$.

 \item $Q_{\Upsilon_P}^I = Q_{\Upsilon_i}^I \cup Q_{\Upsilon_j}^I$,
denotes the combined set of $l_{\Upsilon_P}$ input quality values (levels) used 
by $C_{\Upsilon_i}$ and $C_{\Upsilon_j}$.
%

 \item $R_{\Upsilon_P} = R_{\Upsilon_i} \cup R_{\Upsilon_j}$,
denotes the combined set of all symbolic reliability variables with 
$|R_{\Upsilon_P}| = e_{\Upsilon_P}$.
%

 \item $S_{\Upsilon_P}$ is the set of states (with $s_{\Upsilon_P}^1 \in 
S_{\Upsilon_P}$ being the start state) representing state configurations of 
$\mathcal{Q}_{\Upsilon_P}$.

 \item $\Gamma_{\Upsilon_P}: S_{\Upsilon_P} \rightarrow \{ {\tt 1, 0, X, Y}
\}^{e_{\Upsilon_P}^{}}$ is the $e_{\Upsilon_P}$-length state configuration, 
such that for a state, $s_{\Upsilon_P}^k \in S_{\Upsilon_P}$, the composed 
state configuration, $s_{\Upsilon_P}^k$ is obtained from states 
$s_{\Upsilon_i}^a \in S_{\Upsilon_i}\ (1 \leq a \leq n_{\Upsilon_i})$ 
and $s_{\Upsilon_j}^b \in S_{\Upsilon_j}\ (1 \leq b \leq 
n_{\Upsilon_j})$, denoted as 
$\Gamma_{\Upsilon_P}(s_{\Upsilon_P}^k) = 
\Gamma_{\Upsilon_i}(s_{\Upsilon_i}^a) \odot 
\Gamma_{\Upsilon_j}(s_{\Upsilon_j}^b)$, 
as follows: \quad ($e_{\Upsilon_i} \leq e_{\Upsilon_P} \leq e_{\Upsilon_i} + 
e_{\Upsilon_j}$)
\begin{itemize}
 \item $\Gamma_{\Upsilon_P}(s_{\Upsilon_P}^k)[1..e_{\Upsilon_i}] = 
\Gamma_{\Upsilon_i}(s_{\Upsilon_i}^a)[1..e_{\Upsilon_i}]$, and

 \item $\forall v\ (1 \leq v \leq  e_{\Upsilon_j}),\ \exists u\ (u \leq v)$, 
$\Gamma_{\Upsilon_P}(s_{\Upsilon_P}^k)[e_{\Upsilon_i}+u] = 
\Gamma_{\Upsilon_j}(s_{\Upsilon_j}^b)[v]\ (1 \leq u 
\leq e_{\Upsilon_P} - e_{\Upsilon_i})$ when $r_{\Upsilon_j,v}^{} \in 
R_{\Upsilon_j}$ but $r_{\Upsilon_j,v}^{} \notin R_{\Upsilon_i}$.\\
Intuitively, the last rule prevents the composed configuration to create 
duplicate configuration entries in case of multiple occurrence of the same 
component in both $C_{\Upsilon_i}$ and $C_{\Upsilon_j}$ subsystems.
\end{itemize}

 \item $Q_{\Upsilon_P}^O: S_{\Upsilon_P} \times Q_{\Upsilon_P}^I \rightarrow
\mathbb{R}^{+}$ indicates the function to compute the output quality value.\\
For a state, $s_{\Upsilon_P}^k \in S_{\Upsilon_P}$, where
$\Gamma_{\Upsilon_P}(s_{\Upsilon_P}^k) = 
\Gamma_{\Upsilon_i}(s_{\Upsilon_i}^a) \odot
\Gamma_{\Upsilon_j}(s_{\Upsilon_j}^b)\ (s_{\Upsilon_i}^a \in 
S_{\Upsilon_i},\ s_{\Upsilon_j}^b \in S_{\Upsilon_j})$, we generate:\\
from $q_{\Upsilon_i}^x, q_{\Upsilon_i}^{x'} \in Q_{\Upsilon_i}^I,\ 
q_{\Upsilon_j}^y, q_{\Upsilon_j}^{y'} \in Q_{\Upsilon_j}^I$ (where $1 
\leq 
x,x' \leq l_{\Upsilon_i}$, $1 \leq y,y' \leq l_{\Upsilon_j}$) and $\forall 
w\ (1 \leq w \leq l_{\Upsilon_P})$,

\

$\rhd$ \underline{\em Maximum Quality Output}: In this case, the overall 
quality measure is produced by choosing the maximum quality value from various 
operational modes of the components. Formally, we express this as:\\
\[
Q_{\Upsilon_P}^O(s_{\Upsilon_P}^k, q_{\Upsilon_P}^w) = 
\left\{ \begin{array}{rl}
  {\tt MAX}\big{[} Q_{\Upsilon_i}^O(s_{\Upsilon_i}^a,q_{\Upsilon_i}^x), 
Q_{\Upsilon_j}^O(s_{\Upsilon_j}^b,q_{\Upsilon_j}^y) \big{]}, & \mbox{when 
}
\exists q_{\Upsilon_i}^x,\ \exists q_{\Upsilon_j}^y \mbox{ such that } 
q_{\Upsilon_j}^y \leq q_{\Upsilon_i}^x \leq 
q_{\Upsilon_P}^w, \mbox{ and}\\
& ~~~~~~~~\nexists q_{\Upsilon_j}^{y'} \nexists q_{\Upsilon_i}^{x'}  \mbox{ 
such that } q_{\Upsilon_j}^y < 
q_{\Upsilon_j}^{y'} \leq q_{\Upsilon_i}^x < q_{\Upsilon_i}^{x'} \leq 
q_{\Upsilon_P}^w\\
& ~~~\mbox{ or, } \exists q_{\Upsilon_j}^y,\ \exists q_{\Upsilon_i}^x \mbox{ 
such that } q_{\Upsilon_i}^x < 
q_{\Upsilon_j}^y \leq q_{\Upsilon_P}^w, \mbox{ and}\\
& ~~~~~~~~\nexists q_{\Upsilon_i}^{x'} \nexists q_{\Upsilon_j}^{y'} \mbox{ 
such that } q_{\Upsilon_i}^x < 
q_{\Upsilon_i}^{x'} \leq q_{\Upsilon_j}^y < q_{\Upsilon_j}^{y'} \leq  
q_{\Upsilon_P}^w\\
  Q_{\Upsilon_i}^O(s_{\Upsilon_i}^a,q_{\Upsilon_i}^x), & \mbox{when } 
\forall q_{\Upsilon_j}^y,\ \exists q_{\Upsilon_i}^x \mbox{ such 
that } q_{\Upsilon_i}^x \leq q_{\Upsilon_P}^w < q_{\Upsilon_j}^y, \mbox{ 
and}\\
& ~~~~~~~~\nexists q_{\Upsilon_i}^{x'}  \mbox{ such that } q_{\Upsilon_i}^x 
< q_{\Upsilon_i}^{x'} \leq 
q_{\Upsilon_P}^w\\
  Q_{\Upsilon_j}^O(s_{\Upsilon_j}^b,q_{\Upsilon_j}^y), & \mbox{when } 
\forall q_{\Upsilon_i}^x,\ \exists q_{\Upsilon_j}^y \mbox{ such 
that } q_{\Upsilon_j}^y \leq q_{\Upsilon_P}^w < q_{\Upsilon_i}^x, \mbox{ 
and}\\
& ~~~~~~~~\nexists q_{\Upsilon_j}^{y'}  \mbox{ such that } q_{\Upsilon_j}^y 
< q_{\Upsilon_j}^{y'} \leq 
q_{\Upsilon_P}^w\\
  0, & \mbox{when } \forall q_{\Upsilon_i}^x,\ \forall q_{\Upsilon_j}^y 
\mbox{ such 
that } q_{\Upsilon_P}^w < q_{\Upsilon_i}^x, \mbox{ and } q_{\Upsilon_P}^w < 
q_{\Upsilon_j}^y\\
& ~~~~~~~~\mbox{(otherwise)}
\end{array}\right.
\]
Here, the output quality measure is derived based on the best output value 
possible (given an input level) considering both the output functions of 
$C_{\Upsilon_i}$ and $C_{\Upsilon_j}$.

\

$\rhd$ \underline{\em Ordered Quality Output}: In this case, the out quality is 
governed by a pre-defined ordering among the components/subsystems, say, 
$C_{\Upsilon_i} 
\prec C_{\Upsilon_j}$. Formally, we express this as:
\[
Q_{\Upsilon_P}^O(s_{\Upsilon_P}^k, q_{\Upsilon_P}^w) = 
\left\{ \begin{array}{rl}
  0, & \mbox{when } \forall q_{\Upsilon_i}^x,\ \forall q_{\Upsilon_j}^y 
\mbox{ such 
that } q_{\Upsilon_P}^w < q_{\Upsilon_i}^x, \mbox{ and } q_{\Upsilon_P}^w < 
q_{\Upsilon_j}^y\\
  Q_{\Upsilon_j}^O(s_{\Upsilon_j}^b,q_{\Upsilon_j}^y), & \mbox{when } 
\forall q_{\Upsilon_i}^x,\ \exists q_{\Upsilon_j}^y \mbox{ such 
that } q_{\Upsilon_j}^y \leq q_{\Upsilon_P}^w < q_{\Upsilon_i}^x, \mbox{ 
and}\\
& ~~~~~~~~\nexists q_{\Upsilon_j}^{y'}  \mbox{ such that } q_{\Upsilon_j}^y 
< q_{\Upsilon_j}^{y'} \leq 
q_{\Upsilon_P}^w\\
  Q_{\Upsilon_i}^O(s_{\Upsilon_i}^a,q_{\Upsilon_i}^x), & \mbox{otherwise}
\end{array}\right.
\]
Here, the output quality measure is derived based on the following (ordering) 
rules:
\begin{itemize}
 \item When both the subsystems, $C_{\Upsilon_i}$ and $C_{\Upsilon_j}$, are 
operational then it is 
the output quality measure provided by $C_{\Upsilon_i}$, provided its input 
quality level 
is above the lowest input quality value supported (only otherwise we switch to 
$C_{\Upsilon_j}$'s output quality function),
 \item When the subsystem $C_{\Upsilon_i}$ fails or suspended completely and 
$C_{\Upsilon_j}$ remains 
operational, then it is the output quality measure provided by 
$C_{\Upsilon_j}$, and
 \item It is $0$ (zero) when both $C_{\Upsilon_i}$ and $C_{\Upsilon_j}$ 
fails or suspended completely.
 \end{itemize}

 \item $T_{\Upsilon_P} \subseteq S_{\Upsilon_P} \times S_{\Upsilon_P}$ denotes 
the transition relation and let $T_{\Upsilon_P} = T_{\Upsilon_P}^F \cup
T_{\Upsilon_P}^C$, where $T_{\Upsilon_P}^F$ and $T_{\Upsilon_P}^C$ are the set 
of failure and suspend transitions, respectively. Suppose, 
$T_{\Upsilon_i}(s_{\Upsilon_i}^a,s_{\Upsilon_i}^{a'})$ 
and $T_{\Upsilon_j}(s_{\Upsilon_j}^b,s_{\Upsilon_j}^{b'})$ are transition 
in $\mathcal{Q}_{\Upsilon_i}$ and  
$\mathcal{Q}_{\Upsilon_j}$, respectively where $s_{\Upsilon_i}^a, 
s_{\Upsilon_i}^{a'} \in S_{\Upsilon_i}\ (1 \leq a,a' \leq 
n_{\Upsilon_i})$ and $s_{\Upsilon_j}^b, s_{\Upsilon_j}^{b'} \in 
S_{\Upsilon_j}\ (1 \leq b,b' \leq n_{\Upsilon_j})$. For 
$s_{\Upsilon_P}^k \in S_{\Upsilon_P}$, if
$\Gamma_{\Upsilon_P}(s_{\Upsilon_P}^k) = 
\Gamma_{\Upsilon_i}(s_{\Upsilon_i}^a) \odot 
\Gamma_{\Upsilon_j}(s_{\Upsilon_j}^b)$, 
then $T_{\Upsilon_P}(s_{\Upsilon_P}^k,s_{\Upsilon_P}^{k'})$ is a permissible 
transition ($\exists s_{\Upsilon_P}^{k'} \in S_{\Upsilon_P}$) such that one of 
the following three conditions hold:
\begin{enumerate}
\item[(i)] $\Gamma_{\Upsilon_P}(s_{\Upsilon_P}^{k'}) = 
\Gamma_{\Upsilon_i}(s_{\Upsilon_i}^a) \odot
\Gamma_{\Upsilon_j}(s_{\Upsilon_j}^{b'})$,
~~or~~
(ii) $\Gamma_{\Upsilon_P}(s_{\Upsilon_P}^{k'}) = 
\Gamma_{\Upsilon_i}(s_{\Upsilon_i}^{a'})
\odot \Gamma_{\Upsilon_j}(s_{\Upsilon_j}^b)$, ~~or

\item[(iii)] $\Gamma_{\Upsilon_P}(s_{\Upsilon_P}^{k'}) = 
\Gamma_{\Upsilon_i}(s_{\Upsilon_i}^{a'}) \odot
\Gamma_{\Upsilon_j}(s_{\Upsilon_j}^{b'})$ along with either of the 
following 
two constraints:\\
(a) $T_{\Upsilon_i}(s_{\Upsilon_i}^a, s_{\Upsilon_i}^{a'}) \in 
T_{\Upsilon_i}^F$; $T_{\Upsilon_j}(s_{\Upsilon_j}^b, 
s_{\Upsilon_j}^{b'}) \in T_{\Upsilon_j}^F$, ~~or~~
(b) $T_{\Upsilon_i}(s_{\Upsilon_i}^a, s_{\Upsilon_i}^{a'}) \in 
T_{\Upsilon_i}^C$; $T_{\Upsilon_j}(s_{\Upsilon_j}^b, 
s_{\Upsilon_j}^{b'}) \in T_{\Upsilon_j}^C$.
\end{enumerate}
Further, it is important to note the categorization of transitions (failure or 
suspended) here as follows:
\begin{itemize}
 \item $T_{\Upsilon_P}(s_{\Upsilon_P}^k,s_{\Upsilon_P}^{k'}) \in 
T_{\Upsilon_P}^F$ (is a failure transition), whenever either of the following 
happens:
 \begin{compactenum}[(a)]
  \item $T_{\Upsilon_i}(s_{\Upsilon_j}^b, s_{\Upsilon_j}^{b'}) \in 
T_{\Upsilon_j}^F$ and Condition-(i) / Condition-(iii-a) 
is satisfied from above, or

  \item $T_{\Upsilon_i}(s_{\Upsilon_i}^a, s_{\Upsilon_i}^{a'}) \in 
T_{\Upsilon_i}^F$ and Condition-(ii) / Condition-(iii-a) 
is satisfied from above.
 \end{compactenum}

 \item $T_{\Upsilon_P}(s_{\Upsilon_P}^k,s_{\Upsilon_P}^{k'}) \in 
T_{\Upsilon_P}^C$ (is a suspend transition), whenever either of the following 
happens:
 \begin{compactenum}[(a)]
  \item $T_{\Upsilon_i}(s_{\Upsilon_j}^b, s_{\Upsilon_j}^{b'}) \in 
T_{\Upsilon_j}^C$ and Condition-(i) / Condition-(iii-b) 
is satisfied from above, or

  \item $T_{\Upsilon_i}(s_{\Upsilon_i}^a, s_{\Upsilon_i}^{a'}) \in 
T_{\Upsilon_i}^C$ and Condition-(ii) / Condition-(iii-b) 
is satisfied from above.
 \end{compactenum}
\end{itemize}

 \item ${\tt Expr}_{\Upsilon_P}$ indicates the function for deriving algebraic
reliability expression. For a state, $s_{\Upsilon_P}^k \in S_{\Upsilon_P}$, 
where $\Gamma_{\Upsilon_P}(s_{\Upsilon_P}^k) = 
\Gamma_{\Upsilon_i}(s_{\Upsilon_i}^a) \odot
\Gamma_{\Upsilon_j}(s_{\Upsilon_j}^b)$ with $s_{\Upsilon_i}^a \in 
S_{\Upsilon_i},\ s_{\Upsilon_j}^b \in S_{\Upsilon_j}$, we have:
\[
{\tt Expr}_{\Upsilon_P}(s_{\Upsilon_P}^k) \leftarrow \big{[} {\tt 
Expr}_{\Upsilon_i}(s_{\Upsilon_i}^a) \big{]} . \big{[} {\tt 
Expr}_{\Upsilon_j}(s_{\Upsilon_j}^b) \big{]}\ 
\Big{|}_{r_{\Upsilon_i,k_i}^p = r_{\Upsilon_i,k_i}^{} 
\mbox{ and } r_{\Upsilon_j,k_j}^p = r_{\Upsilon_j,k_j}^{}} (1 \leq k_i \leq 
e_{\Upsilon_i},\ 1 \leq k_j \leq e_{\Upsilon_j}, \text{ and } p \in 
\mathbb{N})
\]
which means that, after making the simplified expression of ${\tt 
Expr}_{\Upsilon_P}(s_{\Upsilon_P}^k)$, all the common symbolic reliability 
variables, leading to the raise in their power of terms, are {\em normalized}. 
Intuitively, this step takes care of the effect of having multiple 
instantiation/invocation of the same component inside the system structure 
while calculating operating probability from the generated 
expressions~\cite{KK2007BOOK}.
\end{itemize}
The following example demonstrates the {\tt QRModel} derivation of a parallel 
system from the given {\tt QRModel} for each of its constituent 
components/subsystems.
\begin{example} \label{ex:parallel_comp}
Let us revisit the 2-component parallel system ($\Upsilon_P$) shown in 
Figure~\ref{fig:series_parallel}(b) and determine the {\tt QRModel} for this, 
say $\mathcal{Q}_{\Upsilon_P}$, comprising of the component $C_1$ connected in 
parallel with the component $C_2$. Formally, $\mathcal{Q}_{\Upsilon_P} = 
\mathcal{Q}_1\ ||\ \mathcal{Q}_2$, where $\mathcal{Q}_3$ and $\mathcal{Q}_2$ 
are already defined in Example~\ref{ex:comp_qual_measure} (with
Figure~\ref{fig:comp_qual} and Table~\ref{tab:comp_qual}).

\begin{figure}
\centering
\includegraphics[width=0.9\textwidth]{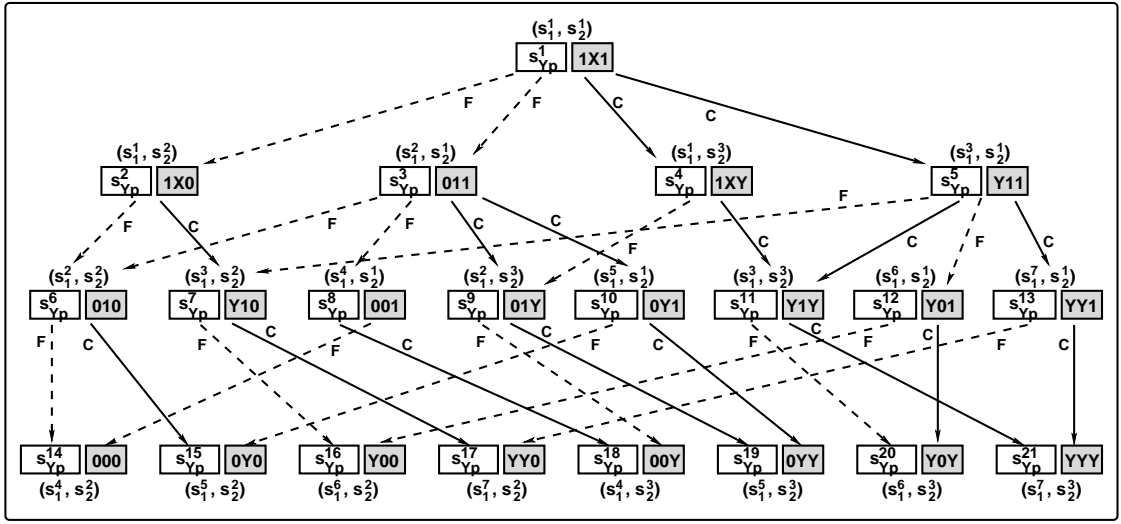}
\caption{State Transition Diagram for the Example {\tt QRModel} after Parallel 
Composition: $\mathcal{Q}_{\Upsilon_P} = \mathcal{Q}_1\ ||\ \mathcal 
{Q}_2$}
\label{fig:compose_qual_par}
\end{figure}

Here for $\Upsilon_P$, the derived {\tt QRModel}, 
$\mathcal{Q}_{\Upsilon_P}^{max}$ and $\mathcal{Q}_{\Upsilon_P}^{ord}$, 
considering both the {\em maximum} and the {\em ordered} quality output, which 
are represented by the state transition diagram and the configuration details 
corresponding to every state, is expressed in 
Figure~\ref{fig:compose_qual_par} with Table~\ref{tab:compose_qual_par_max} and 
Figure~\ref{fig:compose_qual_par} with Table~\ref{tab:compose_qual_par_ord}, 
respectively. Note that, the state transition, mode configuration and the 
operating mode reliability attributes are same for both these 
variants, however the input-output quality measures differ due to their varied 
treatment.
\begin{table}
{\sf
\centering
 \begin{tabular}{ccccccc}
    \cline{1-7}
    State & Composite & Operating Mode & Input Quality & Output Quality & 
\multicolumn{2}{c}{Operating Mode Probability}\\
    \cline{6-7}
    ID & State Pair & Configuration & Values (Levels) & Values (Levels) & 
Algebraic Expression & Value\\
    \cline{1-7}
    $s_{\Upsilon_P}^1$ & $(s_1^1,s_2^1)$ & {\tt 1X1} & $\langle 50,40,30,10 
\rangle$ & $\langle 40,30,25,10 \rangle$ & $r_{1,1}.r_{2,1}$ & $0.760$\\
    $s_{\Upsilon_P}^2$ & $(s_1^1,s_2^2)$ & {\tt 1X0} & $\langle 50,30,20 
\rangle$ & $\langle 40,25,10 \rangle$ & $r_{1,1}.(1-r_{2,1})$ & $0.040$\\
    $s_{\Upsilon_P}^3$ & $(s_1^2,s_2^1)$ & {\tt 011} & $\langle 50,40,30,10 
\rangle$ & $\langle 35,30,25,10 \rangle$& $(1-r_{1,1}).2_{3,2}.r_{2,1}$ & 
$0.133$\\
    $s_{\Upsilon_P}^4$ & $(s_1^1,s_2^3)$ & {\tt 1XY} & $\langle 50,30,20 
\rangle$ & $\langle 40,25,10 \rangle$ & $r_{1,1}$ & $0.800$\\
    $s_{\Upsilon_P}^5$ & $(s_1^3,s_2^1)$ & {\tt Y11} & $\langle 50,40,30,10 
\rangle$ & $\langle 35,30,25,10 \rangle$ & $r_{1,2}.r_{2,1}$ & $0.665$\\
    $s_{\Upsilon_P}^6$ & $(s_1^2,s_2^2)$ & {\tt 010} & $\langle 50,30,20 
\rangle$ & $\langle 35,25,10 \rangle$ & $(1-r_{1,1}).r_{1,2}.(1-r_{2,1})$ & 
$0.007$\\
    $s_{\Upsilon_P}^7$ & $(s_1^3,s_2^2)$ & {\tt Y10} & $\langle 50,30,20 
\rangle$ & $\langle 35,25,10 \rangle$ & $r_{1,2}.(1-r_{2,1})$ & $0.035$\\
    $s_{\Upsilon_P}^8$ & $(s_1^4,s_2^1)$ & {\tt 001} & $\langle 40,10 \rangle$ 
& $\langle 30,10 \rangle$ & $(1-r_{1,1}).(1-r_{1,2}).r_{2,1}$ & $0.057$\\
    $s_{\Upsilon_P}^9$ & $(s_1^2,s_2^3)$ & {\tt 01Y} & $\langle 50,30,20 
\rangle$ & $\langle 35,25,10 \rangle$ & $(1-r_{1,1}).r_{1,2}$ & $0.140$\\
    $s_{\Upsilon_P}^{10}$ & $(s_1^5,s_2^1)$ & {\tt 0Y1} & $\langle 40,10 
\rangle$ & $\langle 30,10 \rangle$ & $(1-r_{1,1}).r_{2,1}$ & $0.190$\\
    $s_{\Upsilon_P}^{11}$ & $(s_1^3,s_2^3)$ & {\tt Y1Y} & $\langle 50,30,20 
\rangle$ & $\langle 35,25,10 \rangle$ & $r_{1,2}$ & $0.700$\\
    $s_{\Upsilon_P}^{12}$ & $(s_1^6,s_2^1)$ & {\tt Y01} & $\langle 40,10 
\rangle$ & $\langle 30,10 \rangle$ & $(1-r_{1,2}).r_{2,1}$ & $0.285$\\
    $s_{\Upsilon_P}^{13}$ & $(s_1^7,s_2^1)$ & {\tt YY1} & $\langle 40,10 
\rangle$ & $\langle 30,10 \rangle$ & $r_{2,1}$ & $0.950$\\
    $s_{\Upsilon_P}^{14}$ & $(s_1^4,s_2^2)$ & {\tt 000} & $\langle 0 \rangle$ & 
$\langle 0 \rangle$ & $(1-r_{1,1}).(1-r_{1,2}).(1-r_{2,1})$ & $0.003$\\
    $s_{\Upsilon_P}^{15}$ & $(s_1^5,s_2^2)$ & {\tt 0Y0} & $\langle 0 \rangle$ & 
$\langle 0 \rangle$ & $(1-r_{1,1}).(1-r_{2,1})$ & $0.010$\\
    $s_{\Upsilon_P}^{16}$ & $(s_1^6,s_2^2)$ & {\tt Y00} & $\langle 0 \rangle$ & 
$\langle 0 \rangle$ & $(1-r_{1,2}).(1-r_{2,1})$ & $0.015$\\
    $s_{\Upsilon_P}^{17}$ & $(s_1^7,s_2^2)$ & {\tt YY0} & $\langle 0 \rangle$ & 
$\langle 0 \rangle$ & $(1-r_2{,1})$ & $0.050$\\
    $s_{\Upsilon_P}^{18}$ & $(s_1^4,s_2^3)$ & {\tt 00Y} & $\langle 0 \rangle$ & 
$\langle 0 \rangle$ & $(1-r_{1,1}).(1-r_{1,2})$ & $0.060$\\
    $s_{\Upsilon_P}^{19}$ & $(s_1^5,s_2^3)$ & {\tt 0YY} & $\langle 0 \rangle$ & 
$\langle 0 \rangle$ & $(1-r_{1,1})$ & $0.200$\\
    $s_{\Upsilon_P}^{20}$ & $(s_1^6,s_2^3)$ & {\tt Y0Y} & $\langle 0 \rangle$ & 
$\langle 0 \rangle$ & $(1-r_{1,2})$ & $0.300$\\
    $s_{\Upsilon_P}^{21}$ & $(s_1^7,s_2^3)$ & {\tt YYY} & $\langle 0 \rangle$ & 
$\langle 0 \rangle$ & $1$ & $1.000$\\
    \cline{1-7}
    \end{tabular}
 \caption{\small {\tt QRModel} Configuration Details after the Example Parallel 
Composition: $\mathcal{Q}_{\Upsilon_P}^{max} = \mathcal{Q}_1\ ||\ \mathcal 
{Q}_2$ {\em (Max. Quality O/P)}}
 \label{tab:compose_qual_par_max}
}
\end{table}
\begin{table}
{\sf
\centering
 \begin{tabular}{ccccccc}
    \cline{1-7}
    State & Composite & Operating Mode & Input Quality & Output Quality & 
\multicolumn{2}{c}{Operating Mode Probability}\\
    \cline{6-7}
    ID & State Pair & Configuration & Values (Levels) & Values (Levels) & 
Algebraic Expression & Value\\
    \cline{1-7}
    $s_{\Upsilon_P}^1$ & $(s_1^1,s_2^1)$ & {\tt 1X1} & $\langle 50,30,20 
\rangle$ & $\langle 40,25,10 \rangle$ & $r_{1,1}.r_{2,1}$ & $0.760$\\
    $s_{\Upsilon_P}^2$ & $(s_1^1,s_2^2)$ & {\tt 1X0} & $\langle 50,30,20 
\rangle$ & $\langle 40,25,10 \rangle$ & $r_{1,1}.(1-r_{2,1})$ & $0.040$\\
    $s_{\Upsilon_P}^3$ & $(s_1^2,s_2^1)$ & {\tt 011} & $\langle 50,30,20 
\rangle$ & $\langle 35,25,10 \rangle$& $(1-r_{1,1}).2_{3,2}.r_{2,1}$ & 
$0.133$\\
    $s_{\Upsilon_P}^4$ & $(s_1^1,s_2^3)$ & {\tt 1XY} & $\langle 50,30,20 
\rangle$ & $\langle 40,25,10 \rangle$ & $r_{1,1}$ & $0.800$\\
    $s_{\Upsilon_P}^5$ & $(s_1^3,s_2^1)$ & {\tt Y11} & $\langle 50,30,20 
\rangle$ & $\langle 35,25,10 \rangle$ & $r_{1,2}.r_{2,1}$ & $0.665$\\
    $s_{\Upsilon_P}^6$ & $(s_1^2,s_2^2)$ & {\tt 010} & $\langle 50,30,20 
\rangle$ & $\langle 35,25,10 \rangle$ & $(1-r_{1,1}).r_{1,2}.(1-r_{2,1})$ & 
$0.007$\\
    $s_{\Upsilon_P}^7$ & $(s_1^3,s_2^2)$ & {\tt Y10} & $\langle 50,30,20 
\rangle$ & $\langle 35,25,10 \rangle$ & $r_{1,2}.(1-r_{2,1})$ & $0.035$\\
    $s_{\Upsilon_P}^8$ & $(s_1^4,s_2^1)$ & {\tt 001} & $\langle 40,10 \rangle$ 
& $\langle 30,10 \rangle$ & $(1-r_{1,1}).(1-r_{1,2}).r_{2,1}$ & $0.057$\\
    $s_{\Upsilon_P}^9$ & $(s_1^2,s_2^3)$ & {\tt 01Y} & $\langle 50,30,20 
\rangle$ & $\langle 35,25,10 \rangle$ & $(1-r_{1,1}).r_{1,2}$ & $0.140$\\
    $s_{\Upsilon_P}^{10}$ & $(s_1^5,s_2^1)$ & {\tt 0Y1} & $\langle 40,10 
\rangle$ & $\langle 30,10 \rangle$ & $(1-r_{1,1}).r_{2,1}$ & $0.190$\\
    $s_{\Upsilon_P}^{11}$ & $(s_1^3,s_2^3)$ & {\tt Y1Y} & $\langle 50,30,20 
\rangle$ & $\langle 35,25,10 \rangle$ & $r_{1,2}$ & $0.700$\\
    $s_{\Upsilon_P}^{12}$ & $(s_1^6,s_2^1)$ & {\tt Y01} & $\langle 40,10 
\rangle$ & $\langle 30,10 \rangle$ & $(1-r_{1,2}).r_{2,1}$ & $0.285$\\
    $s_{\Upsilon_P}^{13}$ & $(s_1^7,s_2^1)$ & {\tt YY1} & $\langle 40,10 
\rangle$ & $\langle 30,10 \rangle$ & $r_{2,1}$ & $0.950$\\
    $s_{\Upsilon_P}^{14}$ & $(s_1^4,s_2^2)$ & {\tt 000} & $\langle 0 \rangle$ & 
$\langle 0 \rangle$ & $(1-r_{1,1}).(1-r_{1,2}).(1-r_{2,1})$ & $0.003$\\
    $s_{\Upsilon_P}^{15}$ & $(s_1^5,s_2^2)$ & {\tt 0Y0} & $\langle 0 \rangle$ & 
$\langle 0 \rangle$ & $(1-r_{1,1}).(1-r_{2,1})$ & $0.010$\\
    $s_{\Upsilon_P}^{16}$ & $(s_1^6,s_2^2)$ & {\tt Y00} & $\langle 0 \rangle$ & 
$\langle 0 \rangle$ & $(1-r_{1,2}).(1-r_{2,1})$ & $0.015$\\
    $s_{\Upsilon_P}^{17}$ & $(s_1^7,s_2^2)$ & {\tt YY0} & $\langle 0 \rangle$ & 
$\langle 0 \rangle$ & $(1-r_2{,1})$ & $0.050$\\
    $s_{\Upsilon_P}^{18}$ & $(s_1^4,s_2^3)$ & {\tt 00Y} & $\langle 0 \rangle$ & 
$\langle 0 \rangle$ & $(1-r_{1,1}).(1-r_{1,2})$ & $0.060$\\
    $s_{\Upsilon_P}^{19}$ & $(s_1^5,s_2^3)$ & {\tt 0YY} & $\langle 0 \rangle$ & 
$\langle 0 \rangle$ & $(1-r_{1,1})$ & $0.200$\\
    $s_{\Upsilon_P}^{20}$ & $(s_1^6,s_2^3)$ & {\tt Y0Y} & $\langle 0 \rangle$ & 
$\langle 0 \rangle$ & $(1-r_{1,2})$ & $0.300$\\
    $s_{\Upsilon_P}^{21}$ & $(s_1^7,s_2^3)$ & {\tt YYY} & $\langle 0 \rangle$ & 
$\langle 0 \rangle$ & $1$ & $1.000$\\
    \cline{1-7}
    \end{tabular}
 \caption{\small {\tt QRModel} Configuration Details after the Example Parallel 
Composition: $\mathcal{Q}_{\Upsilon_P}^{ord} = \mathcal{Q}_1\ ||\ \mathcal 
{Q}_2$ {\em (Ord. Quality O/P)}}
 \label{tab:compose_qual_par_ord}
}
\end{table}
For $\Upsilon_P$, we have:
\begin{itemize}
 \item The set of $2$ participating components as, $C_{\Upsilon_P} = \{ C_1, 
C_2\}$
 
 \item The set of $3$ input quality levels as, $Q_{\Upsilon_P}^I = 
\{50,40,30,20,10\}$.
 
 \item The set of $3$ symbolic reliability variables as, $R_{\Upsilon_P} = \{ 
r_{1,1}, r_{1,2}, r_{2,1} \}$ with values, $0.8, 0.7, 0.95$, respectively.
 
 \item The set of $21$ states, $S_{\Upsilon_P} = \{ s_{\Upsilon_P}^1, 
s_{\Upsilon_P}^2, \ldots, s_{\Upsilon_P}^{21} \}$.

 \item The state configurations are given as,\\ 
$\Gamma_{\Upsilon_P}(s_{\Upsilon_P}^1) = \Gamma_{\Upsilon_1}(s_1^1) \odot 
\Gamma_{\Upsilon_2}(s_2^1) = {\tt 1X1}, \quad 
\Gamma_{\Upsilon_P}(s_{\Upsilon_P}^2) = \Gamma_{\Upsilon_1}(s_1^1) \odot 
\Gamma_{\Upsilon_2}(s_2^2) = {\tt 1X0},\ \ldots$ so on.

 \item Under {\em maximum quality output} criteria, all derived output quality 
values are given in Table~\ref{tab:compose_qual_par_max}. To exemplify the 
derivation,\\
$Q_{\Upsilon_P}^O(s_{\Upsilon_P}^1,50) = {\tt MAX} \big{[} 
Q_1^O(s_1^1,50),Q_2^O(s_2^1,50) \big{]} = {\tt MAX} \big{[} 
Q_1^O(s_1^1,50),Q_2^O(s_2^1,40) \big{]} = {\tt MAX} \big{[} 40, 30 \big{]} = 
40$.\\
(since $\{50\} \notin Q_2^I$ whereas $\{40\} \in Q_2^I$ and $50 > 40$ being 
the immediate next input level)

Under {\em ordered quality output} criteria (with $C_1 \prec C_2$ as given 
order), all derived output quality values are given in 
Table~\ref{tab:compose_qual_par_ord}. To exemplify the 
derivation, let us understand the following two derivation:
\begin{itemize}
 \item $Q_{\Upsilon_P}^O(s_{\Upsilon_P}^1,50) = Q_1^O(s_1^1,50) = 40$
 (since $C_1 \prec C_2$ and $C_1$ has not failed),
 \item $Q_{\Upsilon_P}^O(s_{\Upsilon_P}^8,40) = Q_2^O(s_2^1,40) = 30$ 
(since $C_1$ has completely failed, $C_2$ dictates the output quality)
\end{itemize}

 \item All state transitions (both failure and suspension) are shown in 
Figure~\ref{fig:compose_qual_par}. To exemplify the derivation,
\begin{itemize}
 \item The failure transitions are, $T_{\Upsilon_P}^F = \big{\{} 
(s_{\Upsilon_P}^1, s_{\Upsilon_P}^2), (s_{\Upsilon_P}^1, s_{\Upsilon_P}^3), 
\ldots \big{\}}$.\\
(marked with dotted edges and labeled using {\sf F})

 \item The suspend transitions are, $T_{\Upsilon_P}^C = \big{\{} 
(s_{\Upsilon_P}^1, s_{\Upsilon_P}^4), (s_{\Upsilon_P}^1, s_{\Upsilon_P}^5), 
\ldots \big{\}}$.\\
(marked with solid edges and labeled using {\sf C})
\end{itemize}

 \item All algebraic operational probability expressions are shown in 
Table~\ref{tab:compose_qual_par_max} or in Table~\ref{tab:compose_qual_par_ord} 
(these are same for both variants). To exemplify the derivation,\\
${\tt Expr}_{\Upsilon_P}(s_{\Upsilon_P}^1) = r_{1,1}.r_{2,1};\ {\tt 
Expr}_{\Upsilon_P}(s_{\Upsilon_P}^2) = r_{1,1}.(1-r_{2,1});\ {\tt 
Expr}_{\Upsilon_P}(s_{\Upsilon_P}^3) = (1-r_{1,1}).r_{1,2}.r_{2,1};\ \ldots$ so 
on.
\end{itemize}
There may be multiple choices from an operational state upon its failure or 
suspend and such a choice can lead to any one among these options {\em 
non-deterministically}.
\qed
\end{example}
It may be noted that, the series and parallel composition mechanisms with 
respect to given subsystem structures are similar except the computation of 
output quality values. This is due to the fact that the underlying model of 
computation is a state transition diagram which composes pair of states from 
two subsystems, thereby maintaining identical configurations in both cases.

\subsection{Generic Composition Procedures} \label{subsec:generic_compose}
In order to compute the composed {\tt QRModel} for a generic series/parallel 
system structure, we hierarchically use the series and the parallel composition
operations, as mentioned above.

\subsubsection{Properties of Composition Operations}
Such compositional rules can be applied hierarchically due to the satisfaction 
of some inherent properties of composition operations, which are listed below.
\quad (Assume: $\mathcal{Q}_a, \mathcal{Q}_b, \mathcal{Q}_c$ are three {\tt 
QRModel}s.)
\begin{description}
 \item[{\em Idempotent Properties:}]\
 \begin{enumerate}[(a)]
  \item Series ($~\circ~$) composition operation is {\em not} idempotent, i.e., 
$\mathcal{Q}_a\ \circ\ \mathcal{Q}_a\ \not\equiv\ \mathcal{Q}_a$.

 \item Parallel ($~||~$) composition operation is idempotent, i.e., 
$\mathcal{Q}_a\ ||\ \mathcal{Q}_a\ \equiv\ \mathcal{Q}_a$.
 \end{enumerate}

 \item[{\em Associative Properties:}]\
 \begin{enumerate}[(a)]
  \item Series ($~\circ~$) composition operation is associative, i.e., 
$(\mathcal{Q}_a\ \circ\ \mathcal{Q}_b)\ \circ\ \mathcal{Q}_c\ \equiv\ 
\mathcal{Q}_a\ \circ\ (\mathcal{Q}_b\ \circ\ \mathcal{Q}_c)$.

 \item Parallel ($~||~$) composition operation is associative, i.e., 
$(\mathcal{Q}_a\ ||\ \mathcal{Q}_b)\ ||\ \mathcal{Q}_c\ \equiv\ \mathcal{Q}_a\ 
||\ (\mathcal{Q}_b\ ||\ \mathcal{Q}_c)$.
 \end{enumerate}

 \item[{\em Commutative Properties:}]\
 \begin{enumerate}[(a)]
  \item Series ($~\circ~$) composition operation is {\em not} commutative, 
i.e., $\mathcal{Q}_a\ \circ\ \mathcal{Q}_b\ \not\equiv\ \mathcal{Q}_b\ \circ\ 
\mathcal{Q}_a$.

 \item Parallel ($~||~$) composition operation is commutative, i.e., 
$\mathcal{Q}_a\ ||\ \mathcal{Q}_b\ \equiv\ \mathcal{Q}_b\ ||\ \mathcal{Q}_a$.
 \end{enumerate}

 \item[{\em Distributive Properties:}]\
 \begin{enumerate}[(a)]
  \item Series ($~\circ~$) composition operation is distributive over 
Parallel ($~||~$) composition operation, i.e., $\mathcal{Q}_a\ \circ\ 
(\mathcal{Q}_b\ ||\ \mathcal{Q}_c) \equiv\ (\mathcal{Q}_a\ \circ\ 
\mathcal{Q}_b)\ ||\ (\mathcal{Q}_a\ \circ\ \mathcal{Q}_c)$.

 \item Parallel ($~||~$) composition operation is {\em not} distributive over 
Series ($~\circ~$) composition operation, i.e., $\mathcal{Q}_a\ ||\ 
(\mathcal{Q}_b\ \circ\ \mathcal{Q}_c) \not\equiv\ (\mathcal{Q}_a\ ||\ 
\mathcal{Q}_b)\ \circ\ (\mathcal{Q}_a\ ||\ \mathcal{Q}_c)$.
 \end{enumerate}
The distributive nature of $\circ$ over $||$ establishes the fact that any 
series-parallel composition is equivalent to parallel composition of all paths 
(where each path components are composed in series) from input to output (which 
is basically the path-enumeration based rule).
\end{description}

\subsubsection{Composition Rules}
As an example, the {\tt QRModel} for the system, $\Upsilon_{sp}$ (refer to 
Figure~\ref{fig:system_example}), can be computed in two ways leveraging the 
generic composition techniques proposed above.
\begin{itemize}
 \item {\em Using Series/Parallel-Structure based Composition Rule}:
The {\tt QRModel} for the two subsystems, $\Upsilon_S$ (the system structure 
is shown in Figure~\ref{fig:series_parallel}(a)) and $\Upsilon_P$ (the system 
structure is shown in Figure~\ref{fig:series_parallel}(b)), are 
already computed in Example~\ref{ex:series_comp} and 
Example~\ref{ex:parallel_comp}, respectively. The {\tt QRModel} for the
overall system, $\Upsilon_{sp}$ (shown in Figure~\ref{fig:system_example}), 
i.e. $\mathcal{Q}_{\Upsilon_{sp}}$, is the parallel quality composition of 
$\mathcal{Q}_{\Upsilon_P}$ and $\mathcal{Q}_{\Upsilon_S}$ again, and it can be 
formally derived as: \quad
$\mathcal{Q}_{\Upsilon_{sp}} = (\mathcal{Q}_{\Upsilon_P}\ 
||\ \mathcal{Q}_{\Upsilon_S}) = \Big{(} (\mathcal{Q}_1\ ||\
\mathcal{Q}_2)\ ||\ (\mathcal{Q}_3\ \circ\ \mathcal{Q}_2) \Big{)}$

 \item {\em Using Path-Enumeration based Composition Rule}:
There are three paths from input to output of $\Upsilon_{sp}$, namely through
$C_1$, through $C_2$ and through $C_3$-$C_2$. The quality measure for the
overall system, $\Upsilon_{sp}$, that is $\mathcal{Q}_{\Upsilon_{sp}}$, is the 
parallel quality composition of $C_1$ (say $\mathcal{Q}_1$), $C_2$ (say 
$\mathcal{Q}_2$) and $\Upsilon_S$ (say $\mathcal{Q}_{\Upsilon_S}$), and it 
can be derived expressed as: \quad
$\mathcal{Q}_{\Upsilon_{sp}} = (\mathcal{Q}_1\ ||\ \mathcal{Q}_2\ ||\ 
\mathcal{Q}_{\Upsilon_S}) = \Big{(} \mathcal{Q}_1\ ||\ \mathcal{Q}_2\ ||\ 
(\mathcal{Q}_3\ \circ\ \mathcal{Q}_2) \Big{)}$

\end{itemize}
It is worthy to note two points here as follows:
\begin{itemize}
 \item The order of serial and parallel compositions may be arbitrary in a 
composite expression as long as their associative, commutative and distributive 
properties, as discussed above, remain intact.

 \item The quality configuration model for non-series/parallel systems is 
determined using the {\em path-enumeration} techniques. Here, we can apply the 
series composition rules over all the component structures that reside in every 
path from input to the output of the system. Finally, the parallel composition 
is applied over each of these new composed quality measures for each end-to-end 
path to get the overall system quality.
\end{itemize}

\section{Conformance Check for System-level {\tt QRSpec} from {\tt QRModel} 
Configurations} \label{sec:qual_conformance}
The {\tt QRModel} of a system describes the formal characterization of its
quality attributes using an underlying state-transition model based 
configuration. Such a formalization is useful for assessing quality and 
reliability properties of systems, which we shall be detailing in the next 
section. However, from an user perspective, the architectural {\tt QRSpec} of 
systems forms a foundational basis, as defined in 
Section~\ref{subsec:qual_spec}. Such type of specification, depicting system 
fault-tolerant behavior with only with failure provisions, is na\"{i}ve and 
abstract. So, it is  imperative to infer the high-level specifications 
expressing the system-level quality and reliability measures by refining our 
formal modeling. In the following, we present the steps to automatically 
extract back the {\tt QRSpecs}, 
$\mathcal{P}_{\Upsilon} = \langle C_{\Upsilon}, M_{\Upsilon}, Z_{\Upsilon}, 
Q_{\Upsilon}^I, Q_{\Upsilon}^O \rangle$, from the underlying {\tt QRModel}, 
$\mathcal{Q}_{\Upsilon} = \langle C_{\Upsilon}, Q_{\Upsilon}^I, R_{\Upsilon}, 
S_{\Upsilon}, s_{\Upsilon}^1, \Gamma_{\Upsilon}, Q_{\Upsilon}^O, T_{\Upsilon},
{\tt Expr}_{\Upsilon} \rangle$, of the system, $\Upsilon$.
\begin{description}
 \item[Step-1:] {\em Make abstractions to the {\tt QRModel} representation 
(state-transition system), $\mathcal{Q}_{\Upsilon}$, keeping only the failure 
(marked as {\sf F}) transitions and relevant states to build a new 
configuration, say ${\mathcal{Q}'}_{\Upsilon}$.}
This can be easily obtained by doing a depth-first or breadth-first traversal 
starting from the initial state of $\mathcal{Q}_{\Upsilon}$ and progressing 
only through the {\sf F}-marked transitions.

 \item[Step-2:] {\em Represent the composed system modes with respect to the 
operating modes of each $n$ participating components together as an $n$-tuple.}
If $\Upsilon$ comprises of $n$ components, where each component $C_i\ (1 \leq i 
\leq n)$ operates in $m_i$ number of operational modes (excluding the failure 
mode), then the length every state configuration is $e_{\Upsilon} = 
\sum_{i=1}^{n} d_i$.
Here, if for some state $s \in S_{\Upsilon}$ in $\mathcal{Q}'_{\Upsilon}$, we 
find that,\\
$\exists k_i\ (1 \leq \delta < k_i \leq \delta+d_i \leq e_{\Upsilon}),\ 
\text{ where } \delta = \sum_{j=0}^{i-1} d_j\ (1 \leq i \leq n)$, such that the
following holds:
\[ \Gamma_{\Upsilon}(s)[(\delta+1)..(\delta+k_i-1)] = {\tt 0}^{(k_i-1)}, 
\Gamma_{\Upsilon}(s)[\delta+k_i] = {\tt 1} \text{ and } 
\Gamma_{\Upsilon}(s)[(\delta+k_i+1)..(\delta+d_i)] = {\tt X}^{(d_i-k_i)}, \]
then the operating mode for the participating component, $C_i$, is $m_i^{k_i}$.
The combination of the modes, $\big{(} m_1^{k_1}, m_2^{k_2}, \ldots, m_n^{k_n} 
\big{)}$, for all the components, $C_1, C_2, \ldots, C_n$, gives the 
operating modes for the system, $\Upsilon$, in state, $s$. Formally, 
$m_{\Upsilon}^s = \big{(} m_1^{k_1}, m_2^{k_2}, \ldots, m_n^{k_n} \big{)} \in 
M_{\Upsilon}$.

 \item[Step-3:] {\em Estimate the reliability of each mode from the system 
structure, $\Upsilon = \langle \mathcal{I}, \mathcal{O}, \mathcal{C}, 
\mathcal{V}, \mathcal{E}, \mathcal{L} \rangle$.} To do so, traditional 
reliability estimation methods~\cite{KK2007BOOK} are followed using typical
path-enumeration or series/parallel approaches utilizing the reliability value 
$Z_i(m_i^j)$ for mode $m_i^j$ of component $C_i$ ($1 \leq i \leq n,\ 1 \leq j 
\leq d_i$). 

 \item[Step-4:] {\em Establish the output quality value function 
$Q_{\Upsilon}^O$ at each of the composite operational modes of the abstracted 
{\tt QRModel}, $\mathcal{Q}'_{\Upsilon}$.} The {\tt QRModel} directly lists the 
output quality values corresponding to each state (modes) of $\Upsilon$ with 
respect to the mentioned input level, thereby defining $Q_{\Upsilon}^O$. 
However, $Q_{\Upsilon}^I$ can be found taking the union of input quality levels 
present in every state/mode of $\mathcal{Q}'_{\Upsilon}$.
\end{description}
The above steps reverse synthesize the {\tt QRSpec} model by only keeping the 
component mode failure option in the overall system behavior.
Let us present an example to illustrate these mentioned steps in details.
\begin{example}
First, let us revisit the {\tt QRModel}, $\mathcal{Q}_{\Upsilon_S}$, obtained 
for system, $\Upsilon_S$, in Example~\ref{ex:series_comp} 
(and Figure~\ref{fig:compose_qual_ser} with Table~\ref{tab:compose_qual_ser}). 
{\em Step-1} abstracts $\mathcal{Q}_{\Upsilon_S}$ to form 
$\mathcal{Q}'_{\Upsilon_S}$ which is shown in 
Figure~\ref{fig:abstracted_qrmodel}(a) with Table~\ref{tab:abstracted_qrmodel} 
(for $\Upsilon_S$). According to {\em Step-2}, the generated composite modes 
are labeled against the states of $\mathcal{Q}'_{\Upsilon_S}$ as mentioned in 
Table~\ref{tab:series_compose_mode} (also refer to 
Figure~\ref{fig:abstracted_qrmodel}(a)). Then, from the system structure of 
$\Upsilon_S$ (also schematic is shown in Figure~\ref{fig:series_parallel}(a)), 
we derive in {\em Step-3} the reliability of each composite modes by taking 
product of the two component-level reliability values, i.e. 
$Z_{\Upsilon_S}\big{(} (m_3^i,m_2^j) \big{)} = Z_3(m_3^i).Z_2(m_2^j)\ (0 \leq i 
\leq 2,\ 0 \leq j \leq 1)$. As part of {\em Step-4}, the input-output quality 
levels (values) can be directly found from {\tt QRModel}, 
$\mathcal{Q}_{\Upsilon_S}$ (Figure~\ref{fig:compose_qual_ser} with 
Table~\ref{tab:compose_qual_ser}). We 
find that the outcomes of {\em Step-3} and {\em Step-4}  (presented in 
Tables~\ref{tab:series_compose_mode}-\ref{tab:abstracted_qrmodel}) are similar 
to what has been (already) specified as reliability and input-output quality 
value-pairs in Table~\ref{tab:system_series_qual_rel}.

\begin{figure}
\centering
\includegraphics[width=0.75\textwidth]{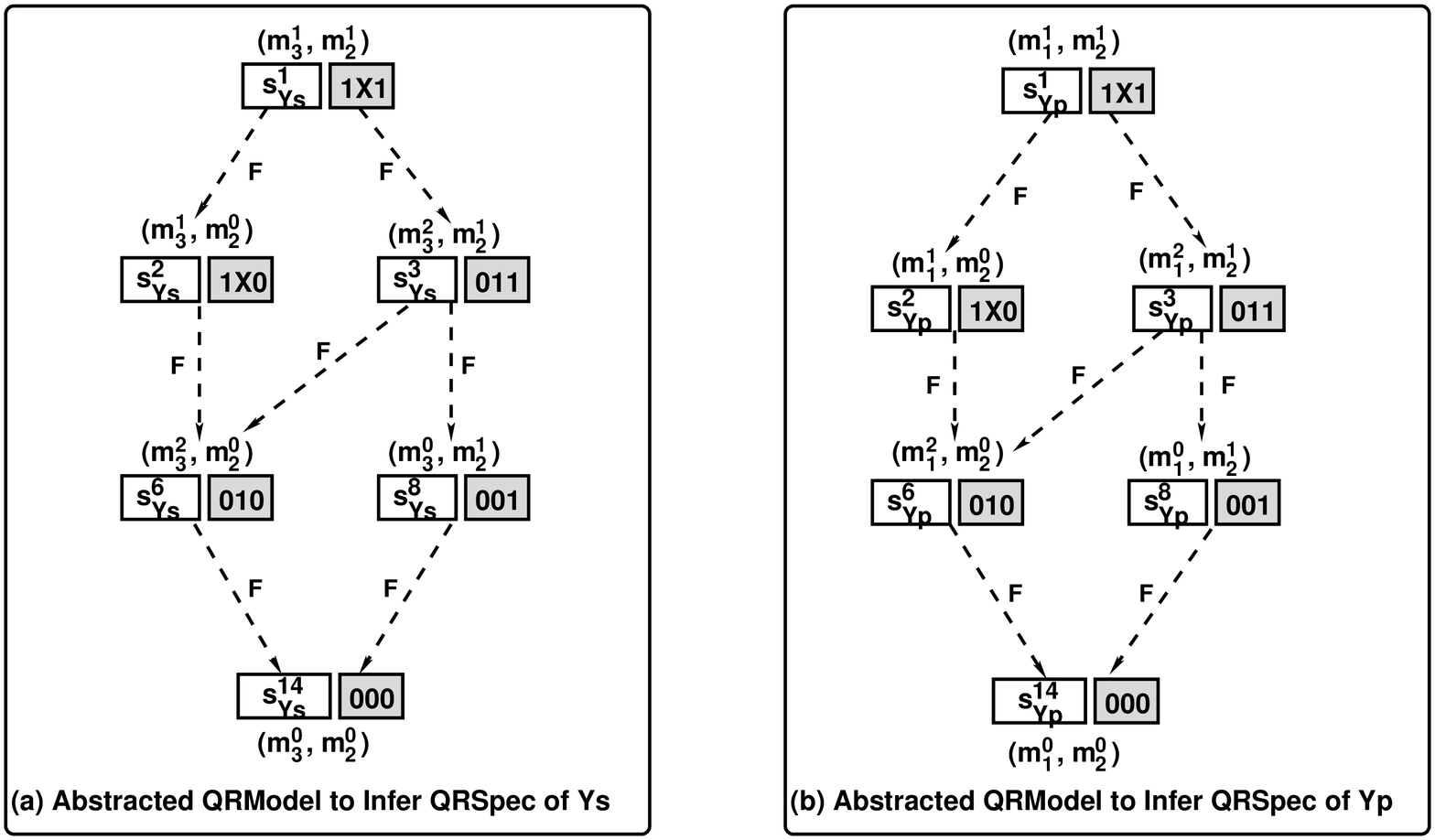}
\caption{\small State Transition Diagram with State Labels as Composite Component-Mode 
Tuples for the Abstracted {\tt QRModel}, $\mathcal{Q}'_{\Upsilon_S}$ and 
${\mathcal{Q}'}_{\Upsilon_P}^{max}$, Inferred from Quality Configurations, 
$\mathcal{Q}_{\Upsilon_S}$ and $\mathcal{Q}_{\Upsilon_P}^{max}$, respectively 
(Introduced in Examples~\ref{ex:series_comp}-\ref{ex:parallel_comp})}
\label{fig:abstracted_qrmodel}
\end{figure}

\begin{table}
{\sf
    \begin{minipage}[b]{0.45\hsize}\centering
        \begin{tabular}{ccc}
            \cline{1-3}
            System & Component Mode    & System\\
            State  & (Composite Tuple) & Mode\\
            \cline{1-3}
            $s_{\Upsilon_S}^{1}$ & $(m_3^1, m_2^1)$ & $m_{\Upsilon_S}^1$\\
            $s_{\Upsilon_S}^{2}$ & $(m_3^1, m_2^0)$ & $m_{\Upsilon_S}^0$\\
            $s_{\Upsilon_S}^{3}$ & $(m_3^2, m_2^1)$ & $m_{\Upsilon_S}^2$\\
            $s_{\Upsilon_S}^{6}$ & $(m_3^2, m_2^0)$ & $m_{\Upsilon_S}^0$\\
            $s_{\Upsilon_S}^{8}$ & $(m_3^0, m_2^1)$ & $m_{\Upsilon_S}^0$\\
            $s_{\Upsilon_S}^{14}$ & $(m_3^0, m_2^0)$ & $m_{\Upsilon_S}^0$\\
            \cline{1-3}
            \end{tabular}
        \caption{\small Composite Operating Modes w.r.t. $\mathcal{Q}'_{\Upsilon_S}$ 
States}
        \label{tab:series_compose_mode}
    \end{minipage}
    \hfill
    \begin{minipage}[b]{0.5\hsize}\centering
        \begin{tabular}{ccc}
            \cline{1-3}
            System & Component Mode    & System\\
            State  & (Composite Tuple) & Mode\\
            \cline{1-3}
            $s_{\Upsilon_P}^{1}$ & $(m_1^1, m_2^1)$ & $m_{\Upsilon_P}^3$\\
            $s_{\Upsilon_P}^{2}$ & $(m_1^1, m_2^0)$ & $m_{\Upsilon_P}^2$\\
            $s_{\Upsilon_P}^{3}$ & $(m_1^2, m_2^1)$ & $m_{\Upsilon_P}^5$\\
            $s_{\Upsilon_P}^{6}$ & $(m_1^2, m_2^0)$ & $m_{\Upsilon_P}^4$\\
            $s_{\Upsilon_P}^{8}$ & $(m_1^0, m_2^1)$ & $m_{\Upsilon_P}^1$\\
            $s_{\Upsilon_P}^{14}$ & $(m_1^0, m_2^0)$ & $m_{\Upsilon_P}^0$\\
            \cline{1-3}
            \end{tabular}
        \caption{\small Composite Operating Modes w.r.t. 
${\mathcal{Q}'}_{\Upsilon_P}^{max}$ States}
        \label{tab:parallel_compose_mode}
    \end{minipage}
}
\end{table}

\begin{table}
{\sf
\centering
 \begin{tabular}{ccccccc}
    \cline{1-7}
    System & State & Operating Mode & Input Quality & Output Quality & 
\multicolumn{2}{c}{Operating Mode Probability}\\
    \cline{6-7}
    Name & ID & Configuration & Values (Levels) & Values (Levels) & Algebraic 
Expression & Value\\
    \cline{1-7}
                 & $s_{\Upsilon_S}^1$ & {\tt 1X1} & $\langle 50,20 \rangle$ & 
$\langle 30,10 \rangle$ & $r_{3,1}.r_{2,1}$ & $0.855$\\
                 & $s_{\Upsilon_S}^2$ & {\tt 1X0} & $\langle 0 \rangle$ & 
$\langle 0 \rangle$ & $r_{3,1}.(1-r_{2,1})$ & $0.045$\\
    $\Upsilon_S$ & $s_{\Upsilon_S}^3$ & {\tt 011} & $\langle 50,20 \rangle$ & 
$\langle 30,10 \rangle$ & $(1-r_{3,1}).2_{3,2}.r_{2,1}$ & $0.076$\\
                & $s_{\Upsilon_S}^6$ & {\tt 010} & $\langle 0 \rangle$ & 
$\langle 0 \rangle$ & $(1-r_{3,1}).r_{3,2}.(1-r_{2,1})$ & $0.004$\\
                & $s_{\Upsilon_S}^8$ & {\tt 001} & $\langle 0 
\rangle$ & $\langle 0 \rangle$ & $(1-r_{3,1}).(1-r_{3,2}).r_{2,1}$ & $0.019$\\
                & $s_{\Upsilon_S}^{14}$ & {\tt 000} & $\langle 0 \rangle$ & 
$\langle 0 \rangle$ & $(1-r_{3,1}).(1-r_{3,2}).(1-r_{2,1})$ & $0.001$\\
    \cline{1-7}
                & $s_{\Upsilon_P}^1$ & {\tt 1X1} & $\langle 50,40,30,10 \rangle$ 
& $\langle 40,30,25,10 \rangle$ & $r_{1,1}.r_{2,1}$ & $0.760$\\
                & $s_{\Upsilon_P}^2$ & {\tt 1X0} & $\langle 50,30,20 \rangle$ & 
$\langle 40,25,10 \rangle$ & $r_{1,1}.(1-r_{2,1})$ & $0.040$\\
   $\Upsilon_P$ & $s_{\Upsilon_P}^3$ & {\tt 011} & $\langle 50,40,30,10 \rangle$ 
& $\langle 35,30,25,10 \rangle$& $(1-r_{1,1}).2_{3,2}.r_{2,1}$ & $0.133$\\
                & $s_{\Upsilon_P}^6$ & {\tt 010} & $\langle 50,30,20 \rangle$ & 
$\langle 35,25,10 \rangle$ & $(1-r_{1,1}).r_{1,2}.(1-r_{2,1})$ & $0.007$\\
                & $s_{\Upsilon_P}^8$ & {\tt 001} & $\langle 40,10 \rangle$ 
& $\langle 30,10 \rangle$ & $(1-r_{1,1}).(1-r_{1,2}).r_{2,1}$ & $0.057$\\
                & $s_{\Upsilon_P}^{14}$ & {\tt 000} & $\langle 0 \rangle$ & 
$\langle 0 \rangle$ & $(1-r_{1,1}).(1-r_{1,2}).(1-r_{2,1})$ & $0.003$\\
    \cline{1-7}
    \end{tabular}
 \caption{\small Abstracted {\tt QRModel} Configuration Details, 
${\mathcal{Q}'}_{\Upsilon_S}$ and ${\mathcal{Q}'}_{\Upsilon_P}^{max}$, obtained 
from $\mathcal{Q}_{\Upsilon_S}$ and $\mathcal{Q}_{\Upsilon_P}^{max}$ for 
$\Upsilon_S$ and $\Upsilon_P$}
 \label{tab:abstracted_qrmodel}
}
\end{table}

Similarly, let us revisit the {\tt QRModel}, $\mathcal{Q}_{\Upsilon_P}^{max}$, 
obtained for system, $\Upsilon_P$, in Example~\ref{ex:parallel_comp} 
(and Figure~\ref{fig:compose_qual_par} with 
Table~\ref{tab:compose_qual_par_max}).
{\em Step-1} abstracts $\mathcal{Q}_{\Upsilon_P}^{max}$ to form 
${\mathcal{Q}'}_{\Upsilon_P}^{max}$ which is shown in 
Figure~\ref{fig:abstracted_qrmodel}(b) with Table~\ref{tab:abstracted_qrmodel} 
(for $\Upsilon_P$).
According to {\em Step-2}, the generated composite modes are labeled against 
the states of ${\mathcal{Q}'}_{\Upsilon_P}^{max}$ as mentioned in 
Table~\ref{tab:parallel_compose_mode} (also refer to 
Figure~\ref{fig:abstracted_qrmodel}(b)).
Then, from the system structure of $\Upsilon_P$ (also schematic is shown in 
Figure~\ref{fig:series_parallel}(b)), we derive in {\em Step-3} the reliability 
of each composite modes by composing the two component-level reliability values 
in parallel, i.e. $Z_{\Upsilon_P}\big{(} (m_1^i,m_2^j) \big{)} = 1 - \big{(} 1 -
Z_1(m_1^i) \big{)} . \big{(} 1 - Z_2(m_2^j) \big{)}\ (0 \leq i \leq 2,\ 0 \leq j 
\leq 1)$.
As part of {\em Step-4}, the input-output quality levels (values) can be 
directly found from {\tt QRModel}, $\mathcal{Q}_{\Upsilon_P}^{max}$ 
(Figure~\ref{fig:compose_qual_par} with Table~\ref{tab:compose_qual_par_max}). 
We find that the outcomes of {\em Step-3} and {\em Step-4} (presented in 
Tables~\ref{tab:parallel_compose_mode}-\ref{tab:abstracted_qrmodel}) are 
similar to 
what has been (already) specified as reliability and input-output quality 
value-pairs in Table~\ref{tab:system_parallel_qual_rel}.

In order to complete the conformance checking of automatically inferred {\tt 
QRSpec}, $\mathcal{P}_{\Upsilon_S}$ and $\mathcal{P}_{\Upsilon_P}$, from {\tt 
QRModel}, $\mathcal{Q}_{\Upsilon_S}$ and $\mathcal{Q}_{\Upsilon_P}^{max}$, 
respectively, the reader may cross-check Table~\ref{tab:system_series_qual_rel} 
(from Example~\ref{ex:system_qual_spec}) with 
Tables~\ref{tab:series_compose_mode}-\ref{tab:abstracted_qrmodel} and 
Figure~\ref{fig:abstracted_qrmodel}(a), as well as 
Table~\ref{tab:system_parallel_qual_rel} 
(from Example~\ref{ex:system_qual_spec}) with 
Tables~\ref{tab:parallel_compose_mode}-\ref{tab:abstracted_qrmodel} and 
Figure~\ref{fig:abstracted_qrmodel}(b).
\qed
\end{example}
It may be noted that, the above mentioned steps provide an generic 
inferencing procedure, aiding to the automatic synthesis of {\tt QRSpec}
($\mathcal{P}_{\Upsilon}$) which may further be checked for conformance against 
the already laid specifications to ensure the correctness of the built system 
in terms of its quality and reliability artifacts. Such conformance checking 
may happen either through manual inspection, or through a systematic 
query-based engine (which is able to provide more detailed information from 
underlying {\tt QRModel}), by which designers may explore the quality and 
reliability measures of a system under various operational setup of its 
constituent components leading to comprehensive coverage. The following section 
presents this part in details.

\section{Query Processing and Analysis over System-level {\tt QRModel}} 
\label{sec:qual_query_analysis}
To explore the system completeness with respect to quality and reliability 
specifications, we first develop a language through which we can interact with 
the system-level {\tt QRModel} in order to extract various relevant parameters 
under different operating conditions of the system. Such an exploration 
can be performed at architectural-level and it helps the designer to determine 
the coverage of the system in terms of its best and worst possible measures of 
quality and reliability attributes attainable with the given compositional
structure having the possibility of component failures and suspensions as part 
of the operating setups.

\subsection{Structured Query Description Language ({\tt SQDL})} 
\label{subsec:query_language}
We propose a structured query description language, named as {\tt SQDL}, to 
express various queries related to a given component-based system structure. 
{\tt SQDL} enjoys similar and simple syntactic sugar as that of SQL (Structured 
Query Language). The detailed syntax for expressing each {\em query-block} is 
given below.
\begin{tcolorbox}
{\color{magenta}
\begin{verbatim}
begin_query [ IDENTIFIER ]
    select
        [ - input_quality ]
        [ - output_quality ]
        [ - operating_mode ]
        [ - reliability ]
        [ - operate_prob ]
        [ - failure ]
        [ - suspend ]
    from
        - system SYSTEM_STRUCTURE_FILE
        - qrspec COMPONENT_QRSPEC_FILE
    where
        [ - input_quality   VALUE_LIST ]
        [ - output_quality  [ - minimum VALUE_LIST ] [ - maximum VALUE_LIST ]  ]
        [ - reliability     [ - minimum VALUE  ] [ - maximum VALUE  ]  ]
        [ - operate_prob    [ - minimum VALUE  ] [ - maximum VALUE  ]  ]
        [ - failure         [ - minimum NUMBER ] [ - maximum NUMBER ]  ]
        [ - suspend         [ - minimum NUMBER ] [ - maximum NUMBER ]  ]
end_query
\end{verbatim}
}
\end{tcolorbox}
It may be noted that, each three-part ({\color{magenta} \tt select - from - 
where}) query specification is encapsulated using ``{\color{magenta} \tt 
begin\_query ... end\_query}'' block and all {\em small-letter} phrases (that 
followed after a `{\color{magenta} --}') indicate the reserved keywords for 
query specification. The syntax, {\color{magenta} \tt [...]}, indicates that 
those attributes are optional in query specification. Each query is given a name 
which is indicated by the {\color{magenta} \tt IDENTIFIER}. The input system 
structure description and the component-level {\tt QRSPEC} definitions are 
provided using {\color{magenta} \tt SYSTEM\_STRUCTURE\_FILE} and 
{\color{magenta} \tt COMPONENT\_QRSPEC\_FILE}. Moreover, using the 
placeholders, {\color{magenta} \tt VALUE}, {\color{magenta} \tt VALUE\_LIST} 
and {\color{magenta} \tt NUMBER}, we indicate a real number, a list of real 
values (within {\color{magenta} $\{...\}$} separated by comma), and an integer, 
respectively.

Semantically, the architectural component-based system structure with the {\tt 
QRSpec} definitions for each component is chosen from the files mentioned in 
{\color{magenta} \tt from} block. The {\color{magenta} \tt select} block 
proposes to extract the following information:
(i) the coverage for input-output quality ranges supported by the system,
(ii) the ranges for its reliability and operating probability,
(iii) the supporting operational modes of the components,
(iv) the minimum and maximum failures or suspensions admissible by the system. 
These extraction are subject to the constraints of the system as specified 
under {\color{magenta} \tt where} block having the options of constraining for 
the following:
(i) the prescribed quality values to be attained,
(ii) the highest and lowest reliability and operating probability allowed,
(iii) the predicted choice of failures and suspensions required during 
operations.

The following example presents two queries and their representations in {\tt 
SQDL} to bring out the essence.
\begin{example} \label{ex:query_sqdl}
Let us first specify (in English) the intent of two example queries over a 
system as follows:
\begin{description}
 \item[Query1:] {\em What are the operating modes with their respective set of 
input-output quality levels (values) and reliability values supported by the 
system, when we try to operate it maintaining the input quality levels at least 
$30$, with the system reliability guarantee be above $0.85$ and allowing a 
maximum of $2$ component failures and without suspending any component?}

 \item[Query2:] {\em How many component failures can be tolerated and in which 
operating modes (with their operating probability) the system can operate in 
those cases, when we want to maintain the minimum output-quality values as 
$\langle 30,25,10,5 \rangle$ corresponding to the given input-quality levels 
$\langle 40,30,15,5 \rangle$, respectively, with the system reliability 
guarantee be above $0.95$ and a maximum of $1$ component may be suspended?}
\end{description}
We may express these queries using our proposed {\tt SQDL} framework as:
\begin{multicols}{2}
\begin{tcolorbox}
\begin{verbatim}
begin_query Query1
    select
        - input_quality
        - output_quality
        - operating_mode
        - reliability
    from
        - system file.sys
        - qrspec spec.qr
    where
        - input_quality { 30 }
        - reliability
            - minimum 0.85
        - failure
            - maximum 2
        - control
            - maximum 0
end_query
\end{verbatim}
\end{tcolorbox}
\columnbreak
\begin{tcolorbox}
\begin{verbatim}
begin_query Query2
    select
        - operating_mode
        - operate_prob
        - failure
    from
        - system file.sys
        - qrspec spec.qr
    where
        - input_quality { 40,30,15,5 }
        - output_quality
            - minimum { 30,25,10,5 }
        - reliability
            - minimum 0.95
        - suspend
            - maximum 1
end_query
\end{verbatim}
\end{tcolorbox}
\end{multicols}
Note that, the system structure will be supplied as a directed acyclic graph 
(as 
defined formally in Section~\ref{subsec:system}) through the formatted file, 
{\color{magenta} \tt file.sys} and the {\tt QRSpec} for each component module 
(as defined formally in Section~\ref{subsec:qual_spec}) will be described 
through the formatted file, {\color{magenta} \tt spec.qr}.
\qed
\end{example}
Interestingly, this query platform can also be used to easily abstract any 
system behavior out of its {\tt QRModel}. So, in order to infer back the {\tt 
QRSpec} (as discussed in Section~\ref{sec:qual_conformance}), we may write a 
simple query only constraining the maximum number of control to be $0$ and 
hence retrieving all possible failure options of the system.

In the next subsection, we discuss how the query is processed to retrieve 
results from the system {\tt QRModel}.

\subsection{Query-driven System Behavior Assessment} 
\label{subsec:query_assess}
The query-driven analysis framework helps the component-based system designer 
by providing a meaningful coverage information related to the quality and 
reliability attributes of systems through extraction of a range of admissible 
design configurations. Thereby, it also helps the designer to explore through 
multiple system setups and finally converge into meaningful compositions of 
reliable and quality system design.
The processing and retrieval of quality measures by our proposed framework are 
done involving three basic steps.
\begin{enumerate}[(i)]
 \item {\em Developing {\tt QRModel}.} From the system architecture description 
and the component {\tt QRSpec} (as specified in {\color{magenta} \tt from} 
part of the query), we follow the component characterization (as proposed in 
Section~\ref{sec:qual_config}) and generic composition procedures (proposed in 
Section~\ref{sec:qual_compose}) to build the system {\tt QRModel} automatically.
 
 \item {\em Constraining {\tt QRModel}.} After parsing {\tt SQDL} queries, we 
impose the restriction over the built {\tt QRModel} adhering to the constraints 
mentioned under the {\color{magenta} \tt where} block of a query.
 
 \item {\em Exploring {\tt QRModel}.} This is a simple search and data 
retrieval step from the underlying state-transition system to gather relevant 
information from {\tt QRModel} as asked in the {\color{magenta} \tt select} 
block of the query.
\end{enumerate}
Let us elucidate the outcome of the above-mentioned three-step process by an 
example.
\begin{example}
We again revisit the parallel system, $\Upsilon_P$ and seek for the answers of 
the two queries ({\tt Query1} and {\tt Query2}) introduced in 
Example~\ref{ex:query_sqdl}. Table~\ref{tab:result_query1} and 
Table~\ref{tab:result_query2} present the snapshots of the information 
retrieved over the $\Upsilon_P$ system.

Some interesting observations can be made for the system, $\Upsilon_P$, from 
Table~\ref{tab:result_query1}. Though we asked for the output quality for the 
input quality of at least $30$ units, but the mode configuration {\tt 001} 
(where $C_1$ fails completely in both modes) can only provide meaningful output 
with input level being at least $40$ and the output quality falls to $0$ 
(zero), when input quality is set within $[30,40)$. Moreover, the system can 
maintain a reliability of at least $0.9$ (higher than given) in the given setup 
with at most $1$ failure admissible. However, the reliability can ranges upto 
$[0.985-0.990]$ under the assumption of non-failing components.

\begin{table}
{\sf \centering
 \begin{tabular}{ccccc}
    \cline{1-5}
    Mode Configuration & Component Operating Modes & Input Quality Levels & 
Output Quality Values & Reliability\\
    \cline{1-5}
    {\tt 1X1} & $C_1 = (m_1^1 : {\tt OP}, m_1^2 : {\tt NA}) \quad C_2 = (m_2^1 
: {\tt OP})$ & $\langle 50,40,30 \rangle$ & $\langle 40,30,25 \rangle$ & 
$0.990$\\
    {\tt 011} & $C_1 = (m_1^1 : {\tt FL}, m_1^2 : {\tt OP}) \quad C_2 = (m_2^1 
: {\tt OP})$ & $\langle 50,40,30 \rangle$ & $\langle 35,30,25 \rangle$ & 
$0.985$\\
    {\tt 001} & $C_1 = (m_1^1 : {\tt FL}, m_1^2 : {\tt FL}) \quad C_2 = (m_2^1 
: {\tt OP})$ & $\langle 40 \rangle$ & $\langle 30 \rangle$ & $0.950$\\
    \cline{1-5}
    \end{tabular}
 \caption{Extracted Results over $\Upsilon_P$ from {\sf Query1} ~~ ({\tt 
OP: Operating, FL: Failed, NA: Not-Availed})}
 \label{tab:result_query1}
}
\end{table}
\begin{table}
{\sf \centering
 \begin{tabular}{cccc}
    \cline{1-4}
    Mode Configuration & Component Operating Modes & Operating Probability & \# 
Failure\\
    \cline{1-4}
    {\tt 1X1} & $C_1 = (m_1^1 : {\tt OP}, m_1^2 : {\tt NA}) \quad C_2 = (m_2^1 
: {\tt OP})$ & $0.760$ & $0$~~\\
    {\tt 011} & $C_1 = (m_1^1 : {\tt FL}, m_1^2 : {\tt OP}) \quad C_2 = (m_2^1 
: {\tt OP})$ & $0.133$ & $0$~~\\
    {\tt 001} & $C_1 = (m_1^1 : {\tt FL}, m_1^2 : {\tt FL}) \quad C_2 = (m_2^1 
: {\tt OP})$ & $0.057$ & $1$*\\
    {\tt 0Y1} & $C_1 = (m_1^1 : {\tt FL}, m_1^2 : {\tt SU}) \quad C_2 = (m_2^1 
: {\tt OP})$ & $0.190$ & $0$~~\\
    {\tt Y11} & $C_1 = (m_1^1 : {\tt SU}, m_1^2 : {\tt OP}) \quad C_2 = (m_2^1 
: {\tt OP})$ & $0.665$ & $0$~~\\
    {\tt Y01} & $C_1 = (m_1^1 : {\tt SU}, m_1^2 : {\tt FL}) \quad C_2 = (m_2^1 
: {\tt OP})$ & $0.285$ & $0$~~\\
    \cline{1-4}
    \end{tabular}
 \caption{\small Extracted Results over $\Upsilon_P$ from {\sf Query2} ~~ ({\tt OP: 
Operating, FL: Failed, SU: Suspended, NA: Not-Availed}) \qquad \qquad \qquad [* Maximum Number of
Failures that can be Tolerated = $1$ (though failure in $C_2$ is not admissible!) ]}
 \label{tab:result_query2}
}
\end{table}

Similarly, from the component operating modes listed in 
Table~\ref{tab:result_query2}, it is evident that indicates that it is {\em 
never} possible in $\Upsilon_P$ to maintain the reliability threshold of $0.95$ 
when the component $C_2$ fails. It is also shown that only a maximum of $1$ 
failure is admissible to guarantee the query constraint here, but it must be 
$C_1$ in the worst case. Moreover, there are few possible provisions to suspend 
$C_1$, if required without violating the given {\tt Query2} constraints.

{\em When the same queries are solved over the series system, $\Upsilon_S$, we 
find that there is only one configuration (i.e. {\tt 1X1}) admissible for {\tt 
Query1} with reliability of $0.855$ and input-output quality of $50 \rightarrow 
30$; but for {\tt Query2}, there is {\em none} adhering to the setup 
constraints.}
\qed
\end{example}
Though we provided a simple example system to illustrate the query-based 
assessment mechanism, but our proposed framework is well-generalized to handle 
any complex systems structure that can be represented by a directed acyclic 
graph, and with generic queries expressed using {\tt SQDL}.

\section{Case-Study} \label{sec:case_study}
Let us revisit the example component-based system structure, $\Upsilon_{sp}$ 
that is introduced in the beginning of this article (refer to 
Example~\ref{ex:comp-based_system} and Figure~\ref{fig:system_example}). We 
show the applicability of our proposed quality characterization and query-based 
assessment framework in details over this system to establish that the method 
can be applied in general to any composite structures.

\begin{figure}
\centering
\includegraphics[width=0.8\textwidth]{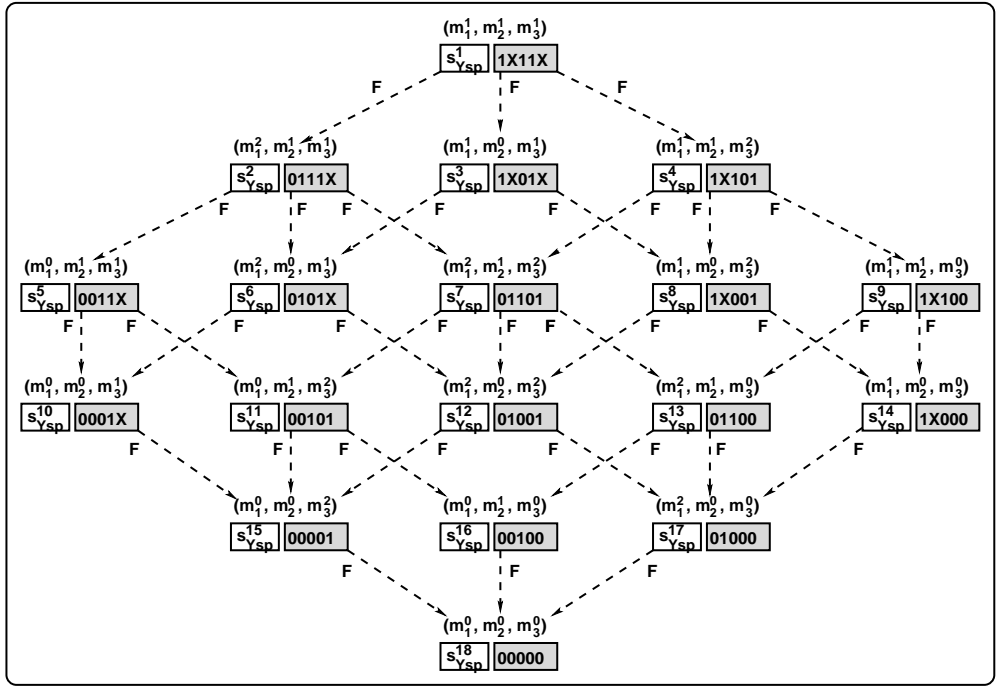}
\caption{\small State Transition Diagram with State Labels as Composite Component-Mode 
Tuples for the Abstracted {\tt QRModel}, $\mathcal{Q}'_{\Upsilon_{sp}}$, 
Inferred from Quality Configurations, $\mathcal{Q}_{\Upsilon_{sp}}$, which has 
a total of $147$ states with $348$ failure and suspend transitions ($174$ 
each)}
\label{fig:SP_qrmodel}
\end{figure}

For $\Upsilon_{sp}$, the automatically derived formal {\tt QRModel} has $147$ 
states and $174$ failure transitions as well as $174$ suspend transitions. When 
we synthesize the formal {\tt QRSpec} by abstracting the {\tt QRModel} 
(keeping only failure transitions and the states reachable by failure 
transitions from the initial state in the underlying state-transition system), 
we get the abstract state-transition system as shown in 
Figure~\ref{fig:SP_qrmodel} with its specification configurations given in 
Table~\ref{tab:SP_qrmodel}. In this {\tt QRSpec} model formed, we have $18$ 
states (combined component modes) and $33$ transitions.

\begin{table}
{\scriptsize \sf \centering
 \begin{tabular}{ccccccc}
    \cline{1-7}
    State & Operating & Component Mode & Input Quality & Output Quality & 
\multicolumn{2}{c}{Operating Mode Probability}\\
    \cline{6-7}
    ID & Mode Config. & (Composite Tuple) & Values (Levels) & Values (Levels) 
& Algebraic Expression & Value\\
    \cline{1-7}
    $s_{\Upsilon_{sp}}^1$ & {\tt 1X11X} & $(m_1^1, m_2^1, m_3^1)$ & $\langle 
50,40,30,10 \rangle$ & $\langle 40,30,25,10 \rangle$ & 
{\tiny $r_{1,1}.r_{2,1}.r_{3,1}$} & $0.68400$\\
    $s_{\Upsilon_{sp}}^2$ & {\tt 0111X} & $(m_1^2, m_2^1, m_3^1)$ & $\langle 
50,40,30,10 \rangle$ & $\langle 35,30,25,10 \rangle$ & 
{\tiny $(1-r_{1,1}).r_{1,2}.r_{2,1}.r_{3,1}$} & $0.11970$\\
    $s_{\Upsilon_{sp}}^3$ & {\tt 1X01X} & $(m_1^1, m_2^0, m_3^1)$ & $\langle 
50,30,20 \rangle$ & $\langle 40,25,10 \rangle$ & 
{\tiny $r_{1,1}.(1-r_{2,1}).r_{3,1}$} & $0.03600$\\
    $s_{\Upsilon_{sp}}^4$ & {\tt 1X101} & $(m_1^1, m_2^1, m_3^2)$ & $\langle 
50,40,30,10 \rangle$ & $\langle 40,30,25,10 \rangle$ & 
{\tiny $r_{1,1}.r_{2,1}.(1-r_{3,1}).r_{3,2}$} & $0.06080$\\
    $s_{\Upsilon_{sp}}^5$ & {\tt 0011X} & $(m_1^0, m_2^1, m_3^1)$ & $\langle 
40,10 \rangle$ & $\langle 30,10 \rangle$ & 
{\tiny $(1-r_{1,1}).(1-r_{1,2}).r_{2,1}.r_{3,1}$} & $0.05130$\\
    $s_{\Upsilon_{sp}}^6$ & {\tt 0101X} & $(m_1^2, m_2^0, m_3^1)$ & $\langle 
50,30,20 \rangle$ & $\langle 35,25,10 \rangle$ & 
{\tiny $(1-r_{1,1}).r_{1,2}.(1-r_{2,1}).r_{3,1}$} & $0.00630$\\
    $s_{\Upsilon_{sp}}^7$ & {\tt 01101} & $(m_1^2, m_2^1, m_3^2)$ & $\langle 
50,40,30,10 \rangle$ & $\langle 35,30,25,10 \rangle$ & 
{\tiny $(1-r_{1,1}).r_{1,2}.r_{2,1}.(1-r_{3,1}).r_{3,2}$} & $0.01064$\\
    $s_{\Upsilon_{sp}}^8$ & {\tt 1X001} & $(m_1^1, m_2^0, m_3^2)$ & $\langle 
50,30,20 \rangle$ & $\langle 40,25,10 \rangle$ & 
{\tiny $r_{1,1}.(1-r_{2,1}).(1-r_{3,1}).r_{3,2}$} & $0.00320$\\
    $s_{\Upsilon_{sp}}^9$ & {\tt 1X100} & $(m_1^1, m_2^1, m_3^0)$ & $\langle 
50,40,30,10 \rangle$ & $\langle 40,30,25,10 \rangle$ & 
{\tiny $r_{1,1}.r_{2,1}.(1-r_{3,1}).(1-r_{3,2})$} & $0.01520$\\
    $s_{\Upsilon_{sp}}^{10}$ & {\tt 0001X} & $(m_1^0, m_2^0, m_3^1)$ & $\langle 
0 \rangle$ & $\langle 0 \rangle$ & 
{\tiny $(1-r_{1,1}).(1-r_{1,2}).(1-r_{2,1}).r_{3,1}$} & $0.00270$\\
    $s_{\Upsilon_{sp}}^{11}$ & {\tt 00101} & $(m_1^0, m_2^1, m_3^2)$ & $\langle 
50,20 \rangle$ & $\langle 30,10 \rangle$ & 
{\tiny $(1-r_{1,1}).(1-r_{1,2}).r_{2,1}.(1-r_{3,1}).r_{3,2}$} & $0.04560$\\
    $s_{\Upsilon_{sp}}^{12}$ & {\tt 01001} & $(m_1^2, m_2^0, m_3^2)$ & $\langle 
50,30,20 \rangle$ & $\langle 35,25,10 \rangle$ & 
{\tiny $(1-r_{1,1}).r_{1,2}.(1-r_{2,1}).(1-r_{3,1}).r_{3,2}$} & $0.00056$\\
    $s_{\Upsilon_{sp}}^{13}$ & {\tt 01100} & $(m_1^2, m_2^1, m_3^0)$ & $\langle 
50,40,30,10 \rangle$ & $\langle 35,30,25,10 \rangle$ & 
{\tiny $(1-r_{1,1}).r_{1,2}.r_{2,1}.(1-r_{3,1}).(1-r_{3,2})$} & $0.00266$\\
    $s_{\Upsilon_{sp}}^{14}$ & {\tt 1X000} & $(m_1^1, m_2^0, m_3^0)$ & $\langle 
50,30,20 \rangle$ & $\langle 40,25,10 \rangle$ & 
{\tiny $r_{1,1}.(1-r_{2,1}).(1-r_{3,1}).(1-r_{3,2})$} & $0.00080$\\
    $s_{\Upsilon_{sp}}^{15}$ & {\tt 00001} & $(m_1^0, m_2^0, m_3^2)$ & $\langle 
0 \rangle$ & $\langle 0 \rangle$ & 
{\tiny $(1-r_{1,1}).(1-r_{1,2}).(1-r_{2,1}).(1-r_{3,1}).r_{3,2}$} & $0.00024$\\
    $s_{\Upsilon_{sp}}^{16}$ & {\tt 00100} & $(m_1^0, m_2^1, m_3^0)$ & $\langle 
40,10 \rangle$ & $\langle 30,10 \rangle$ & 
{\tiny $(1-r_{1,1}).(1-r_{1,2}).r_{2,1}.(1-r_{3,1}).(1-r_{3,2})$} & $0.00114$\\
    $s_{\Upsilon_{sp}}^{17}$ & {\tt 01000} & $(m_1^2, m_2^0, m_3^0)$ & $\langle 
50,30,20 \rangle$ & $\langle 35,25,10 \rangle$ & 
{\tiny $(1-r_{1,1}).r_{1,2}.(1-r_{2,1}).(1-r_{3,1}).(1-r_{3,2})$} & $0.00014$\\
    $s_{\Upsilon_{sp}}^{18}$ & {\tt 00000} & $(m_1^0, m_2^0, m_3^0)$ & $\langle 
0 \rangle$ & $\langle 0 \rangle$ & 
{\tiny $(1-r_{1,1}).(1-r_{1,2}).(1-r_{2,1}).(1-r_{3,1}).(1-r_{3,2})$} & 
$0.00006$\\
    \cline{1-7}
    \end{tabular}
 \caption{\small Abstracted {\tt QRModel} Configuration Details, 
${\mathcal{Q}'}_{\Upsilon_{sp}}$ (presenting the {\tt QRSpec} for 
$\Upsilon_{sp}$), as obtained from $\mathcal{Q}_{\Upsilon_{sp}}$}
 \label{tab:SP_qrmodel}
}
\end{table}

\begin{table}
{ \sf \centering
 \begin{tabular}{ccccc}
    \cline{1-5}
    Mode & Component-level Operating Mode Combinations & Input Quality & Output 
Quality & Reliability\\
    Config. & & Levels (Values) & 
Levels (Values) & Value \\
    \cline{1-5}
    {\tt 1X11X} & $C_1 = (m_1^1 : {\tt OP},\ m_1^2 : {\tt NA}) \quad C_2 = 
(m_2^1 : {\tt OP}) \quad C_3 = (m_3^1 : {\tt OP},\ m_3^2 : {\tt NA})$ & 
$\langle 50,40,30 \rangle$ & $\langle 40,30,25 \rangle$ & 
$0.990$\\
    {\tt 0111X} & $C_1 = (m_1^1 : {\tt FL},\ m_1^2 : {\tt OP}) \quad C_2 = 
(m_2^1 : {\tt OP}) \quad C_3 = (m_3^1 : {\tt OP},\ m_3^2 : {\tt NA})$ & 
$\langle 50,40,30 \rangle$ & $\langle 35,30,25 \rangle$ & 
$0.985$\\
    {\tt 0011X} & $C_1 = (m_1^1 : {\tt FL},\ m_1^2 : {\tt FL}) \quad C_2 = 
(m_2^1 : {\tt OP}) \quad C_3 = (m_3^1 : {\tt OP},\ m_3^2 : {\tt NA})$ & 
$\langle 40 \rangle$ & $\langle 30 \rangle$ & 
$0.950$\\
    {\tt 01101} & $C_1 = (m_1^1 : {\tt FL},\ m_1^2 : {\tt OP}) \quad C_2 = 
(m_2^1 : {\tt OP}) \quad C_3 = (m_3^1 : {\tt FL},\ m_3^2 : {\tt OP})$ & 
$\langle 50,40,30 \rangle$ & $\langle 35,30,25 \rangle$ & 
$0.985$\\
    {\tt 1X101} & $C_1 = (m_1^1 : {\tt OP},\ m_1^2 : {\tt NA}) \quad C_2 = 
(m_2^1 : {\tt OP}) \quad C_3 = (m_3^1 : {\tt FL},\ m_3^2 : {\tt OP})$ 
& $\langle 50,40,30 \rangle$ & $\langle 40,30,25 \rangle$ & 
$0.990$\\
    {\tt 1X100} & $C_1 = (m_1^1 : {\tt OP},\ m_1^2 : {\tt NA}) \quad C_2 = 
(m_2^1 : {\tt OP}) \quad C_3 = (m_3^1 : {\tt FL},\ m_3^2 : {\tt FL})$ 
& $\langle 50,40,30 \rangle$ & $\langle 40,30,25 \rangle$ & 
$0.990$\\
    \cline{1-5}
    \end{tabular}
 \caption{Extracted Results over $\Upsilon_{sp}$ from {\sf Query1} ~~ ({\tt 
OP: Operating, FL: Failed, NA: Not-Availed})}
 \label{tab:result_SP_query1}
}
\end{table}
\begin{table}
{ \sf \centering
 \begin{tabular}{cccc}
    \cline{1-4}
    Mode & Component-level Operating Mode Combinations & Operating & No. of\\
    Config. & & Probability & Failure\\
    \cline{1-4}
    {\tt 1X11X} & $C_1 = (m_1^1 : {\tt OP},\ m_1^2 : {\tt NA}) \quad C_2 = 
(m_2^1 : {\tt OP}) \quad C_3 = (m_3^1 : {\tt OP},\ m_3^2 : {\tt NA})$ & 
$0.68400$ & $0$~~\\
    {\tt 0111X} & $C_1 = (m_1^1 : {\tt FL},\ m_1^2 : {\tt OP}) \quad C_2 = 
(m_2^1 : {\tt OP}) \quad C_3 = (m_3^1 : {\tt OP},\ m_3^2 : {\tt NA})$ & 
$0.11970$ & $0$~~\\
    {\tt 0011X} & $C_1 = (m_1^1 : {\tt FL},\ m_1^2 : {\tt FL}) \quad C_2 = 
(m_2^1 : {\tt OP}) \quad C_3 = (m_3^1 : {\tt OP},\ m_3^2 : {\tt NA})$ & 
$0.05130$ & $1$~~\\
    {\tt 00101} & $C_1 = (m_1^1 : {\tt FL},\ m_1^2 : {\tt FL}) \quad C_2 = 
(m_2^1 : {\tt OP}) \quad C_3 = (m_3^1 : {\tt FL},\ m_3^2 : {\tt OP})$ & 
$0.00456$ & $1$~~\\
    {\tt 00100} & $C_1 = (m_1^1 : {\tt FL},\ m_1^2 : {\tt FL}) \quad C_2 = 
(m_2^1 : {\tt OP}) \quad C_3 = (m_3^1 : {\tt FL},\ m_3^2 : {\tt FL})$ & 
$0.00114$ & $2$*\\
    {\tt 0010Y} & $C_1 = (m_1^1 : {\tt FL},\ m_1^2 : {\tt FL}) \quad C_2 = 
(m_2^1 : {\tt OP}) \quad C_3 = (m_3^1 : {\tt FL},\ m_3^2 : {\tt SU})$ & 
$0.00570$ & $1$~~\\
    {\tt 001Y1} & $C_1 = (m_1^1 : {\tt FL},\ m_1^2 : {\tt FL}) \quad C_2 = 
(m_2^1 : {\tt OP}) \quad C_3 = (m_3^1 : {\tt SU},\ m_3^2 : {\tt OP})$ & 
$0.04560$ & $1$~~\\
    {\tt 001Y0} & $C_1 = (m_1^1 : {\tt FL},\ m_1^2 : {\tt FL}) \quad C_2 = 
(m_2^1 : {\tt OP}) \quad C_3 = (m_3^1 : {\tt SU},\ m_3^2 : {\tt FL})$ & 
$0.01140$ & $1$~~\\
    {\tt 0Y11X} & $C_1 = (m_1^1 : {\tt FL},\ m_1^2 : {\tt SU}) \quad C_2 = 
(m_2^1 : {\tt OP}) \quad C_3 = (m_3^1 : {\tt OP},\ m_3^2 : {\tt NA})$ & 
$0.17100$ & $0$~~\\
    {\tt 0Y101} & $C_1 = (m_1^1 : {\tt FL},\ m_1^2 : {\tt SU}) \quad C_2 = 
(m_2^1 : {\tt OP}) \quad C_3 = (m_3^1 : {\tt FL},\ m_3^2 : {\tt OP})$ & 
$0.01520$ & $0$~~\\
    {\tt 0Y100} & $C_1 = (m_1^1 : {\tt FL},\ m_1^2 : {\tt SU}) \quad C_2 = 
(m_2^1 : {\tt OP}) \quad C_3 = (m_3^1 : {\tt FL},\ m_3^2 : {\tt FL})$ & 
$0.00380$ & $1$~~\\
    {\tt 01101} & $C_1 = (m_1^1 : {\tt FL},\ m_1^2 : {\tt OP}) \quad C_2 = 
(m_2^1 : {\tt OP}) \quad C_3 = (m_3^1 : {\tt FL},\ m_3^2 : {\tt OP})$ & 
$0.01064$ & $0$~~\\
    {\tt 01100} & $C_1 = (m_1^1 : {\tt FL},\ m_1^2 : {\tt OP}) \quad C_2 = 
(m_2^1 : {\tt OP}) \quad C_3 = (m_3^1 : {\tt FL},\ m_3^2 : {\tt FL})$ & 
$0.00266$ & $1$~~\\
    {\tt 0110Y} & $C_1 = (m_1^1 : {\tt FL},\ m_1^2 : {\tt OP}) \quad C_2 = 
(m_2^1 : {\tt OP}) \quad C_3 = (m_3^1 : {\tt FL},\ m_3^2 : {\tt SU})$ & 
$0.01330$ & $0$~~\\
    {\tt 011Y1} & $C_1 = (m_1^1 : {\tt FL},\ m_1^2 : {\tt OP}) \quad C_2 = 
(m_2^1 : {\tt OP}) \quad C_3 = (m_3^1 : {\tt SU},\ m_3^2 : {\tt OP})$ & 
$0.10640$ & $0$~~\\
    {\tt 011Y0} & $C_1 = (m_1^1 : {\tt FL},\ m_1^2 : {\tt OP}) \quad C_2 = 
(m_2^1 : {\tt OP}) \quad C_3 = (m_3^1 : {\tt SU},\ m_3^2 : {\tt FL})$ & 
$0.02660$ & $0$~~\\
    {\tt Y111X} & $C_1 = (m_1^1 : {\tt SU},\ m_1^2 : {\tt OP}) \quad C_2 = 
(m_2^1 : {\tt OP}) \quad C_3 = (m_3^1 : {\tt OP},\ m_3^2 : {\tt NA})$ & 
$0.59850$ & $0$~~\\
    {\tt Y011X} & $C_1 = (m_1^1 : {\tt SU},\ m_1^2 : {\tt FL}) \quad C_2 = 
(m_2^1 : {\tt OP}) \quad C_3 = (m_3^1 : {\tt OP},\ m_3^2 : {\tt NA})$ & 
$0.25650$ & $0$~~\\
    {\tt Y0101} & $C_1 = (m_1^1 : {\tt SU},\ m_1^2 : {\tt FL}) \quad C_2 = 
(m_2^1 : {\tt OP}) \quad C_3 = (m_3^1 : {\tt FL},\ m_3^2 : {\tt OP})$ & 
$0.02280$ & $0$~~\\
    {\tt Y0100} & $C_1 = (m_1^1 : {\tt SU},\ m_1^2 : {\tt FL}) \quad C_2 = 
(m_2^1 : {\tt OP}) \quad C_3 = (m_3^1 : {\tt FL},\ m_3^2 : {\tt FL})$ & 
$0.00570$ & $1$~~\\
    {\tt Y1101} & $C_1 = (m_1^1 : {\tt SU},\ m_1^2 : {\tt OP}) \quad C_2 = 
(m_2^1 : {\tt OP}) \quad C_3 = (m_3^1 : {\tt FL},\ m_3^2 : {\tt OP})$ & 
$0.05320$ & $0$~~\\
    {\tt Y1100} & $C_1 = (m_1^1 : {\tt SU},\ m_1^2 : {\tt OP}) \quad C_2 = 
(m_2^1 : {\tt OP}) \quad C_3 = (m_3^1 : {\tt FL},\ m_3^2 : {\tt FL})$ & 
$0.01330$ & $1$~~\\
    {\tt 1X101} & $C_1 = (m_1^1 : {\tt OP},\ m_1^2 : {\tt NA}) \quad C_2 = 
(m_2^1 : {\tt OP}) \quad C_3 = (m_3^1 : {\tt FL},\ m_3^2 : {\tt OP})$ & 
$0.06080$ & $0$~~\\
    {\tt 1X100} & $C_1 = (m_1^1 : {\tt OP},\ m_1^2 : {\tt NA}) \quad C_2 = 
(m_2^1 : {\tt OP}) \quad C_3 = (m_3^1 : {\tt FL},\ m_3^2 : {\tt FL})$ & 
$0.01520$ & $1$~~\\
    {\tt 1X10Y} & $C_1 = (m_1^1 : {\tt OP},\ m_1^2 : {\tt NA}) \quad C_2 = 
(m_2^1 : {\tt OP}) \quad C_3 = (m_3^1 : {\tt FL},\ m_3^2 : {\tt SU})$ & 
$0.07600$ & $0$~~\\
    {\tt 1X1Y1} & $C_1 = (m_1^1 : {\tt OP},\ m_1^2 : {\tt NA}) \quad C_2 = 
(m_2^1 : {\tt OP}) \quad C_3 = (m_3^1 : {\tt SU},\ m_3^2 : {\tt OP})$ & 
$0.60800$ & $0$~~\\
    {\tt 1X1Y0} & $C_1 = (m_1^1 : {\tt OP},\ m_1^2 : {\tt NA}) \quad C_2 = 
(m_2^1 : {\tt OP}) \quad C_3 = (m_3^1 : {\tt SU},\ m_3^2 : {\tt FL})$ & 
$0.15200$ & $0$~~\\
    \cline{1-4}
    \end{tabular}
 \caption{\small Extracted Results over $\Upsilon_{sp}$ from {\sf Query2} ~~ ({\tt 
OP: Operating, FL: Failed, SU: Suspended, NA: Not-Availed}) \qquad \qquad \qquad [* Maximum Number
of Failures that can be Tolerated = $2$ (though failure in $C_2$ is not admissible!) ]}
 \label{tab:result_SP_query2}
}
\end{table}
Table~\ref{tab:result_SP_query1} and Table~\ref{tab:result_SP_query2} present 
the results extracted if we execute the same two queries as proposed in 
Example~\ref{ex:query_sqdl}. For {\tt Query1}, it may be found (from 
Table~\ref{tab:result_SP_query1}) that there is no meaningful (guaranteed) 
output quality when input quality is in the range $[30,40)$ and we have 
complete failure of $C_1$. However, the reliability can ranges upto 
$[0.985-0.999]$ under the no component failure assumption. Similarly, for 
{\tt Query2}, it is evident (from Table~\ref{tab:result_SP_query2}) that 
non-failure of $C_2$ is must with respect to the given constraints (primarily 
keeping the system reliability above $0.95$) in the query. Additionally, there 
are still some provisions of suspending at most one component yet satisfying 
the query requirements, even in case of failure in one of the components.

It may be further noted that, for $\Upsilon_{sp}$, the reliability computation 
is primarily dictated by $C_1$ and $C_2$ components. This is because of the 
instantiation/invocation of same component, $C_2$, present in two parallel 
paths from input to output -- which makes the system to only operate through 
$C_1$ in case of failure/suspension of $C_2$ (since $C_2$ failure closing the 
$C_3-C_2$ path too). Therefore, the computed reliability value of the overall 
system always comes from the computation of, $\big{[} Z_1(m_1^i) + 
Z_2(m_2^j) - Z_1(m_1^i).Z_2(m_2^j) \big{]}$, depending on the modes, $m_1^i$ 
and $m_2^j$ of $C_1$ and $C_2$ ($0 \leq i \leq 2,\ 0 \leq j \leq 1$), 
respectively, independent of the involvement of reliability values from 
$C_3$\footnote{The reliability of $\Upsilon_{sp}$ in every mode is computed as: 
(Here, $0 \leq i,k \leq 2$ and $0 \leq j \leq 1$)
\[ Z_{\Upsilon_{sp}}\big{(}m_1^i,m_2^j,m_3^k\big{)} ~~=~~ 1 - \big{[} 1 - 
Z_1(m_1^i) \big{]} . \big{[} 1 - Z_2(m_2^j) \big{]} . \big{[} 1 - 
Z_3(m_3^k).Z_2(m_2^j) \big{]} ~~=~~ \big{[} Z_1(m_1^i) + Z_2(m_2^j) - 
Z_1(m_1^i).Z_2(m_2^j) \big{]} \]}. 
This is also the reason why $C_2$ is critical to maintain high reliability for 
the operability of $\Upsilon_{sp}$, as also evident from the outcomes of {\tt 
Query1} and {\tt Query2}.

\

\begin{figure}
\centering
\includegraphics[width=\textwidth]{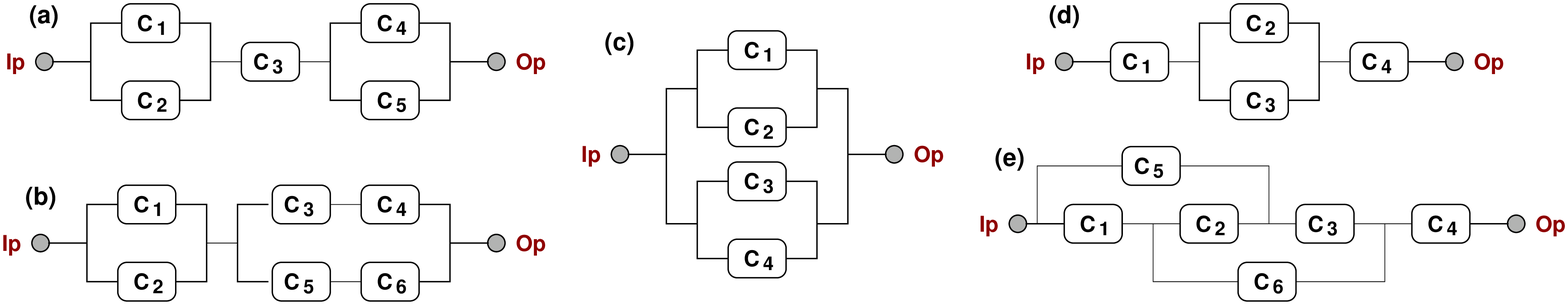}
\caption{Some General Composite System Structures Explored by our Framework}
\label{fig:system_explore}
\end{figure}

\noindent {\em Exploration over Miscellaneous System Structures:}
We also explored our proposed approach over some general system structures, 
five such composite system variations are shown in 
Figure~\ref{fig:system_explore}. We also tried different queries on these 
variations and found that our framework helps in analyzing the non-functional 
quality requirements. Moreover, each of the building block of these systems can 
also be a subsystem by itself and our framework is capable of handling such 
compositions, which can be primarily derived in a hierarchical manner. 

\section{Conclusion} \label{sec:conclusion}
In this work, we propose a novel framework (Figure~\ref{fig:framework}) 
to formally define and characterize the non-functional specifications, such as 
quality and reliability, of a component-based system and analyze the same. From 
the component-level quality configurations, we deduce system-level quality 
measures following a generic series and parallel composition techniques. We 
have further shown how to synthesize back the system-level non-functional 
specifications from the derived quality models of the composite system. We find 
interesting ramifications of such an extraction of non-functional 
specifications from an architectural level, since it may be used to assess the 
functional coverage that the quality measure achieves and it may also be 
correlated with the design intent (if specified) to check for its conformance 
with the desired and laid specifications. We also establish a query-driven 
analysis framework to extract relevant non-functional features of the overall 
system in greater details leveraging the internally derived quality 
configuration model. To the best of our knowledge, this is the first work that 
tries to provide a formal and automated analysis for compositional quality 
requirements from an architectural level of system design. We believe that such 
an assessment framework will be useful in formally certifying the 
non-functional requirements of component-based systems.

Additionally, the query-driven quality assessment counterpart equips the system 
designers presenting them deeper insights while exploring through the possible 
component-level configurations and thereby the designers may find it easier to 
explore and converge into the best possible system structures. In case that the 
desired system-level specification is not met by the derived composite 
structure, such an assessment tool, adding to our proposed compositional 
framework, will always help to understand the gap in terms of non-admissible 
quality configurations and/or prohibitive operational setups from its 
constituent components. As part of the future work, we shall explore the 
possibility to automatically bridge the gap by minimal addition or alteration 
to the existing composite structure. Besides, this being the maiden work, we 
have considered a rudimentary notion of quality which can be defined 
generically (and simply) in terms of some levels (as real values). In future, 
we plan to extend this definition of quality also as a function of component 
activity and interactions by possibly modeling the same involving logical 
properties. We also plan to integrate our proposed framework with industrial 
system design and validation flow and devise suitable metrics for quality and 
reliability coverage for comprehensive certification of component-based systems 
from architectural level.

\bibliographystyle{plain}
\bibliography{reference}

%
%
%

\end{document}